\documentclass[%
 reprint,
 amsmath,amssymb,
 aps,
]{revtex4-2}
\usepackage{subcaption}
\usepackage{color,soul}
\usepackage{graphicx}
\usepackage{dcolumn}
\usepackage{bm}
\usepackage{framed,graphicx,xcolor}

\begin{document}
\definecolor{shadecolor}{rgb}{1.0,1.0,0} 
\preprint{APS/123-QED}

\title{Geometric Thermodynamics of Strain-Induced \\Crystallization in Polymers}

\author{Sanhita~Das}
 \email{sanhita.das@weizmann.ac.il}
 \affiliation{Department of Chemical and Biological Physics, \\Weizmann Institute of Science, Israel}
\author{Asif~Raza}%
 \email{asifraza@iisc.ac.in}
\author{Debasish~Roy}%
 \email{royd@iisc.ac.in}
 \altaffiliation[Also at ]{Centre of Excellence on Advanced Mechanics of Materials, Indian Institute of Science.}
\affiliation{%
 Computational Mechanics Laboratory, Department of Civil Engineering,\\
Indian Institute of Science, Bangalore, 
}%

\date{\today}

\begin{abstract}
Going beyond the classical Gaussian approximation of Einstein's fluctuation theory, Ruppeiner gave it a Riemannian geometric structure with an entropic metric. This yielded a fundamental quantity -- the Riemannian curvature, which was used to extract information on the nature of interactions between molecules in fluids, ideal gases and other open systems. In this article, we examine the implications of this curvature in a non-equilibrium thermodynamic system where relaxation is sufficiently slow so as not to invalidate the local equilibrium hypothesis. The non-equilibrium system  comprises of a rubbery polymer undergoing strain induced crystallization. The curvature is found to impart information on a spurious isochoric energy arising from the conformational stretching of already crystallized segments. This unphysical component perhaps arises as the crystallized manifold is considered Euclidean with the stretch measures defined via the Euclidean metric.  The thermodynamic state associated with curvature is the key to determine the isochoric stretch and hence the spurious energy. We determine this stretch and propose a form for the spurious free energy that must be removed from the total energy in order that the correct stresses are recovered. 
\end{abstract}

\maketitle


\section{Introduction}\label{SIC_Introduction}

{Contrary to the notion that thermodynamics describes only the macroscopic material behaviour, it also affords information on the microstates through the fluctuation theory} (Ref.~\cite{ruppeiner2010thermodynamic}). The Gaussian approximation to the thermodynamic fluctuation theory yields the probability of finding a system in a given thermodynamic state close to equilibrium (Ref.~\cite{ruppeiner2010thermodynamic}).
 However, the Gaussian approximation holds if the volume of the system is very large vis-\'a-vis fluctuations in the 
 intensive variables. As pointed out by  \cite{ruppeiner1995riemannian}, when such conditions are not met, there arises issues of inconsistency and lack of covariance (i.e. coordinate dependence). Depending on the scenario, there is thus a need to go beyond the Gaussian approximation  (Refs.~\cite{ruppeiner1995riemannian,ruppeiner2010thermodynamic}). Distributions with characteristics free of these shortcomings are shown to satisfy a diffusion equation involving coefficients, which necessarily satisfy the metric transformation rules of Riemannian geometry. This indicates that the thermodynamic manifold generated by the possible microstates is Riemannian. 

\cite{ruppeiner1979thermodynamics} has devised an approach to introduce a Riemannian structure to the classical fluctuation theory, thereby non-trivially modifying it. It also highlights a new quantity: the thermodynamic Riemannian curvature, which emerges from the Riemannian structure of the thermodynamic manifold. This quantity contains rich information on the underlying fluctuations. Subsequently, \cite{ruppeiner2010thermodynamic,ruppeiner2015thermodynamic,may2013thermodynamic,ruppeiner2008thermodynamic,ruppeiner2021thermodynamic,ruppeiner2020thermodynamic} employ this geometrically motivated thermodynamic fluctuation theory to several open thermodynamic systems such as ideal gases, paramagnets, Van der Waals gases, fluids with single or multiple components. In fluids, the curvature is strongly correlated with the nature of interactions with a reversal in sign through critical values.
 
 The implications of the thermodynamic fluctuation theory in polymers have always been of interest to mechanicians (Ref.~\cite{bu2016scattering}) especially for semiflexible polymers. For flexible polymers where chains can be microscopically idealised as comprising of pin-jointed freely rotating rigid links, we may specifically adopt the affine three-chain network model by \cite{arruda1993three} which relates the microscopic and macroscopic free energies. In this case, the macroscopic free energy in response to the hyperelastic deformation typically observed in such polymers assumes a neo-Hookean form. The thermodynamic fluctuation theory in such a case predicts a Gaussian probability distribution for the effective average macroscopic stretch. The case is however not so simple for semi-flexible polymers where the constituting monomers impart bending, extensional or torsional stiffnesses to the links that were assumed rigid for flexible polymers. Also constraints such as entanglements or cross-links demand the use of non-affine network models. Moreover, if the macroscopic deformation induces additional inelastic mechanisms such as plasticity, damage or strain-induced crystallization, additional free energy components due to the formation of defects or appearance of inelastic dissipation need to be considered. The total free energy is then not only a function of the macroscopic or microscopic stretch, but also of other variables, e.g. internal variables, describing the particular inelastic phenomenon. The information afforded by a fluctuation theory on the underlying microstates should shed light on the underlying mechanics. It would also be of interest to see what additional information the scalar curvature described in \cite{ruppeiner2010thermodynamic,ruppeiner2015thermodynamic,may2013thermodynamic,ruppeiner2008thermodynamic,ruppeiner2021thermodynamic,ruppeiner2020thermodynamic} provides. If it were to show criticality at some particular themodynamic state, then the natural question is what we could say about the process itself. This might offer insights into whether we might put this criticality to some advantage. Finally, Ruppeiner limited his analysis to purely equilibrium systems. The important issue of extending this useful construct to systems undergoing non-equilibrium changes also remains unresolved at this time. 
 
 In an attempt to partially answer some of these questions pertaining to polymer inelasticity, we explore the implications of a Riemannian structure to the fluctuation theory consistent with \cite{ruppeiner1979thermodynamics} and others. Specifically, we rely on the emerging curvature to reveal the microscale information of a thermodynamic system describing strain induced crystallization (SIC) in a rubber specimen. SIC is a non-equilibrium phenomenon in which regions of a rubber sample undergo crystallization when stretched. At the molecular level, polymer chain segments in these regions orient themselves in the direction of locally developed tension and weak intermolecular bonds are formed in between adjacent straightened segments, establishing a relatively long range order in the region. On removal of deformation or in the event of a reversal of the local tensile stress, the segments fall into a state of disorder and crystallization is lost. 
 
 Strain-induced crystallization is intrinsically a non-equilibrium phenomenon; but it is still possible to exploit Ruppeiner's approach as the characteristic timescale of the process is sufficiently large. In  \cite{tosaka2006, tosaka2007}, it is observed that the total timescale associated with crystallization is 4 sec. The total timescale includes timescale for nucleation and that for the growth of crystallites. Furthermore,  \cite{bruning2015kinetics} predict the total timescale in vulcanised natural crepe rubber to range from 10-100 secs depending on the ambient temperature. According to them, the growth phase in crystallization is strongly dependent on the mobility of the chains and therefore the temperature and the strain rate. Temperature and strain-rate therefore have a strong influence on the overall rate of crystallization and only for very low temperatures (lower than -25$^{\circ}$C) and large strain rates, the rate of crystallization is high which translates to a low characteristic timescale. In the range of temperatures considered in this article, the characteristic timescale is orders above that for atomistic vibrations. Also, since the crystallized segments constitute mesoscale structures which are orders larger than molecular dimensions, we can characterise the partially crystallized material using local thermodynamic states. Both these conditions permit us (see ~\cite{tadmor2012continuum, beris1994thermodynamics, ottinger2005beyond,j2007statistical}) to assume that the local equilibrium hypothesis is valid, thus allowing for an exploitation of the underlying axioms of equilibrium thermodynamics whilst modelling the evolving crystallization using internal state variables. We further assume that crystallization takes place isotropically so every chain in the RVE (Ref.~\cite{arruda1993three}) undergoing SIC and macroscopically represented by a material point has the same 'degree of crystallization' (defined as the fraction of chain segments already crystallized). The free energy of such a partially crystallized specimen is derived on the lines of \cite{rastak2018non,flory1947thermodynamics, mistry2014micro}. 
 
We then define the thermodynamic manifold for the system in local equilibrium, where the coordinates are given by the state variables that fluctuate. Then the metric, the affine connection and the invariant scalar curvature are derived. The variation in scalar curvature over the manifold is analyzed for any critical points and the physical significance of these points is assessed from their relevance to the total free energy density and the strain energy density. Finally, we investigate the possible exploitation of this kinematically relevant information to redefine the free energy density. The critical points as it turns out correspond to states with unphysical conformational stretches. We infer that, by not accounting for the non-Euclidean geometric aspects imparted by crystallization, earlier constitutive theories (\cite{rastak2018non,mistry2014micro}) yield stretch measures incapable of eliminating components due to the already crystallized segments. We identify these stretches and thus propose a form for the spurious energy to be eliminated from the total free energy.

The rest of the article is organized as follows. We provide definitions and some background information on Riemannian thermodynamic manifold, its metric and curvature in section \ref{SIC_RiemannianGeometry}. This is followed in section \ref{SIC_FreeEnergy} by the definition of the free energy of a polymer undergoing SIC. The free energy is then used to determine the metric and the curvature in section \ref{SIC_CurvatureTensionShear}. Analytical computation of geometric quantities in a seven or eight dimensional manifold is immensely complicated. Hence, to simplify calculations, we consider two cases of deformation - uniaxial tension and simple shear. Coordinates corresponding to the components of the right Cauchy-Green tensor may thus be expressed as functions of fewer coordinates representing the uniaxial tensile stretch in an incompressible or a compressible material or the magnitude of shear strain in an incompressible material. This section also describes the physical implication of curvature through a comparison of its variation over the manifold with those of free energy and strain energy densities. Section \ref{SIC_Modified_Free_Energy} defines the spurious isochoric free energy component as a function of the strain determined in the earlier section. The total free energy is also redefined after a removal of this spurious energy. The article is finally concluded in section \ref{Conclusion}. 

\section{Geometry of Riemannian Thermodynamic Manifolds}\label{SIC_RiemannianGeometry}

The basic geometry is based on two fundamental axioms of equilibrium thermodynamics (Ref.~\cite{ruppeiner1979thermodynamics}). 

\begin{itemize}
\item In any thermodynamic system with a fixed scale,
there exist equilibrium states which can be represented by
points on a higher-dimensional manifold which is differentiable everywhere considering no phase transitions and critical points. The coordinates associated with the manifold are the independent state variables describing the equilibrium states.
\item On the manifold, there exists a positive definite Riemannian metric $\mathbf{g}$, which is determined (at any point on the manifold) by the condition that its components in a particular coordinate system are the
second moments of fluctuations of the thermodynamic states. 
\end{itemize}.

Therefore, for a general thermodynamic system described by the state variables $x_i$s, a  metric tensor $\mathbf{g}$ which satisfies the requirements of covariance and consistency, must have its components of the form, 
\begin{equation}\label{MetricinEntropy}
g_{ij}=-\dfrac{1}{k_{B}}\frac{\partial^{2}{\eta}}{\partial{x_{i}}\partial{x_{j}}}
\end{equation}
$\eta$ in the above expression is the total entropy. The metric may also be expressed via the free energy. For a three dimensional thermodynamic system described by states ($x_1$, $x_2$, T), in terms of the specific free energy $\Psi$, the metric takes the form \cite{ruppeiner2020thermodynamic}

\begin{equation}\label{MetricinFreeEnergy}
g_{ij} = \dfrac{1}{k_{B}T}
\begin{bmatrix}
-\dfrac{\partial^2{\Psi}}{\partial{T}^2} & 0 & 0 \\
0 & \dfrac{\partial^2{\Psi}}{\partial{x_1}^2} & \dfrac{\partial^2{\Psi}}{\partial{x_1}\partial{x_2}} \\
0 & \dfrac{\partial^2{\Psi}}{\partial{x_2}\partial{x_1}} & \dfrac{\partial^2{\Psi}}{\partial{x_2}^2}
\end{bmatrix}
\end{equation}
where $T$ is the absolute temperature. 
The entropic metric may also be given by,

\begin{equation}\begin{split}\label{Entropicmetric}
(\Delta l)^2 = &-\dfrac{1}{k_{B}T}\left(\dfrac{\partial^2{\Psi}}{\partial{T}^2}\right)(\Delta T)^{2} 
\\&+ \dfrac{1}{k_{B}T}\sum_{i,j = 1}^{2}\left(\dfrac{\partial^2{\Psi}}{\partial{x_{i}}\partial{x_{j}}}\right)\Delta x_{i}\Delta x_{j}
\end{split}\end{equation}
The Christoffel symbols $\mathbf{\Gamma}_{ij}^{k}$ associated with the affine connection are given by,

\begin{equation}\label{ChristoffelSymbols}
\mathbf{\Gamma}_{ij}^{k} = g^{km}\frac{1}{2}\left[\frac{\partial{g_{im}}}{\partial{x^{j}}}+\frac{\partial{g_{jm}}}{\partial{x^{i}}}-\frac{\partial{g_{ij}}}{\partial{x^{m}}}\right]
\end{equation}
These are then used to calculate the Riemannian curvature tensor $\Tilde{\mathbf{R}}$ whose components may be expressed as,

\begin{equation}\label{RiemannianCurvatureTensor}
\Tilde{R}_{ijk}^{l} = \frac{\partial}{\partial X^{i}}\mathbf{\Gamma}_{jk}^{l}-\frac{\partial}{\partial X^{j}}\mathbf{\Gamma}_{ik}^{l}+\mathbf{\Gamma}_{im}^{l}\mathbf{\Gamma}_{jk}^{m}-\mathbf{\Gamma}_{jm}^{l}\mathbf{\Gamma}_{ik}^{m}
\end{equation}
Upon contraction, we have the symmetric Ricci tensor $\hat{\mathbf{R}}$,
\begin{equation}\label{RicciCurvatureTensor}
\hat{R}_{ij} = \Tilde{R}_{ikj}^{k}
\end{equation}
as well as the Ricci scalar curvature $R$:
\begin{equation}\label{RicciScalar}
R=g^{ij}\Tilde{R}^{k}_{ikj}
\end{equation}
where $g^{ij}$ is the inverse (or the contravariant form) of the metric $g_{ij}$.

\section{Free Energy of a Partially Crystallized Polymer}\label{SIC_FreeEnergy}

On a continuum level, let $\mathbf{\varphi : X \mapsto x}$ be a diffeomorphism that maps material points $\mathbf{X}\ \in\ \mathcal{B}$ of the reference configuration $\mathcal{B} \subset \mathbb{R}^{3}$ to points $\mathbf{x = \varphi(X;t)} \ \in \ \mathcal{S} $ of the current configuration $\mathcal{S} \subset \mathbb{R}^{3}$. Define the deformation gradient as $\mathbf{F} := \mathbf{\nabla\varphi(X;t)}$ with Jacobian $J := \det{\mathbf{F}} > 0$. The polymer is considered slightly compressible; therefore  $\mathbf{F}$ is decomposed into volumetric and isochoric parts denoted respectively by $\mathbf{F}_{vol}$ and $\mathbf{\bar{F}}$: (Ref.~\cite{rastak2018non}). 
\begin{equation}
\mathbf{F} = \mathbf{F}_{vol}\mathbf{\bar{F}} \ \ \text{where}\ \mathbf{F}_{vol} = J^{1/3} \mathbf{I} \ \ \text{and} \  \mathbf{\bar{F}} = J^{-1/3}\mathbf{F}
\end{equation}
If $\mathbf{C}$ denotes the right Cauchy-Green tensor, its isochoric component is given by
\begin{equation}
    \mathbf{\bar{C}} = J^{-2/3}\mathbf{C}
\end{equation}
The total free energy density is considered as a function of $\mathbf{C}$, the ambient temperature $T$ and the internal variable $\Omega$, which is a macroscopic measure of the degree of crystallinity in the material:
\begin{equation}
\Psi = \Psi(\mathbf{C},\Omega,T)
\end{equation}
In particular, $\Omega$ is the ratio of the volume of crystallized polymeric segments to the total volume of the RVE. Thus, if $N_T$ is the total number of polymeric segments in the RVE, the total number of crystallized segments in the RVE is $\Omega N_T$. 
The total free energy density may be written as the sum of a strain energy consisting of volumetric and isochoric parts, a crystalline free energy and the energy due to surrounding heat interactions,
\begin{equation} \label{FreeEnergyDecomposition}
\Psi = \Psi_{strain}(\mathbf{\bar{C}},\Omega,T) + \Psi_{cr}(\Omega,T) + \Psi_{surr}(T)
\end{equation}
As noted, the strain energy $\Psi_{strain}$ comprises of a deviatoric component and a volumetric component:
\begin{equation} \label{StrainEnergyDecomposition}
\Psi_{strain} = \Psi_{vol}(J)+ \Psi_{dev}(\mathbf{\bar{C}},\Omega,T) 
\end{equation}
where the volumetric free energy $\Psi_{vol}$ is given by,
\begin{equation}\label{volumetricfreeenergy}
\Psi_{vol} = \frac{1}{2}B(J-1)^{2}
\end{equation}
and $B$ is the bulk modulus.
The isochoric strain energy component is the average of the strain energy of all the chains in the RVE. Mesocopically speaking, the chains might have different stretches and orientations in response to the deformation $\mathbf{C}$. Ideally, a cubic network of eight chains with orientations along the diagonals and the sides of the cube is used to compute an average stretch $\lambda_a$, which is assigned to every chain in the RVE. If $n$ is the number of the equivalent chains per unit volume of the RVE with stretch $\lambda_a$, the total isochoric strain energy density $\Psi_{dev}$ may be related to the strain energy $\Psi$ in each equivalent chain by the following relation.
\begin{equation} \label{eq:micro-macro isochoric free energy}
\Psi_{dev}(\mathbf{C},\Omega,T) = n\Psi(\lambda_a, \Omega,T)
\end{equation}
The average stretch $\lambda_a$ of the equivalent chain is given by,
\begin{equation} \label{eq:trace of cbar}
\lambda_a = \sqrt{I_{1}/3} \ \ \text{where} \ \ I_{1} = \text{tr}[\bar{\mathbf{C}}]
\end{equation}
The strain energy $\Psi$ in each equivalent chain is given by
\begin{equation}\begin{split}\label{isochoricfreeenergychain}
\Psi = &k_{B}TN(1-\Omega)\\&\times \left[\dfrac{3}{2}\left(\frac{\lambda_a}{\sqrt{N}}\right)^2+\dfrac{9}{20}\left(\frac{\lambda_a}{\sqrt{N}}\right)^4+\dfrac{99}{350}\left(\frac{\lambda_a}{\sqrt{N}}\right)^6\right]
\end{split}\end{equation}
where $N$ is the number of segments in each equivalent chain. This particular form is derived from the entropy generated due to segmental conformations in the chain (Ref.~\citep{ward2012mechanical}). It is important to note that $\lambda_a$ in the above expression is the stretch in an uncrystallized chain and needs be modified to include the effects of crystallization. If the chain is partially crystallized, only the segments in the uncrystallized zone are free to entropically stretch or compress in response to deformation. In accordance with \cite{mistry2014micro, rastak2018non}, we introduce a modified total stretch $\Lambda_a$ in a partially crystallized equivalent chain as $\Lambda_a = \frac{\lambda_a/\sqrt{N}-\Omega}{1-\Omega}$.
Considering the modified stretch $\Lambda_a$ and the relation in Eq.~\eqref{eq:micro-macro isochoric free energy}, the isochoric strain energy assumes the form.
\begin{equation}\begin{split}\label{Totalfreeenergyisochoric}
\Psi_{dev} &= k_{B}TNn(1-\Omega)\left(\dfrac{3}{2}\Lambda_a^2+\dfrac{9}{20}\Lambda_a^4+\dfrac{99}{350}\Lambda_a^6\right)
\end{split}\end{equation}

Now we focus on the remaining components of the free energy density. The crystalline free energy is directly adopted from Refs.~\cite{rastak2018non, mistry2014micro, flory1953principles}. 

\begin{equation}\begin{split}\label{Crystallinefreeenergy}
\Psi_{cr} &= -n c\left(1-\frac{T}{T_m}\right)\Omega-n\xi\left[\frac{\Omega}{\Omega_{max}}+\ln{\left(1-\frac{\Omega}{\Omega_{max}}\right)}\right]
\end{split}\end{equation}
The first term on the right hand side above is the enthalpy of fusion of crystallized segments (Ref.~\cite{flory1947thermodynamics}), where $c$ is the heat of fusion per unit segment and $T_m$ is the melting point of the segments. The second term is a penalty term that prevents the crystallization from growing unbounded, i.e. beyond $\Omega_{max}$. The remaining component, i.e. the surrounding energy, is taken from \cite{anand1996constitutive}.
\begin{equation}\begin{split}\label{Surroundingfreeenergy}
\Psi_{surr} &= c_{V}\left(T-T_{0}-T\ln{\dfrac{T}{T_{0}}}\right)
\end{split}\end{equation}
$c_V$ is the volumetric heat capacity and $T_0$ a reference temperature. $c_V$ in rubbers may be considered temperature independent \cite{anand1996constitutive, guo2018thermo,chadwick1974thermo}. Moreover since the ambient temperatures considered in the numerical experiments are above the glass transition temperature of the material, such an assumption does make sense.  
The total free energy density therefore assumes the following form,
\begin{widetext}
\begin{equation}\begin{split}\label{Totalfreeenergy}
\Psi = \frac{1}{2}B(J-1)^{2}+k_{B}TnN(1-\Omega)\left(\dfrac{3}{2}\Lambda_a^2+\dfrac{9}{20}\Lambda_a^4+\dfrac{99}{350}\Lambda_a^6\right)  - n c \left(1-\frac{T}{T_m}\right)\Omega-n\xi \left[\frac{\Omega}{\Omega_{max}}+\ln{\left(1-\frac{\Omega}{\Omega_{max}}\right)}\right]\\+c_{V}\left(T-T_{0}-T\ln{\dfrac{T}{T_{0}}}\right)
\end{split}\end{equation}
\end{widetext}
The constitutive equation for the second Piola-Kirchhoff stress is given as 
\begin{widetext}
\begin{equation}\begin{split}\label{SecondPiolaKirchhoff}
\mathbf{S} = 2\dfrac{\partial{\Psi}}{\partial{\mathbf{{C}}}} = \frac{B}{2}(J-1)J\mathbf{C}^{-1} + \frac{k_B T n \sqrt{N}}{3\lambda_a} J^{-2/3}   (1-\Omega) \left(3 \Lambda_a +\frac{9}{5}\Lambda_a^3 +\frac{594}{350}\Lambda_a^5 \right) \left(\mathbf{I}-\dfrac{\text{tr}[\mathbf{C}]\mathbf{C}^{-1}}{3} \right) 
\end{split}\end{equation}
\end{widetext}
The Cauchy stress is related to $\mathbf{S}$ by,
\begin{equation}
    \mathbf{\sigma} = \frac{\mathbf{F} \mathbf{S} \mathbf{F}^T}{J}
\end{equation}

\begin{figure*}
     \centering
    \includegraphics[width=0.98\textwidth]{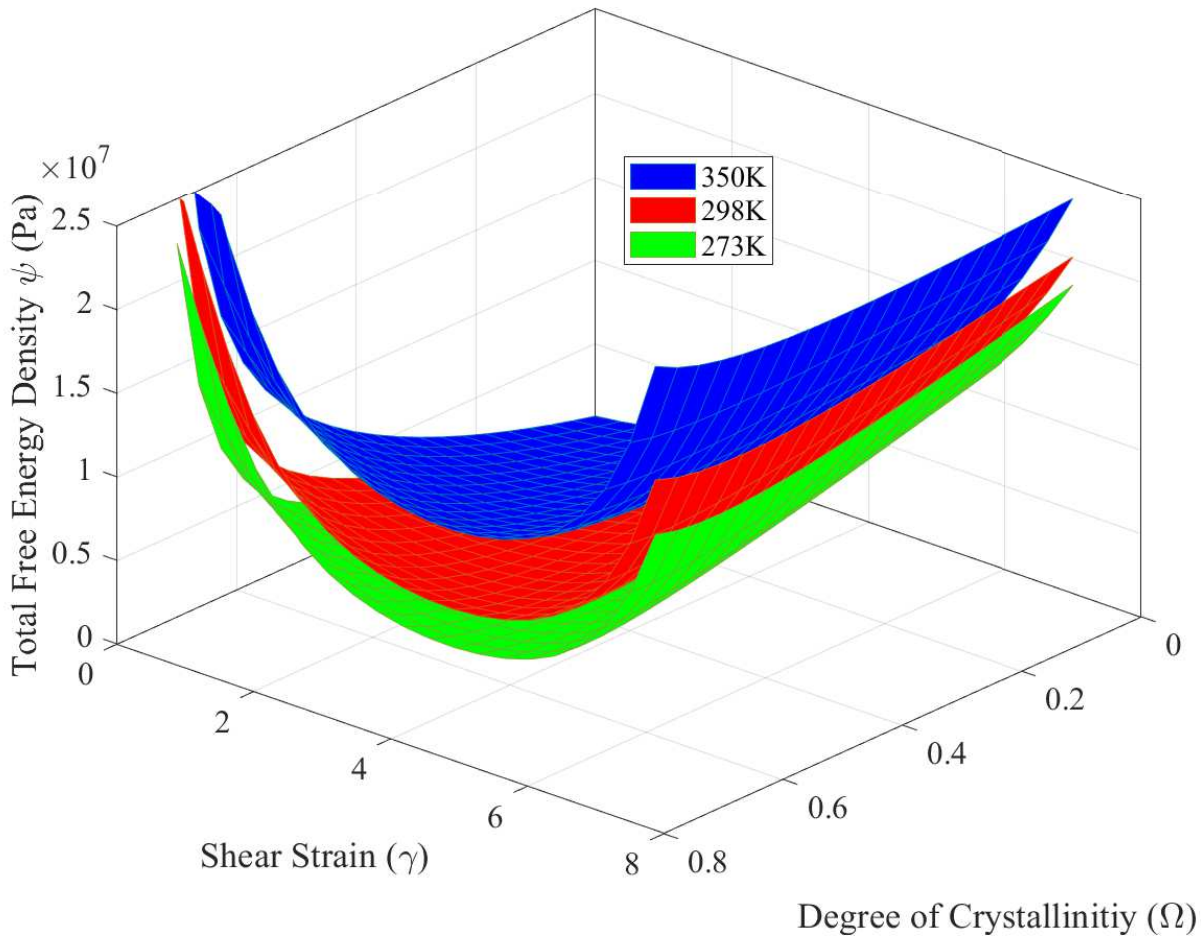}
    \caption{Free energy surfaces for incompressible material under simple shear and different temperatures $T$}
    \label{FreeEnergy3D}
\end{figure*}

\section{Thermodynamic Geometry of Crystallized Polymer}\label{SIC_CurvatureTensionShear}

Let us consider the temporally fluctuating thermodynamic state $(\mathbf{C}, \Omega, T)$ of the deformed, crystallized polymer. It is important to note that the timescale associated with the fluctuations is orders larger than that associated with crystallization. This enables us to invoke the principle of local equilibrium. The thermodynamic manifold is eight dimensional with coordinates $(C_{11},C_{12},C_{13},C_{22},C_{23},C_{33},\Omega,T)$. The metric and curvature of the system given by Eq.~\eqref{MetricinFreeEnergy} and Eq.~\eqref{RicciScalar} may be expressed in the above coordinates.

An attempt to consider fluctuations in all the eight quantities faces the difficulty that the metric and curvature tensors are prohibitively complex -- both to compute and to interpret. This necessitates a reduction in the dimension of the thermodynamic phase space. Thus we analyze the system for three specific states of deformation - simple shear, uniaxial tension with material incompressibility and uniaxial tension allowing for a compressible material. The values for the following material parameters appearing in equations \eqref{SecondPiolaKirchhoff} have been adopted from \cite{rastak2018non} for sulphur cured natural rubber; $N=18.5$, $n=8.99\times10^{25}$m$^{-3}$, $c_v=1.8\times10^3$Nm$^{-2}$K$^{-1}$, $T_0=273$K, $\Omega_{max}=0.78$, $c=3.5409\times10^{-19} N-m$ and $\xi=1.11\times10^{-20}N-m$. $B=3.576\times10^{7}$Pa is the value obtained considering the Poisson's ratio to be 0.45. The value of $T_m=245$K has been taken from \cite{kobayashi2015encyclopedia}.   

\subsection{Simple Shear Deformation in \\a Hyperelastic Solid}\label{CurvatureShear}

\begin{figure*}
\begin{tabular}{cc}
     \centering
     \begin{subfigure}[b]{0.47\textwidth}
         \centering
         \includegraphics[width=0.7\textwidth]{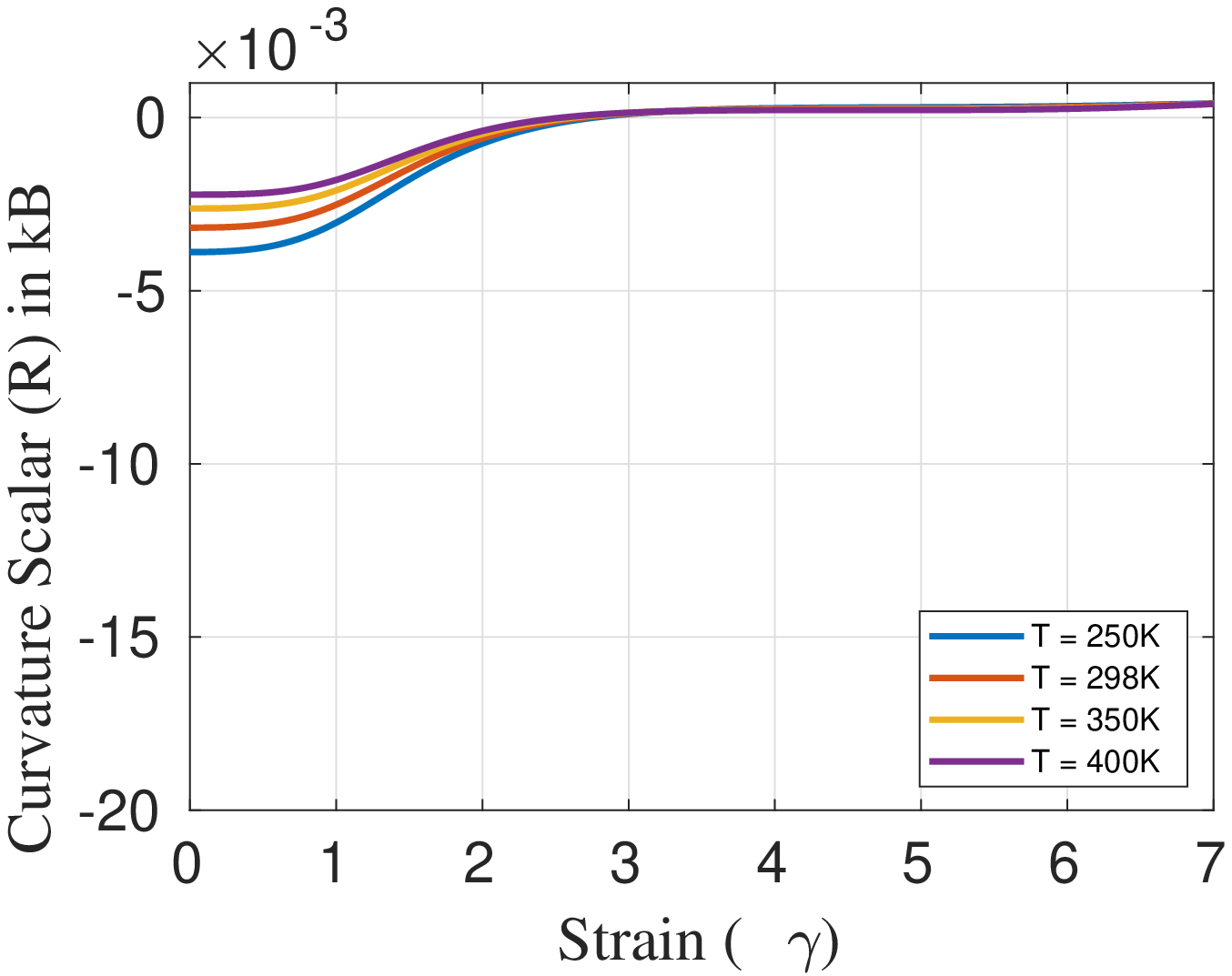}
         \caption{$\Omega = 0$}
         \label{Curvaturevsshearfortemperaturea}
     \end{subfigure} & 
     \begin{subfigure}[b]{0.47\textwidth}
         \centering
         \includegraphics[width=0.7\textwidth]{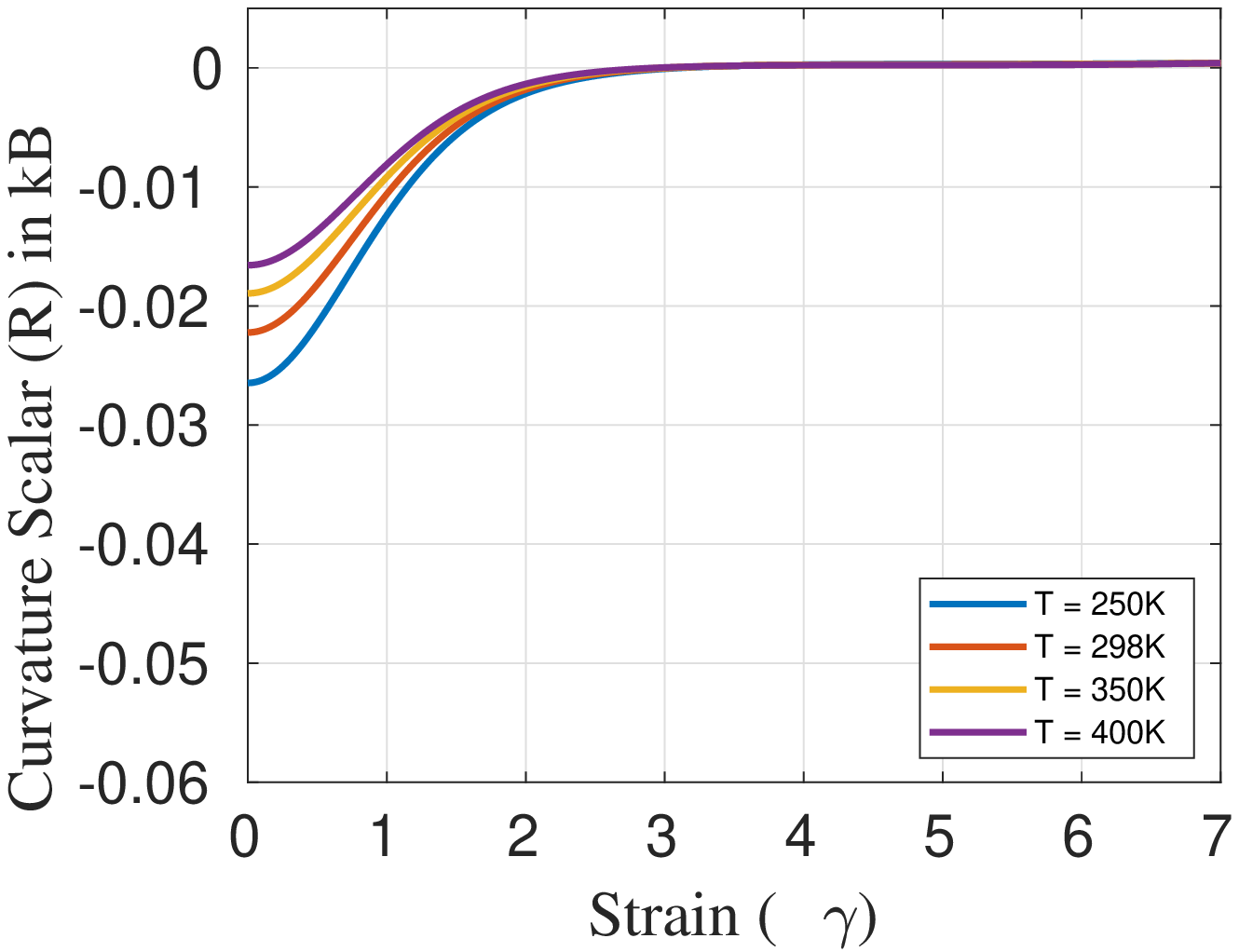}
         \caption{$\Omega = 0.15$}
         \label{Curvaturevsshearfortemperatureb}
     \end{subfigure} \\
     \begin{subfigure}[b]{0.47\textwidth}
         \centering
         \includegraphics[width=0.7\textwidth]{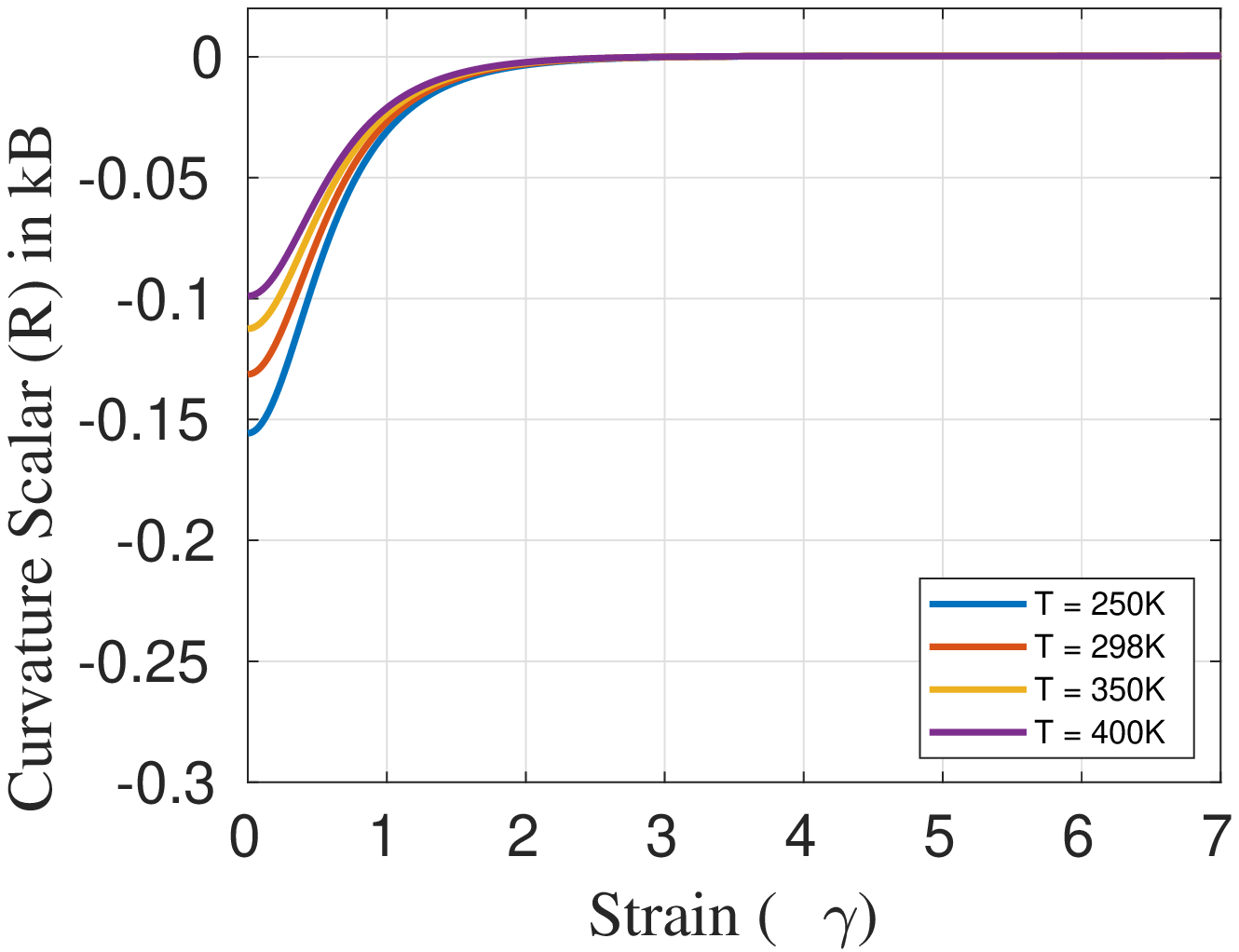}
         \caption{$\Omega = 0.20$}
         \label{Curvaturevsshearfortemperaturec}
     \end{subfigure} &
     \begin{subfigure}[b]{0.47\textwidth}
         \centering
         \includegraphics[width=0.7\textwidth]{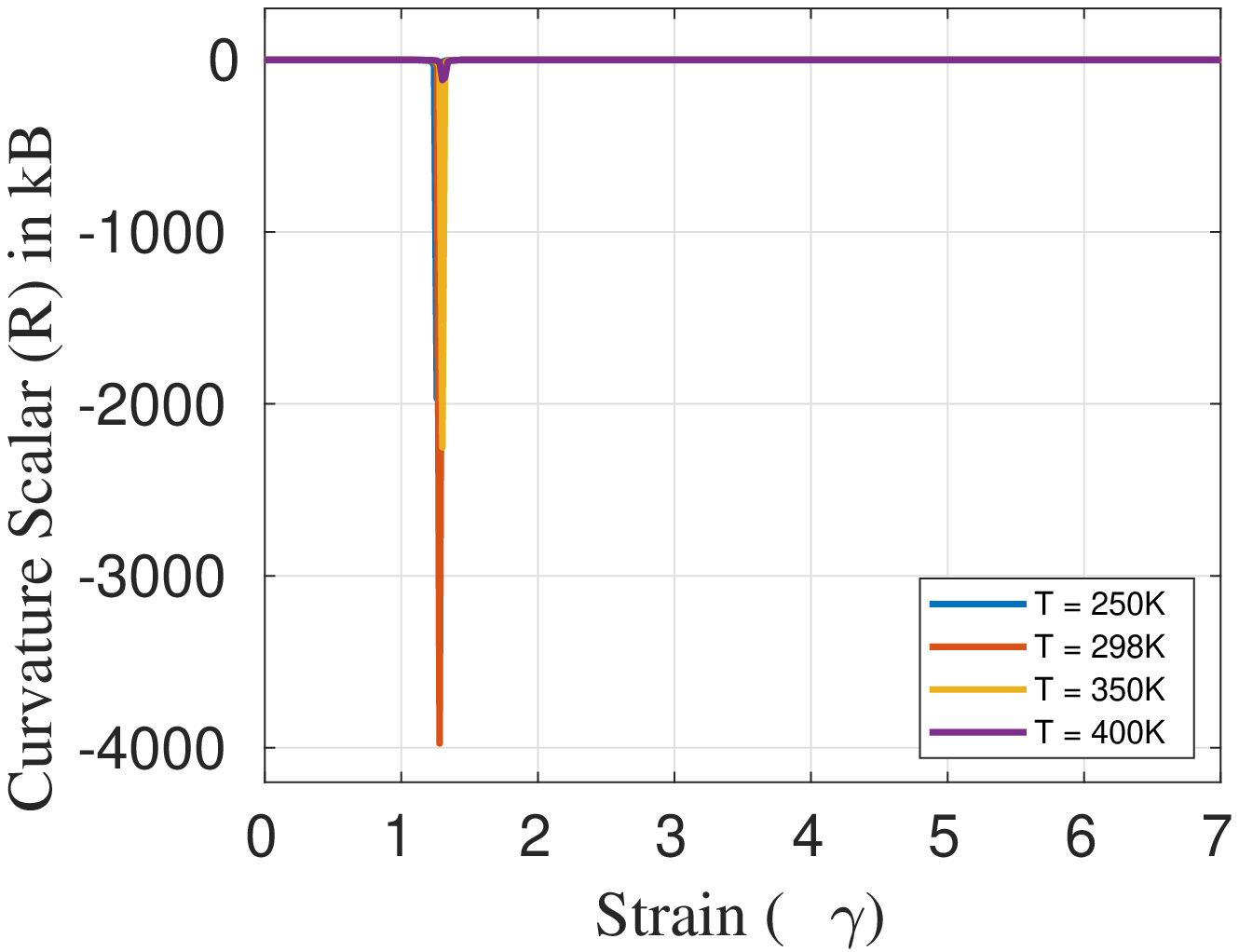}
         \caption{$\Omega = 0.30$}
         \label{Curvaturevsshearfortemperatured}
     \end{subfigure} \\
     \begin{subfigure}[b]{0.47\textwidth}
         \centering
         \includegraphics[width=0.7\textwidth]{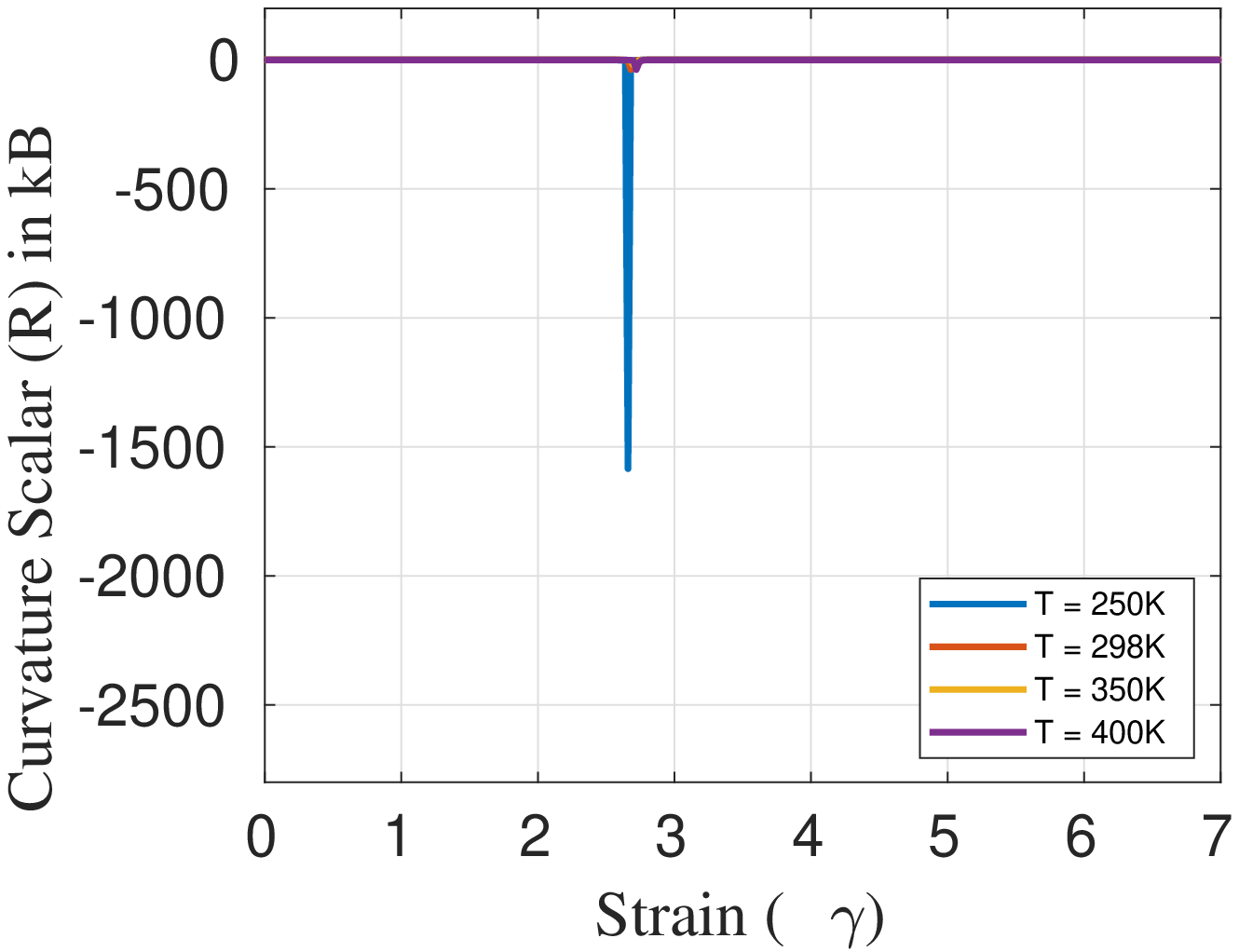}
         \caption{$\Omega = 0.45$}
         \label{Curvaturevsshearfortemperaturee}
     \end{subfigure}   &
     \begin{subfigure}[b]{0.47\textwidth}
         \centering
         \includegraphics[width=0.7\textwidth]{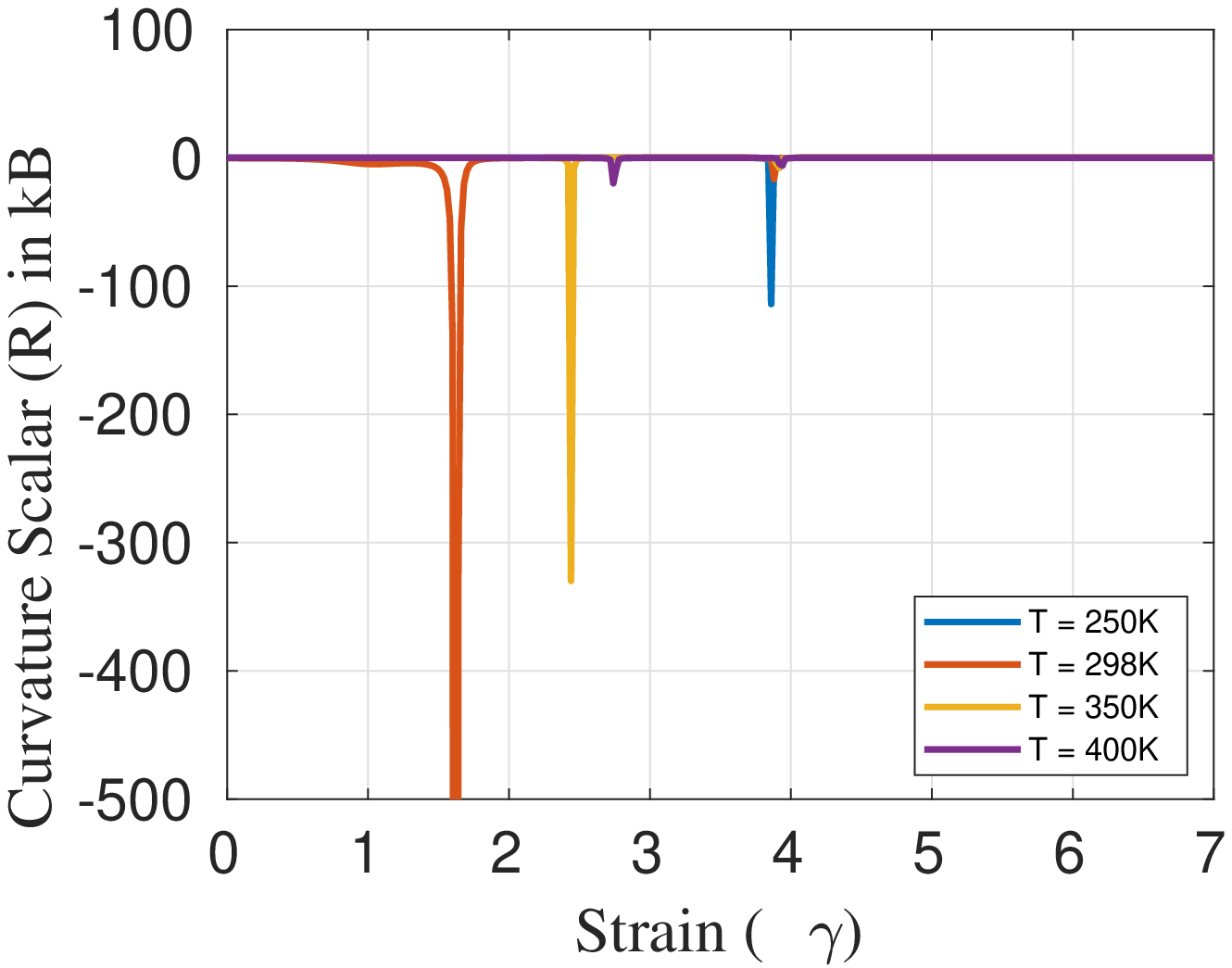}
         \caption{$\Omega = 0.60$}
         \label{Curvaturevsshearfortemperaturef}
     \end{subfigure}
\end{tabular}
     \begin{subfigure}[b]{0.47\textwidth}
         \centering
         \includegraphics[width=0.7\textwidth]{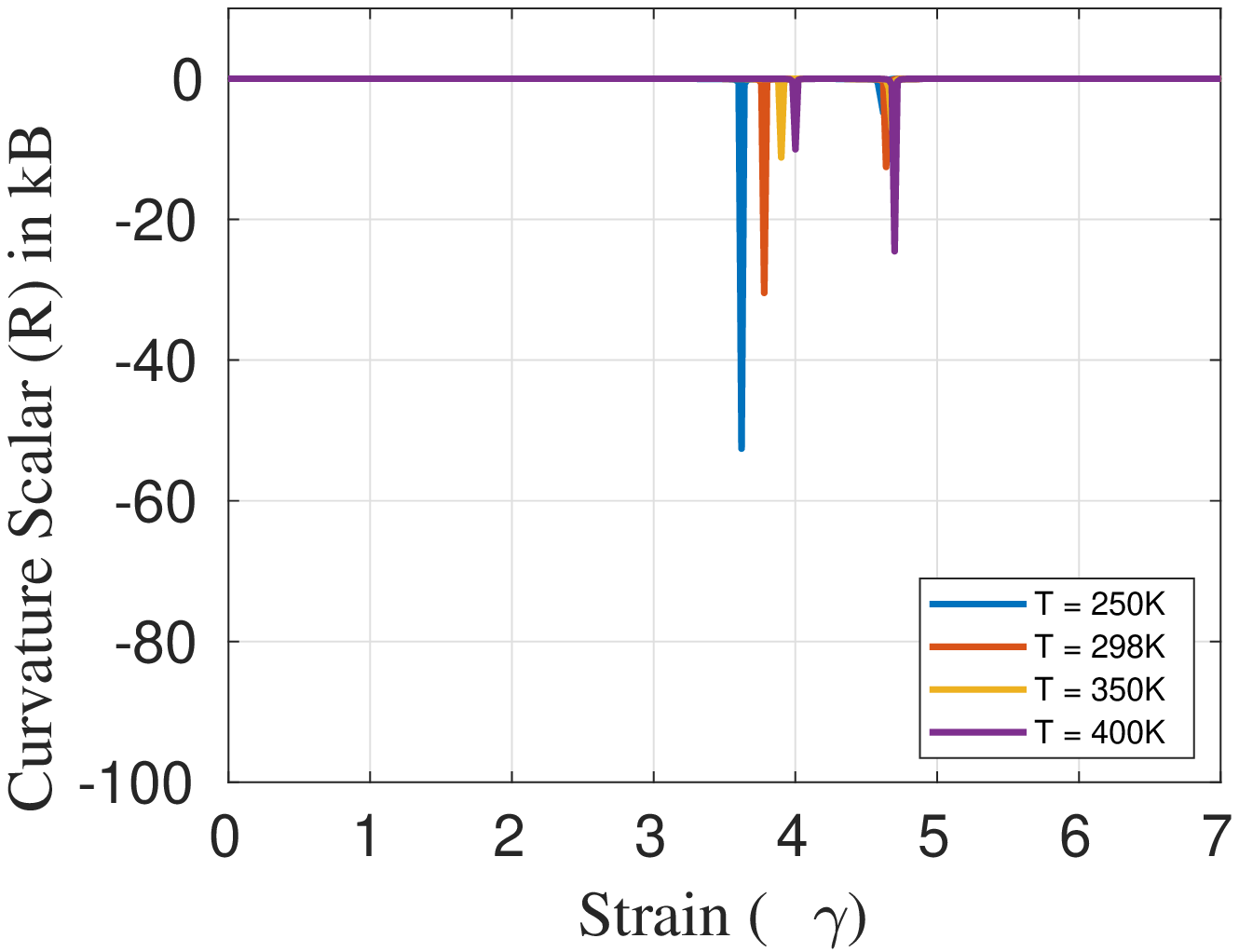}
         \caption{$\Omega = 0.70$}
         \label{Curvaturevsshearfortemperatureg}
     \end{subfigure} 
\caption{Plots of scalar curvature $R$ with shear $\gamma$ for varying temperature and crystallization ratio}
\label{CurvaturNonisothermalSheartemperature}
\end{figure*}

\begin{figure*}
\begin{tabular}{cc}
     \centering
     \begin{subfigure}[b]{0.47\textwidth}
         \centering
         \includegraphics[width=0.7\textwidth]{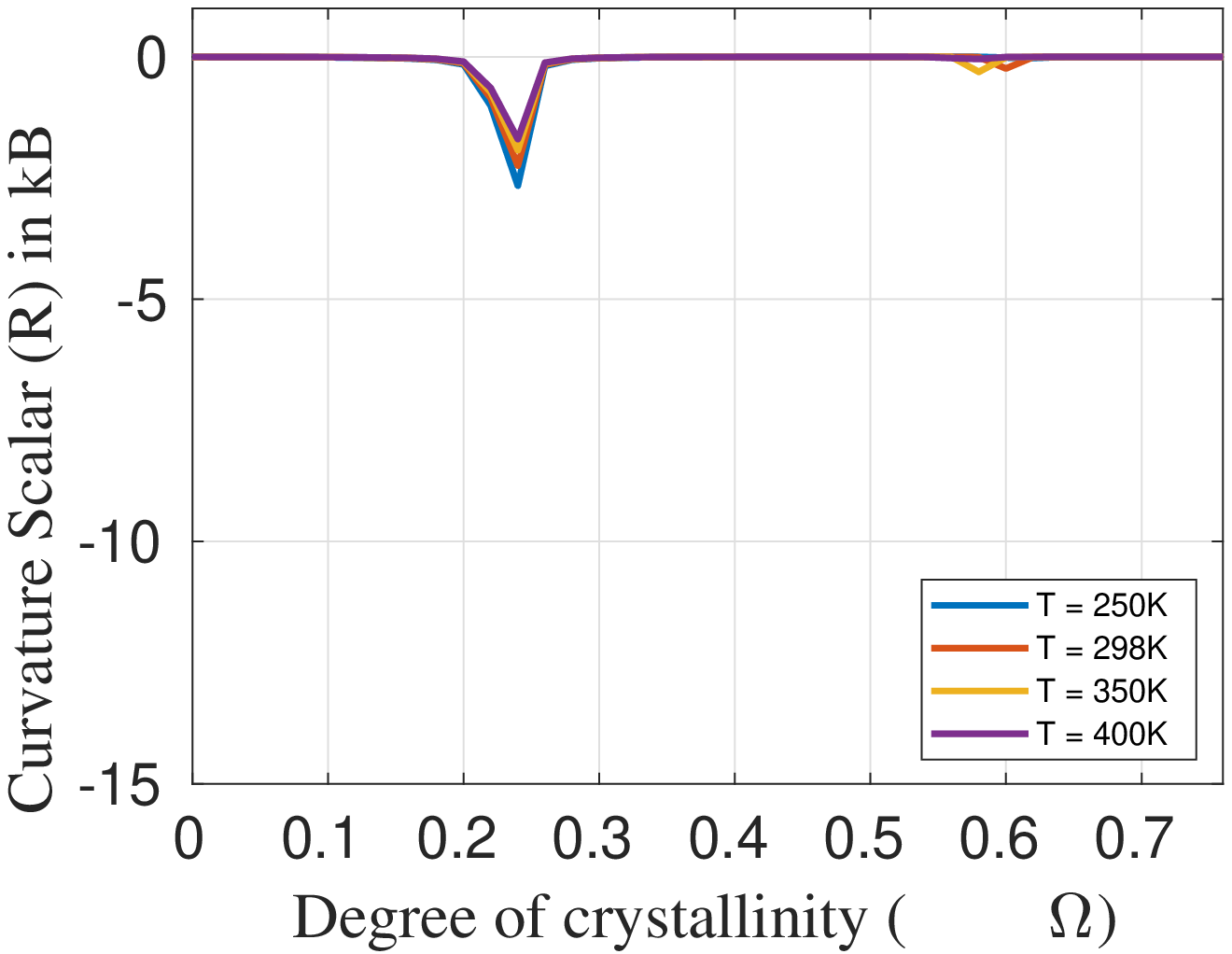}
         \caption{$\gamma = 0$}
         \label{fig:CurvaturNonisothermalShearcrystallizationa}
     \end{subfigure} & 
     \begin{subfigure}[b]{0.47\textwidth}
         \centering
         \includegraphics[width=0.7\textwidth]{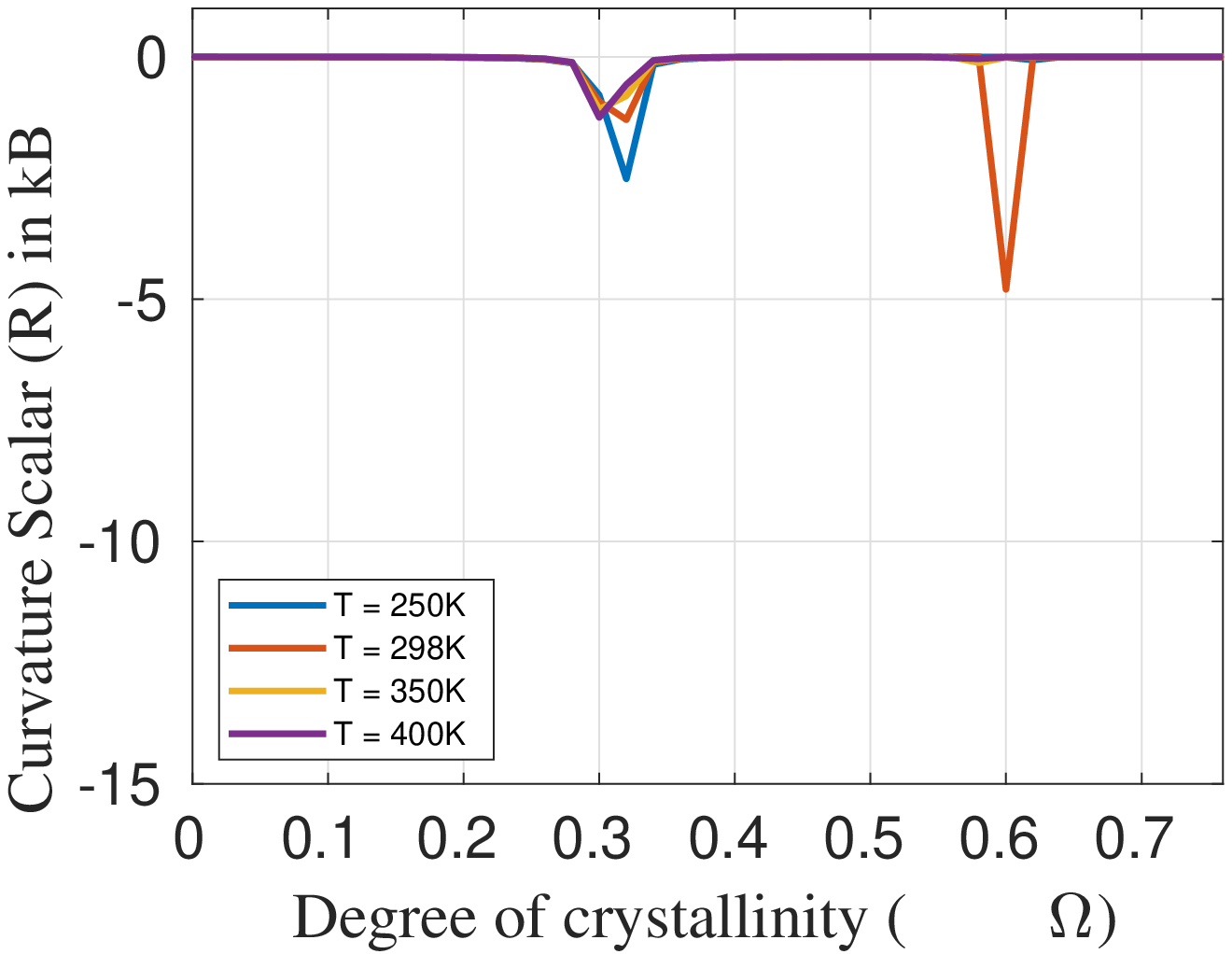}
         \caption{$\gamma = 1.4$}
         \label{fig:CurvaturNonisothermalShearcrystallizationb}
     \end{subfigure}
\end{tabular}
     \begin{subfigure}[b]{0.47\textwidth}
         \centering
         \includegraphics[width=0.7\textwidth]{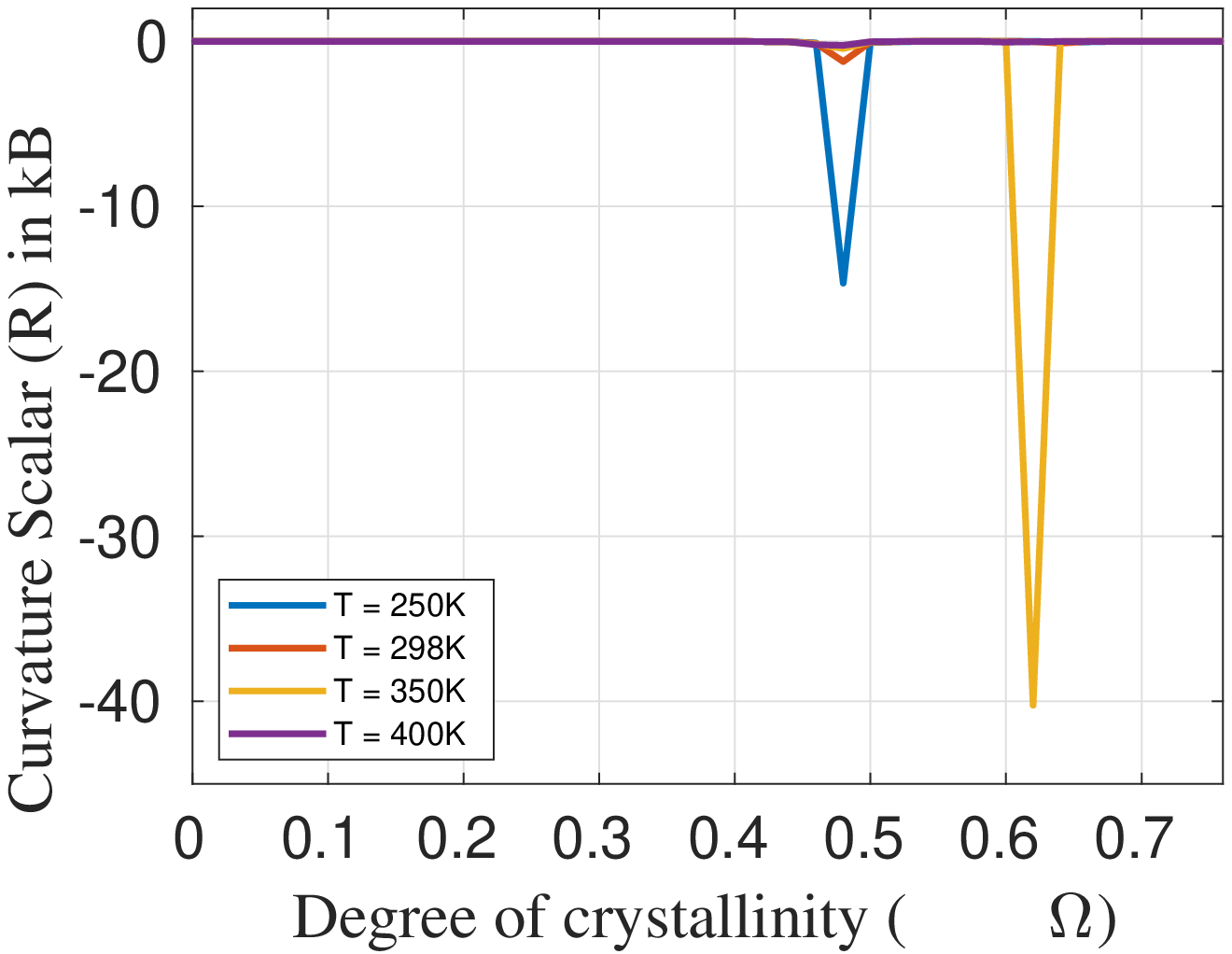}
         \caption{$\gamma = 2.9$}
         \label{fig:CurvaturNonisothermalShearcrystallizationc}
     \end{subfigure} 
\caption{Plots of scalar curvature $R$ with $\Omega$ for varying values of temperature and shear strains}
\label{CurvaturNonisothermalShearcrystallization}
\end{figure*}

\begin{figure*}
    \centering
    \includegraphics[width=0.7\textwidth]{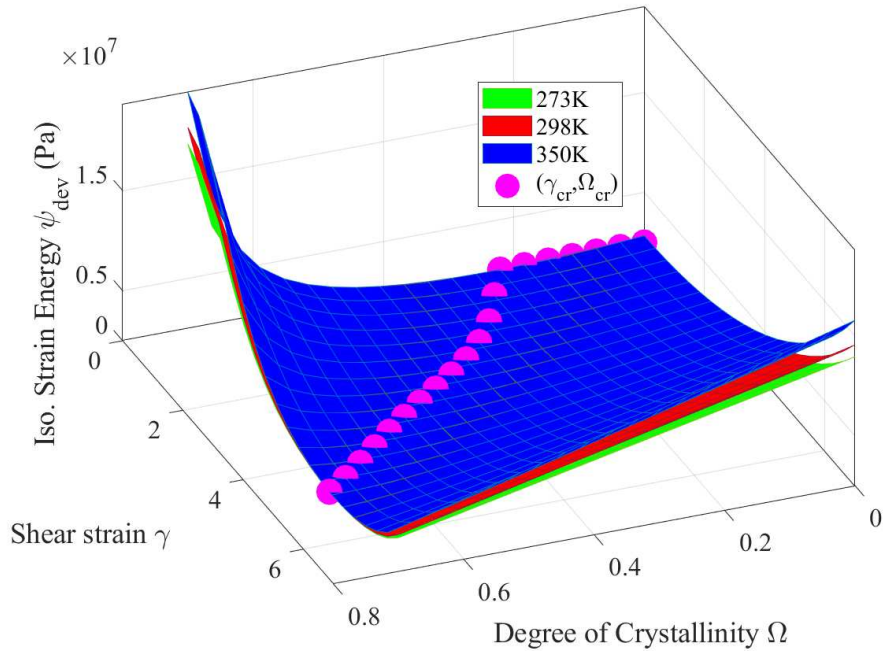}
    \caption{Strain energy surfaces for different temperatures $T$}
    \label{DeviatoricFreeEnergy3D}
\end{figure*}

\begin{figure*}
 \begin{tabular}{cc}
     \centering
     \begin{subfigure}[b]{0.47\textwidth}
         \centering
         \includegraphics[width=0.75\textwidth]{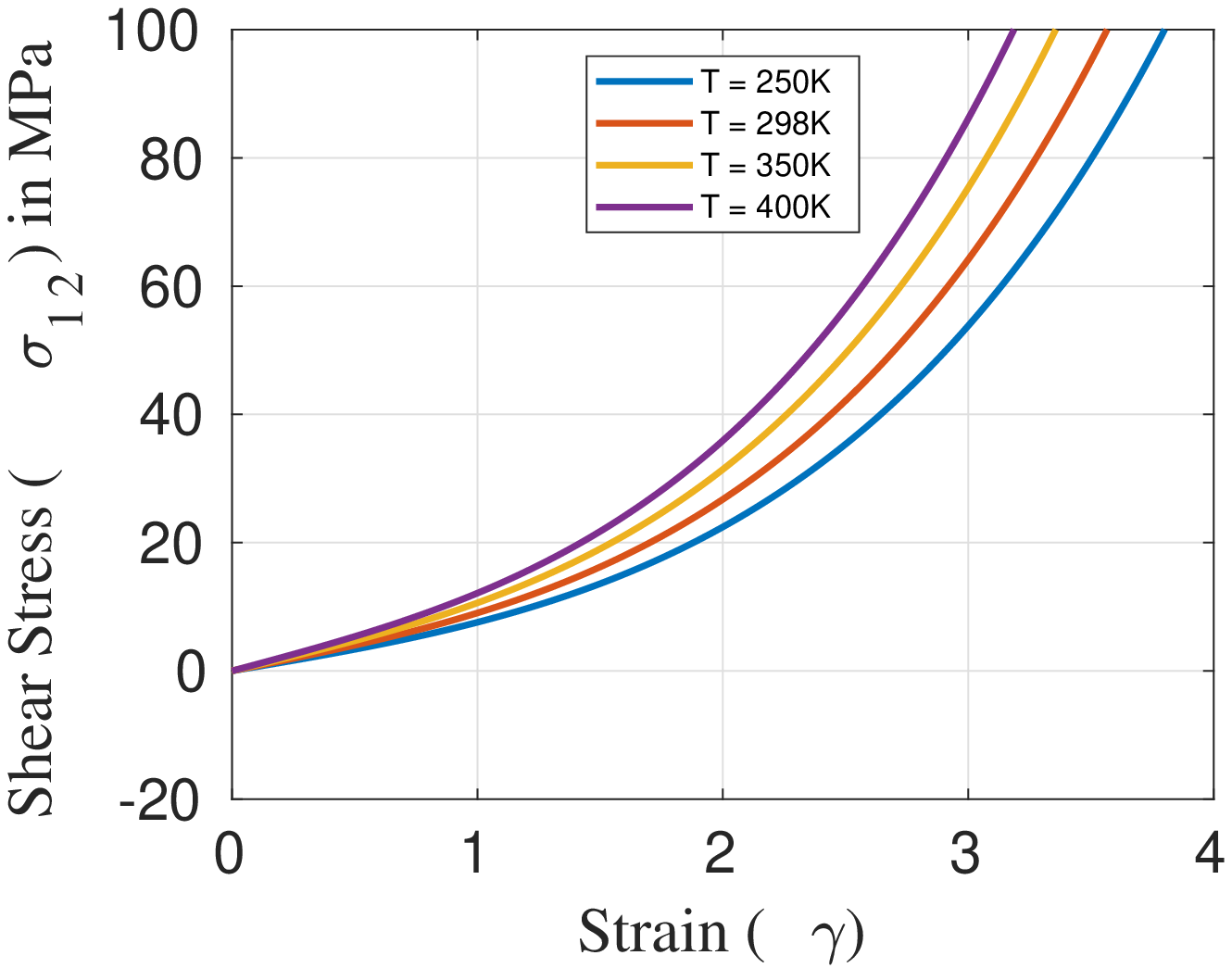}
         \caption{$\Omega = 0$}
     \end{subfigure} &
     \begin{subfigure}[b]{0.47\textwidth}
         \centering
         \includegraphics[width=0.75\textwidth]{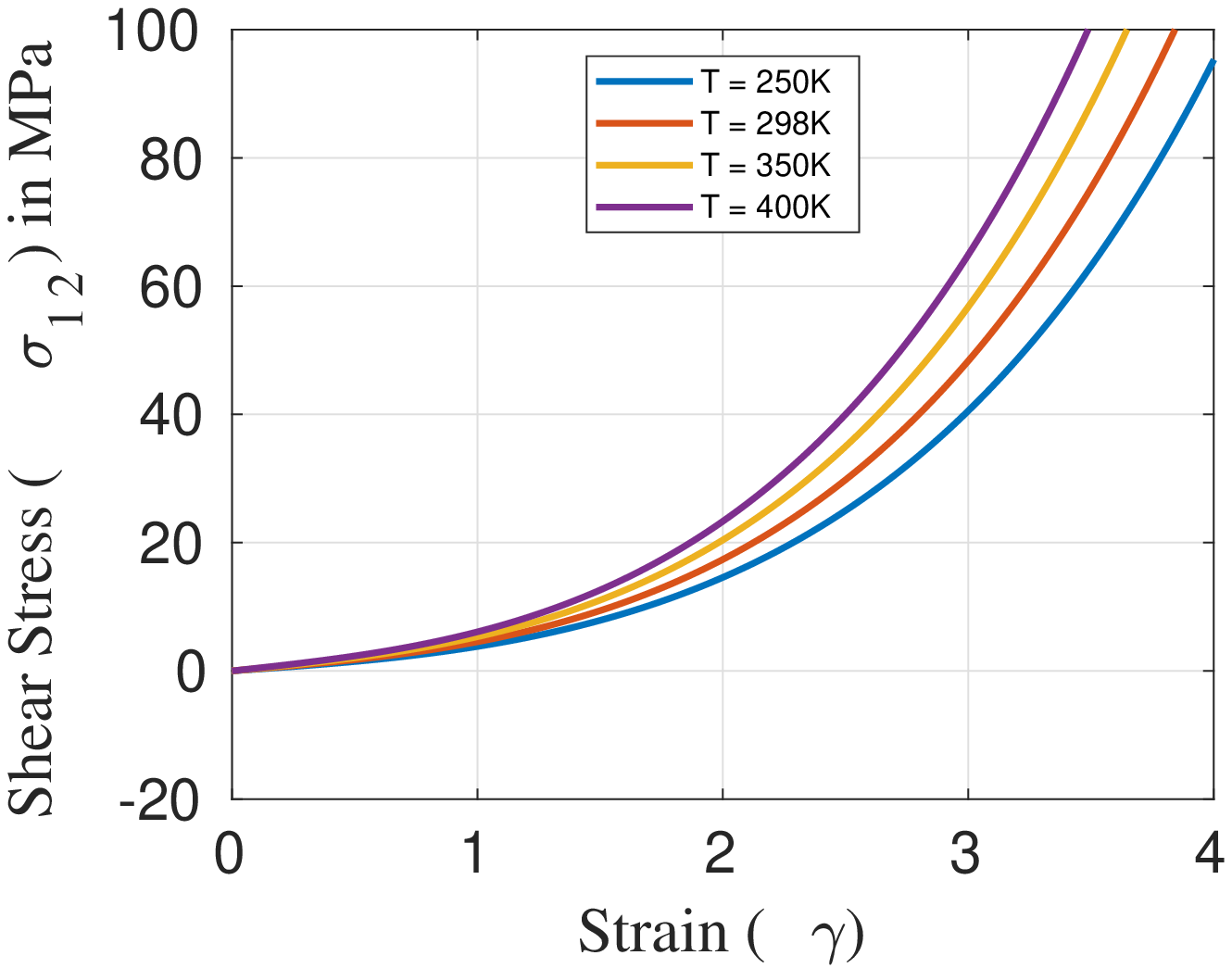}
         \caption{$\Omega = 0.15$}
     \end{subfigure} \\
     \begin{subfigure}[b]{0.47\textwidth}
         \centering
         \includegraphics[width=0.75\textwidth]{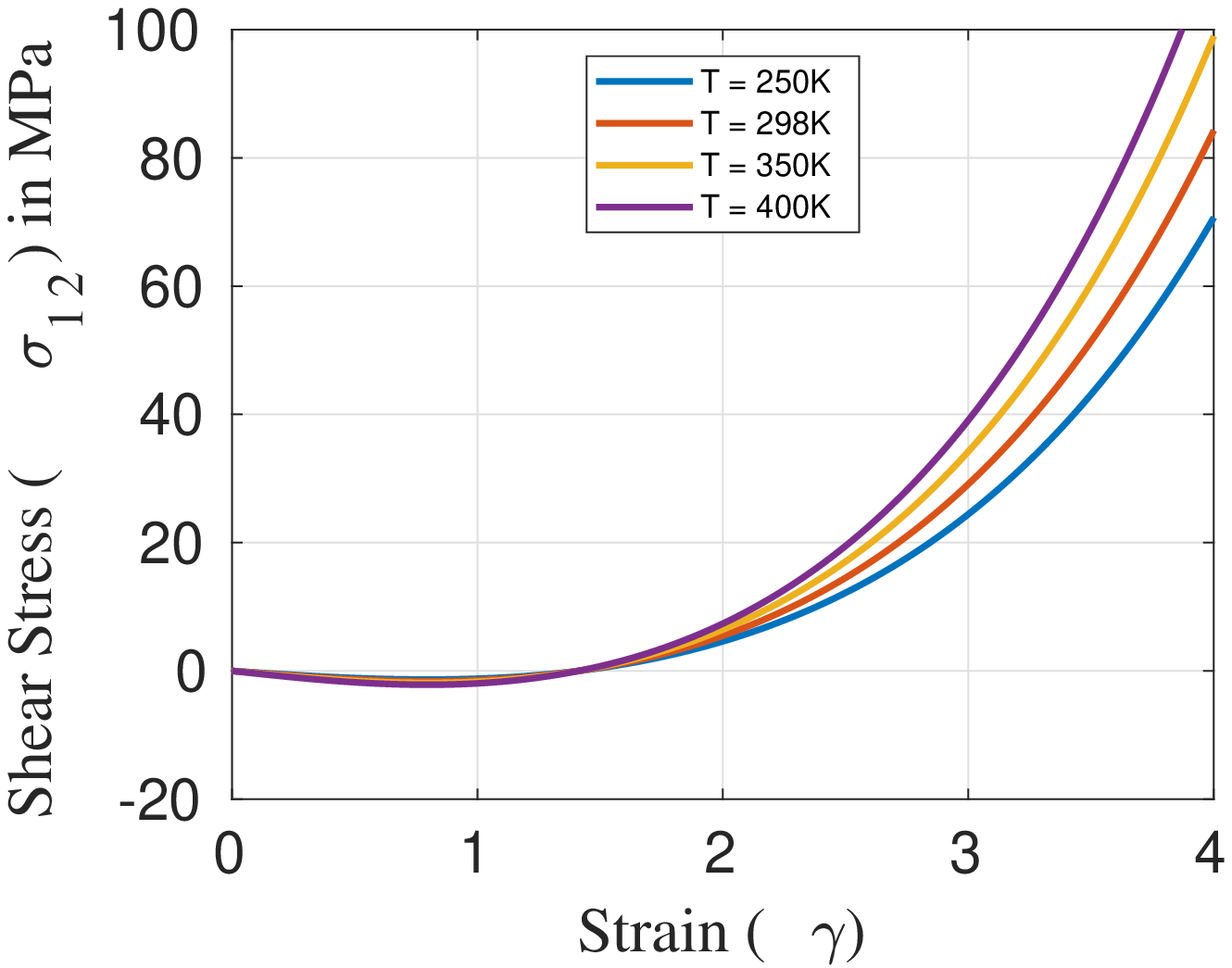}
         \caption{$\Omega = 0.30$}
     \end{subfigure} &
     \begin{subfigure}[b]{0.47\textwidth}
         \centering
         \includegraphics[width=0.75\textwidth]{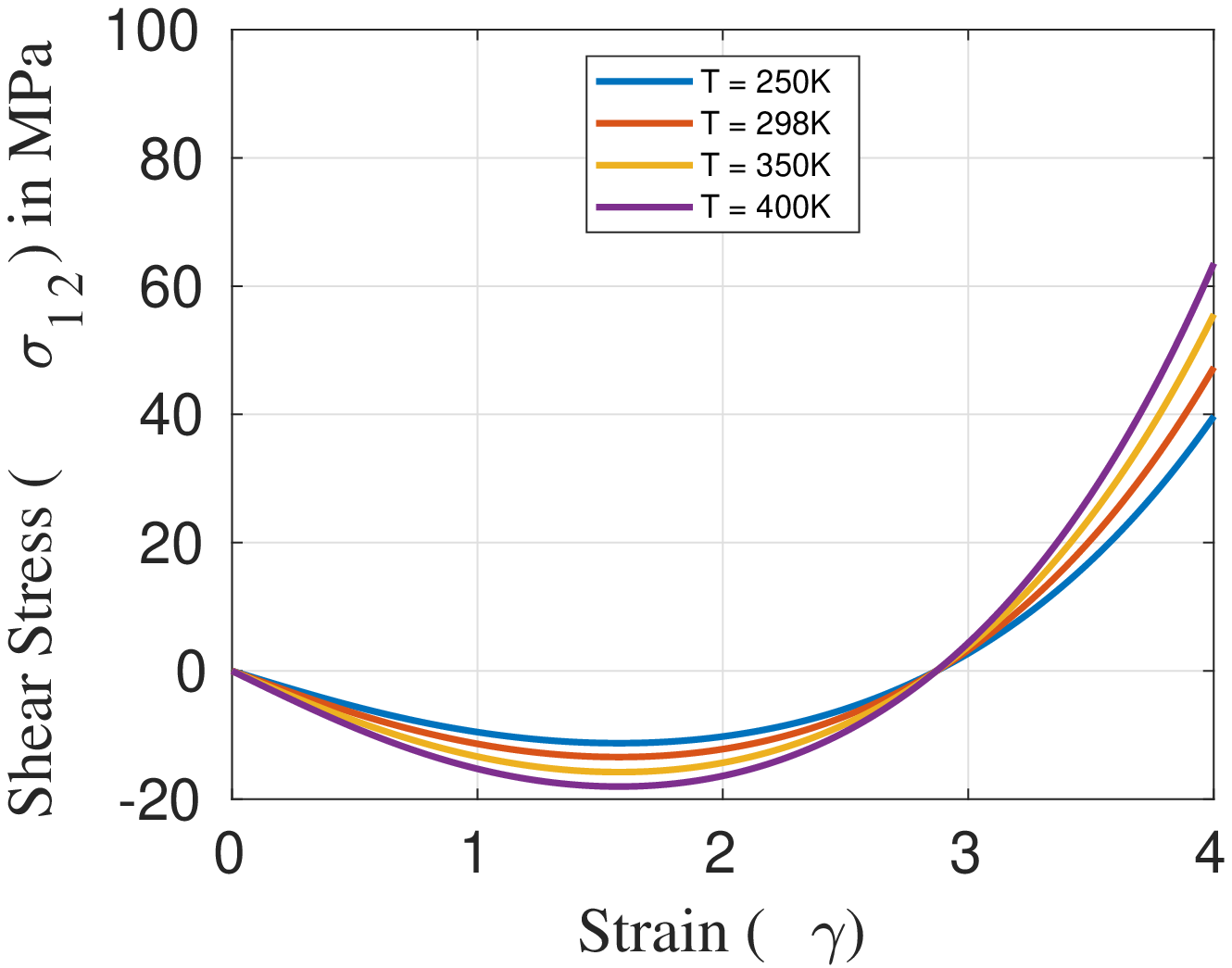}
         \caption{$\Omega = 0.45$}
     \end{subfigure} 
 \end{tabular}
\caption{Plots of $\mathbf{\sigma_{12}}$ with shear strain $\gamma$ for varying temperature and crystallization ratio}
\label{sigma12vsshearfortemperaturenoniso}
\end{figure*}

If $\gamma$ denotes the shear strain, the deformation gradient for the case of simple shear is given by,
\begin{equation}
\mathbf{F} = 
\begin{bmatrix}
1 & \gamma & 0 \\
0 & 1 & 0 \\
0 & 0 & 1
\end{bmatrix}
\end{equation}
The only non-zero components of the right Cauchy-Green tensor are, $C_{11} = C_{33} = 1$, $C_{22} = 1+ \gamma^2$ and $C_{12} = C_{21} = \gamma$. The reduced thermodynamic phase space thus may be represented with just two coordinates, $\gamma$ and $\Omega$. The average stretch $\lambda_{a}$ as obtained from Eq.~\eqref{eq:trace of cbar} is $\sqrt{\frac{3+\gamma^{2}}{3}}$. Using these expressions, we obtain the total free energy from Eq.~\eqref{Totalfreeenergy} as a function of $\gamma$, $\Omega$ and $T$. The surfaces generated by the free energy expression for different temperatures are shown in Fig.~\ref{FreeEnergy3D}. 
Using the free energy, the metric and the curvature can be determined. The metric $\mathbf{g}$ in each case is scaled by $k_B$ so as to ensure an accurate inversion of matrices by the software Mathematica \cite{wolfram2010mathematica}. The resulting curvature is therefore expressed in $k_B$ units.

To maintain expositional brevity, we avoid writing out the explicit expressions for these quantities. 

The scalar curvature $R$ defines a hypersurface in $\mathbb{R}^4$ which is difficult to visualise and interpret. To simplify the analysis, we generate two sets of plots. The first set comprises of plots of $R \text{ vs } \gamma$ at various $\Omega$'s. Each plot contains curves generated at various ambient temperatures $T$. In the second set, there are plots of $R \text{ vs } \Omega$ for certain critical values of $\gamma$, which are determined from the first set of plots. Temperature variation is captured in the this set as well. The first set of plots are shown in Fig.~\ref{CurvaturNonisothermalSheartemperature}. Curves corresponding to $\Omega = 0$, 0.15 and 0.2 monotonically increase from an initial curvature,  till they reach a steady state, whereas for $\Omega > 0.2$ sharp negative peaks are observed breaking the monotonocity. Even though some of these plots have multiple peaks, we can single out a characteristic peak in each of these plots, the strain corresponding to which is independent of temperature. The existence of such a consistent break in the monotony hints at some underlying criticality and perhaps a change of response beyond a certain thermodynamic state. Let the critical shear strain values be denoted $\gamma_{cr}$ corresponding to the chosen $\Omega$.

Across the plots, a temperature invariant peak is observed for all $\Omega$ values greater than 0.2. In the second set of plots shown in Fig.~\ref{CurvaturNonisothermalShearcrystallization}, the curvature is plotted against $\Omega$. These plots are generated for certain values of $\gamma$. To assess the consistency between plots in Fig.~\ref{CurvaturNonisothermalSheartemperature} and Fig.~\ref{CurvaturNonisothermalShearcrystallization}, we choose some $\gamma_{cr}$ as values of $\gamma$. As expected, we observe negative peaks at certain $\Omega$ values for each $\gamma$ chosen, which we denote as $\Omega_{cr}$. The $\Omega_{cr}$ values obtained for $\gamma=1.4$ and $\gamma=2.9$ are the same $\Omega$s at which Fig.~\ref{Curvaturevsshearfortemperatured} and Fig.~\ref{Curvaturevsshearfortemperaturee} respectively were plotted.  
 From Fig.~\ref{CurvaturNonisothermalShearcrystallization}, specifically the plot corresponding to $\gamma=0$, we obtain $\Omega_{cr}$ to be 0.23 which is nothing but $\frac{1}{\sqrt{N}}$. This confirms that for all $\Omega_{cr} < \frac{1}{\sqrt{N}}$ , $\gamma_{cr}$ is always 0.  

Combining our observations from Fig.~\ref{CurvaturNonisothermalSheartemperature}, and Fig.~\ref{CurvaturNonisothermalShearcrystallization}, we conclude that some change in physical response occurs at states corresponding to  ($\Omega_{cr}$, $\gamma_{cr}$), and the values of these states may be obtained graphically. We determine their values to be: ($0,0$), ($0,0.15$), ($0,0.2$), ($1.4,0.3$), ($2.7,0.45$), ($3.9,0.6$), ($4.7,0.7$).  

Towards understanding the physical significance of the critical states, we analyze the variation in free energy in the neighbourhood of these states. From the Eqs.~\eqref{MetricinFreeEnergy}, \eqref{ChristoffelSymbols}, \eqref{RiemannianCurvatureTensor} and \eqref{RicciScalar}, we may infer that the only components of free energy contributing to the curvature would be those involving a strong coupling among the thermodynamic states. Hence it suffices to analyze the variation of $\Psi_{strain}$ across the thermodynamic space. From Fig.~\ref{DeviatoricFreeEnergy3D}, we observe that at ($\gamma_{cr}$,$\Omega_{cr}$) denoted by the pink markers, this component reduces to zero. The total free energy however remains non-zero (see Fig.~\ref{FreeEnergy3D} as well as Eq.~\eqref{FreeEnergy3D}) due to the other components $\Psi_{cr}$ and $\Psi_{surr}$ given by Eqs.~\eqref{Crystallinefreeenergy} and \eqref{Surroundingfreeenergy} respectively. 

Upon analysing the graph and the expression for $\Psi_{strain}$, all the pairs of ($\gamma_{cr}$, $\Omega_{cr}$) satisfy,
\begin{equation}\label{conditionshear}
(\Omega_{cr} \sqrt{N}-\lambda_{acr})H(\Omega_{cr}-\frac{1}{\sqrt{N}})+(1-\lambda_{acr})H(\frac{1}{\sqrt{N}}-\Omega_{cr})=0
\end{equation}
where $\lambda_{acr} = \sqrt{\frac{3+\gamma_{cr}^2}{3}}$.

Let us not limit the set \{($\gamma_{cr}$, $\Omega_{cr}$)\} to just the pairs obtained graphically, but to any pair satisfying the condition above. Physically, at this state, the applied shear strain $\gamma$ and the degree of crystallization $\Omega$ are such that the effective strain from the conformational changes in the uncrystallized links is 0 and no elastic stress is generated. Also, $\gamma_{cr}$ may be considered as the internal strain generated solely due to the crystallized segments. A shear strain can only give rise to any elastic conformational stress if it is greater than $\gamma_{cr}$ for a given $\Omega_{cr}$. However, for any $\gamma < \gamma_{cr}$, the presence of a finite strain energy is unphysical. Here, the only components contributing to the total free energy should be $\Psi_{cr}$ and $\Psi_{surr}$. Hence the total free energy should be reduced by a spurious conformational strain energy generated by $\gamma<\gamma_{cr}$ for a given $\Omega_{cr}$.

That the critical state is a stress-free state is corroborated by the shear stress-strain plots shown in Fig.~\ref{sigma12vsshearfortemperaturenoniso}.

To investigate the influence of loading condition on the critical state, we study curvature plots in the case of uniaxial tension on an incompressible, hyperelastic solid and under a general strain state that involves slight compression.

\begin{figure*}
\begin{tabular}{cc}
     \centering
     \begin{subfigure}[b]{0.47\textwidth}
         \centering
         \includegraphics[width=0.7\textwidth]{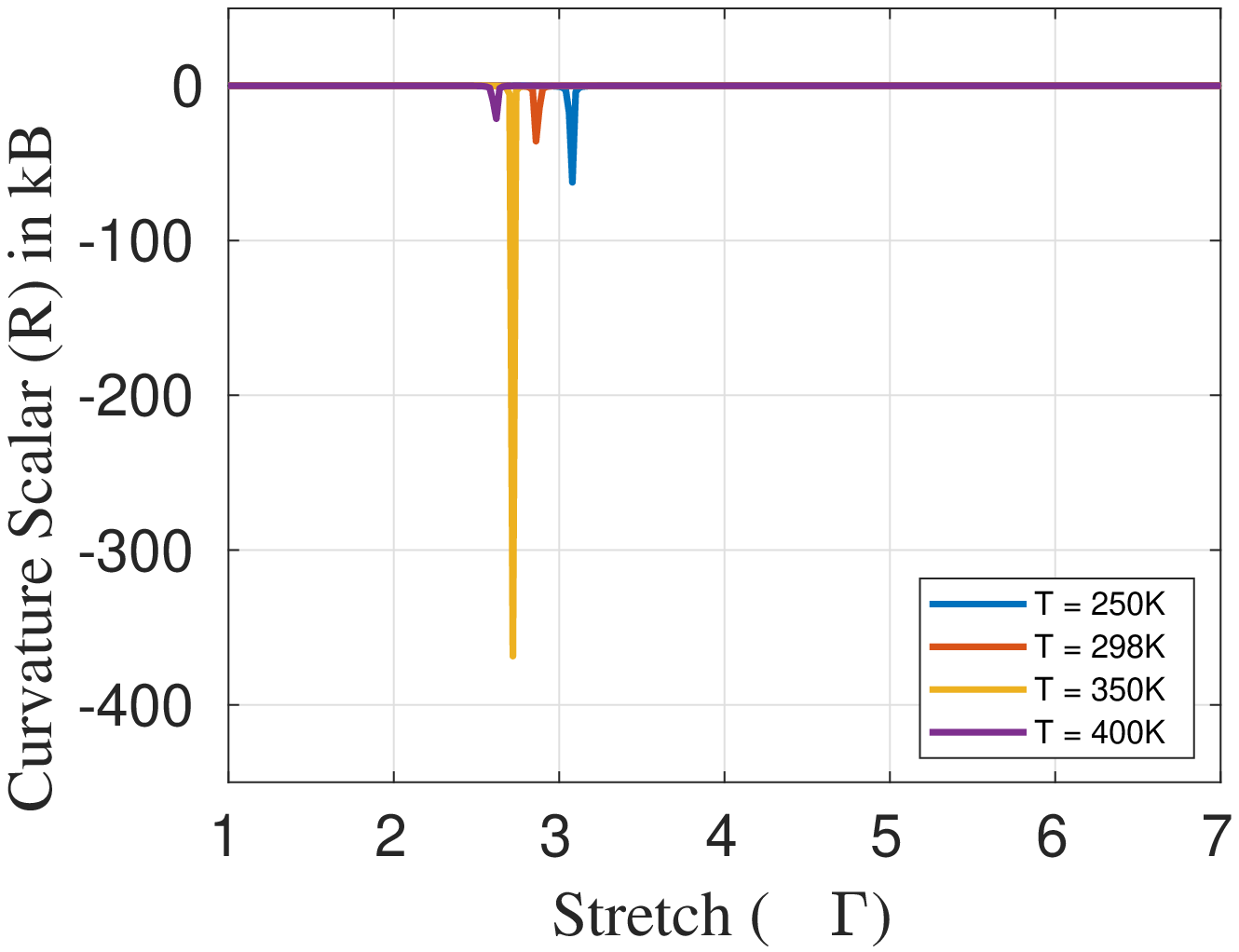}
         \caption{$\Omega = 0$}
         \label{Curvaturevsstretchfortemperaturea}
     \end{subfigure}
&     
     \begin{subfigure}[b]{0.47\textwidth}
         \centering
         \includegraphics[width=0.7\textwidth]{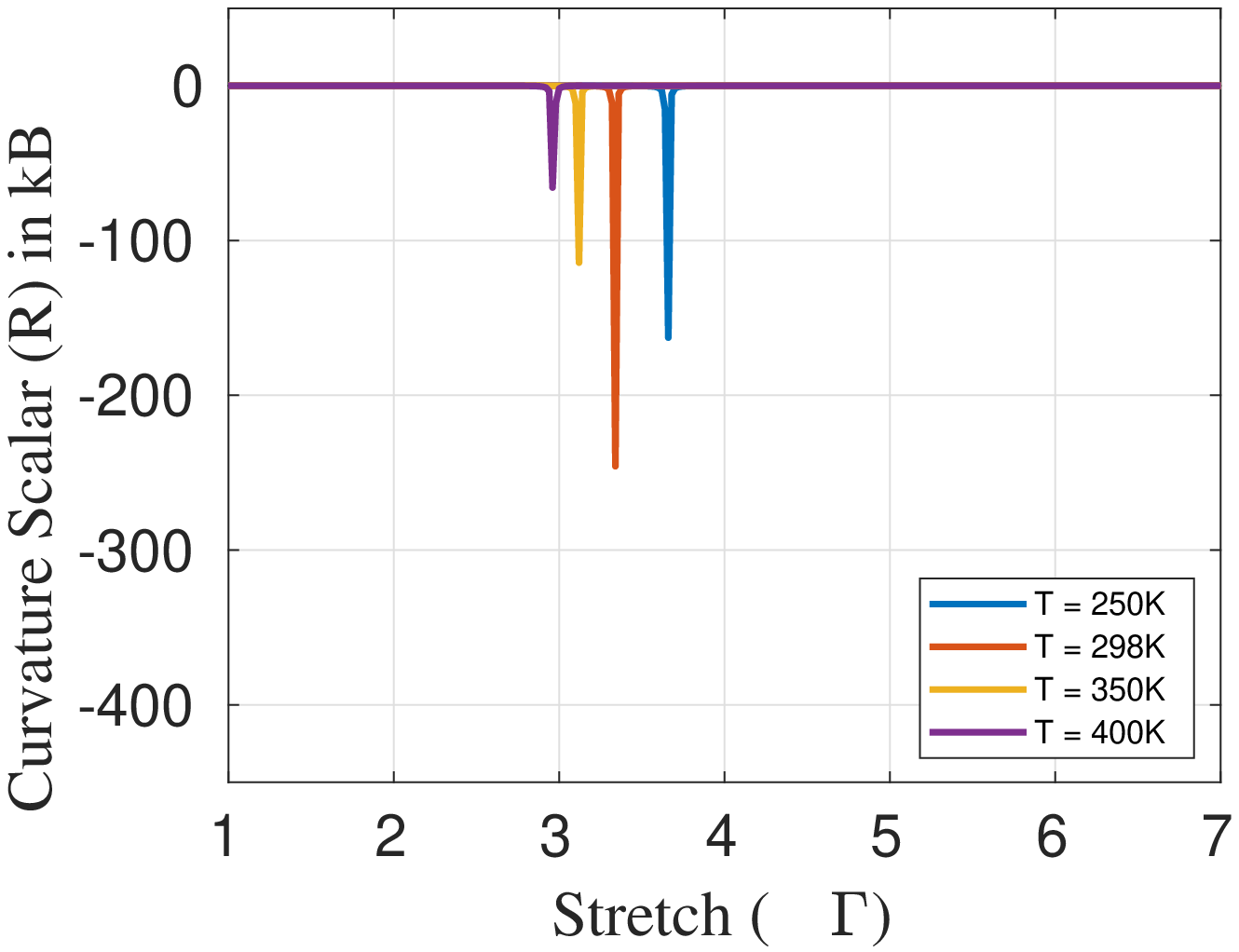}
         \caption{$\Omega = 0.15$}
         \label{Curvaturevsstretchfortemperatureb}
     \end{subfigure}
\\     
     \begin{subfigure}[b]{0.47\textwidth}
         \centering
         \includegraphics[width=0.7\textwidth]{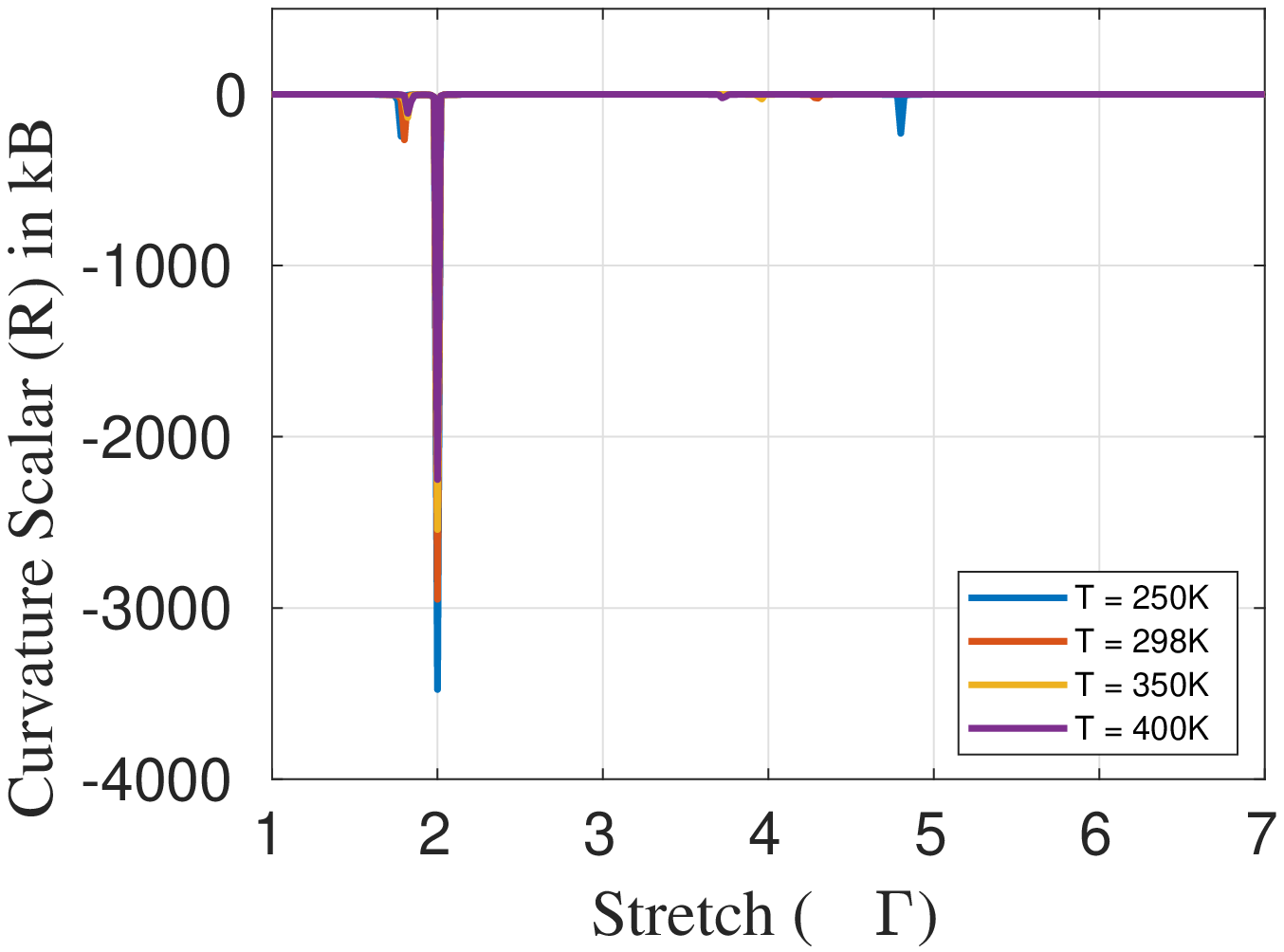}
         \caption{$\Omega = 0.30$}
         \label{Curvaturevsstretchfortemperaturec}
     \end{subfigure}
&    
     \begin{subfigure}[b]{0.47\textwidth}
         \centering
         \includegraphics[width=0.7\textwidth]{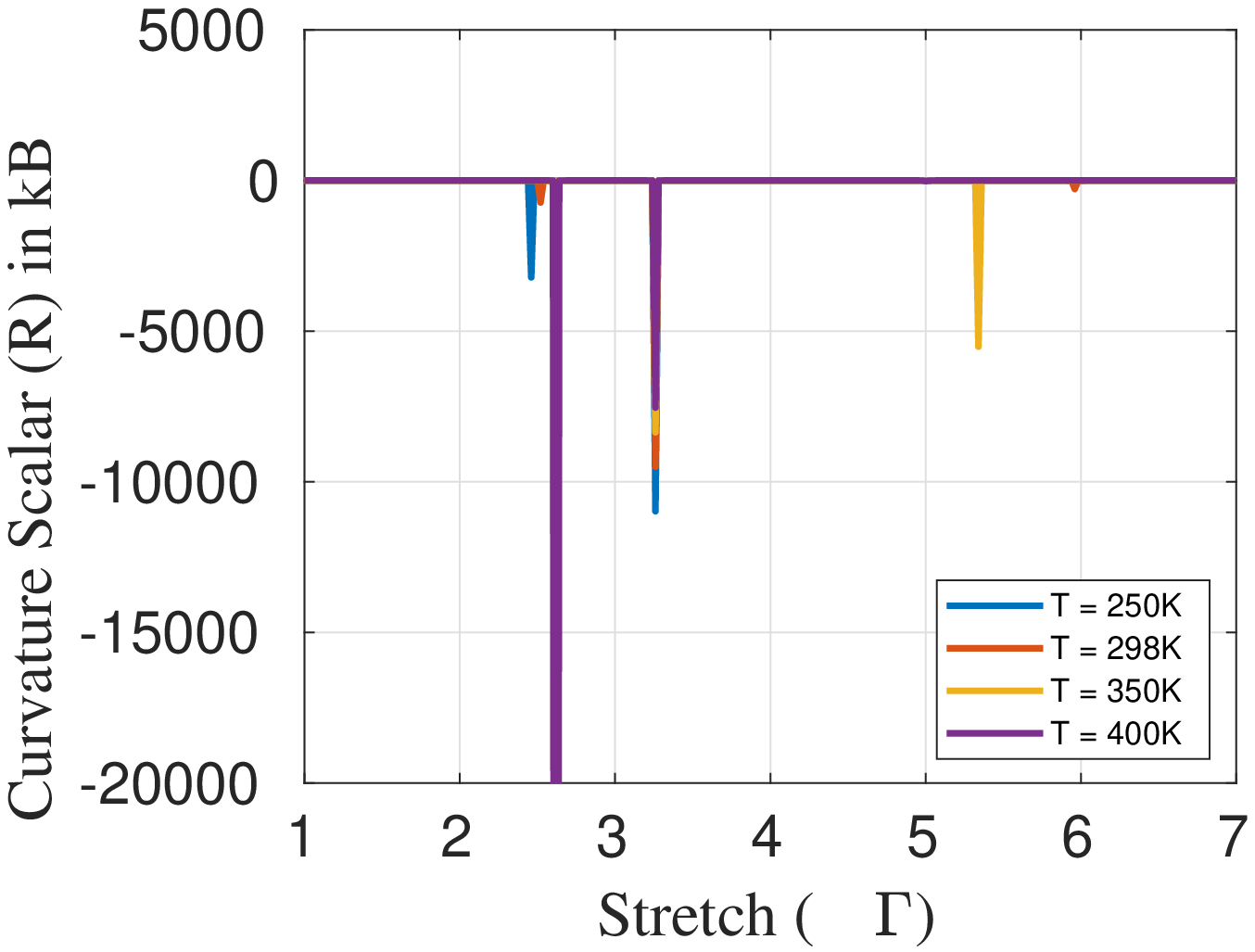}
         \caption{$\Omega = 0.45$}
         \label{Curvaturevsstretchfortemperatured}
     \end{subfigure}   
\\   
     \begin{subfigure}[b]{0.47\textwidth}
         \centering
         \includegraphics[width=0.7\textwidth]{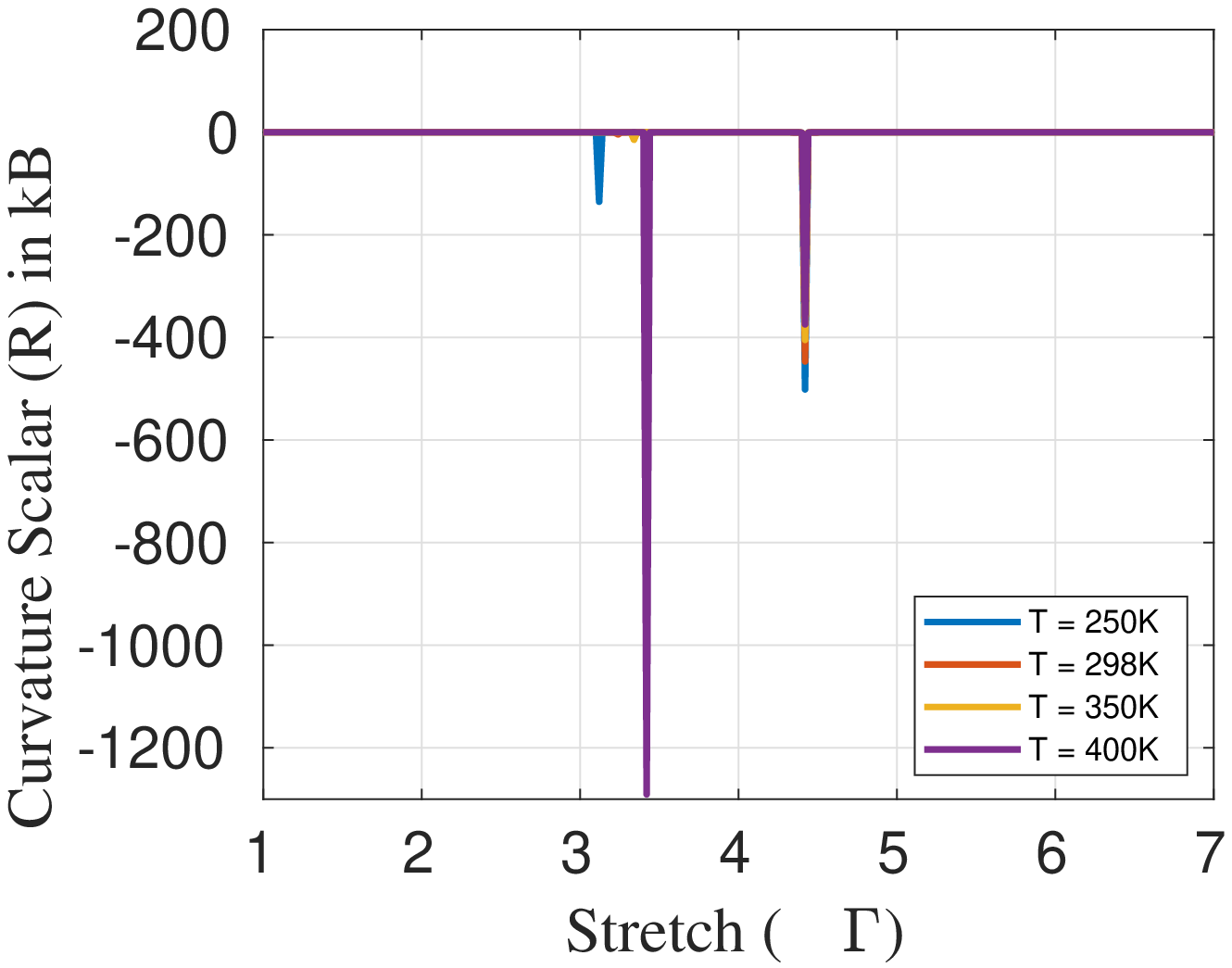}
         \caption{$\Omega = 0.60$}
         \label{Curvaturevsstretchfortemperaturee}
     \end{subfigure} 
&   
     \begin{subfigure}[b]{0.47\textwidth}
         \centering
         \includegraphics[width=0.7\textwidth]{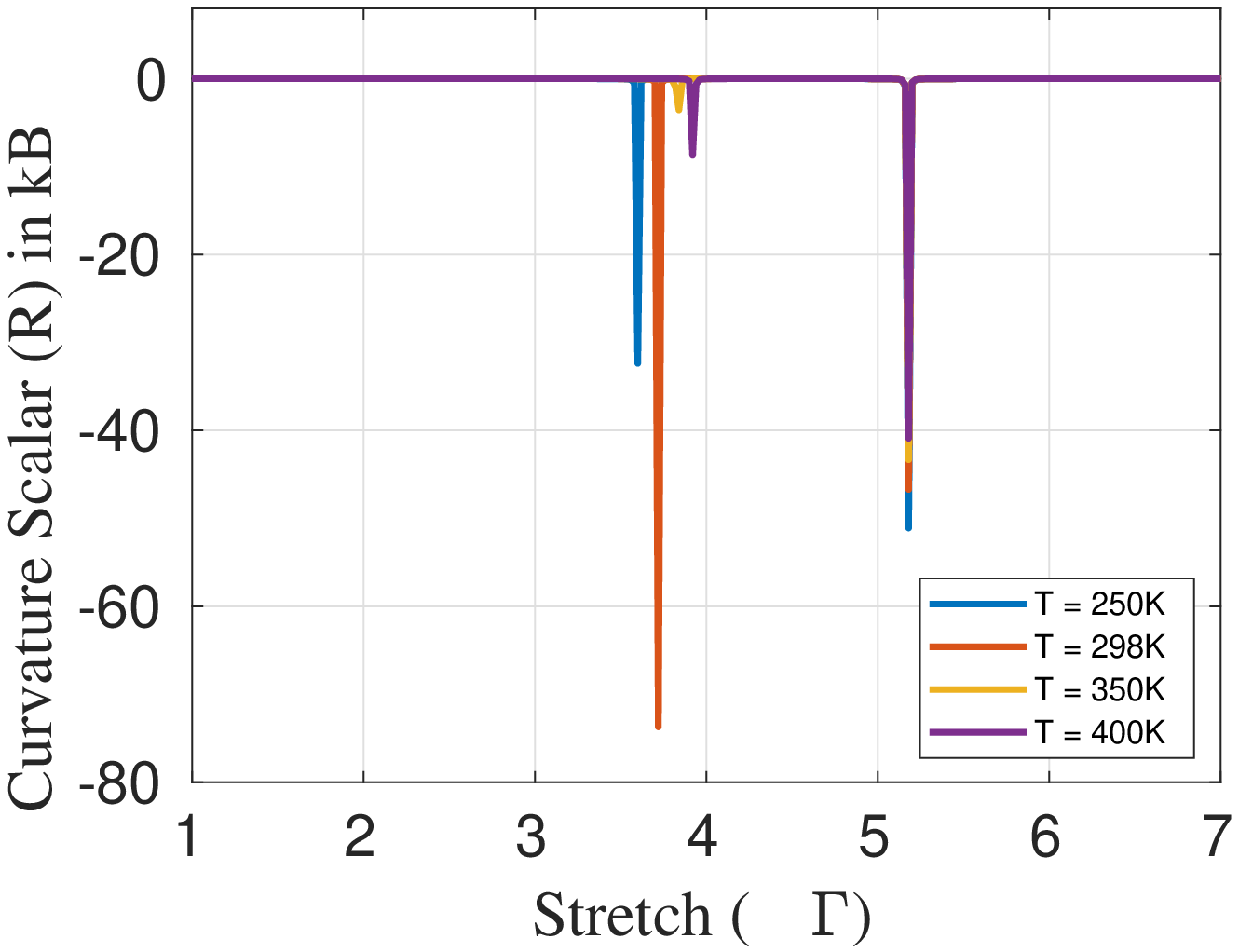}
         \caption{$\Omega = 0.70$}
         \label{Curvaturevsstretchfortemperaturef}
     \end{subfigure} 
\end{tabular}
        \caption{Plots of scalar curvature $R$ with stretch $\Gamma$ for varying temperature and crystallization ratio}
        \label{CurvaturNonisothermalTensiontemperature}
\end{figure*}

\subsection{Uniaxial Tension in an Incompressible\\ Hyperelastic Solid}\label{CurvatureTension}

If the tensile stretch is $\Gamma$, the deformation gradient assumes the form,

\begin{equation}
\mathbf{F} = 
\begin{bmatrix}
\Gamma & 0 & 0 \\
0 & \Gamma^{-1/2} & 0 \\
0 & 0 & \Gamma^{-1/2}
\end{bmatrix}
\end{equation}

The 8-dimensional space is reduced to a 3-dimensional manifold by the following relations among the components of the right Cauchy-Green tensor and the tensile stretch: $\Gamma$ - $C_{11} = \Gamma^2$, $C_{22} = C_{33} = \Gamma^{-1}$, the other components of the tensor being $0$. The thermodynamic states or the coordinates of the manifold are therefore ($\Gamma$, $\Omega$, $T$). 

The scalar curvature is plotted against $\Gamma$ for various values of $\Omega$ and $T$ in Fig.~\ref{CurvaturNonisothermalTensiontemperature}. 

\begin{figure*}
     \centering
     \begin{subfigure}[b]{0.47\textwidth}
         \centering
         \includegraphics[width=0.7\textwidth]{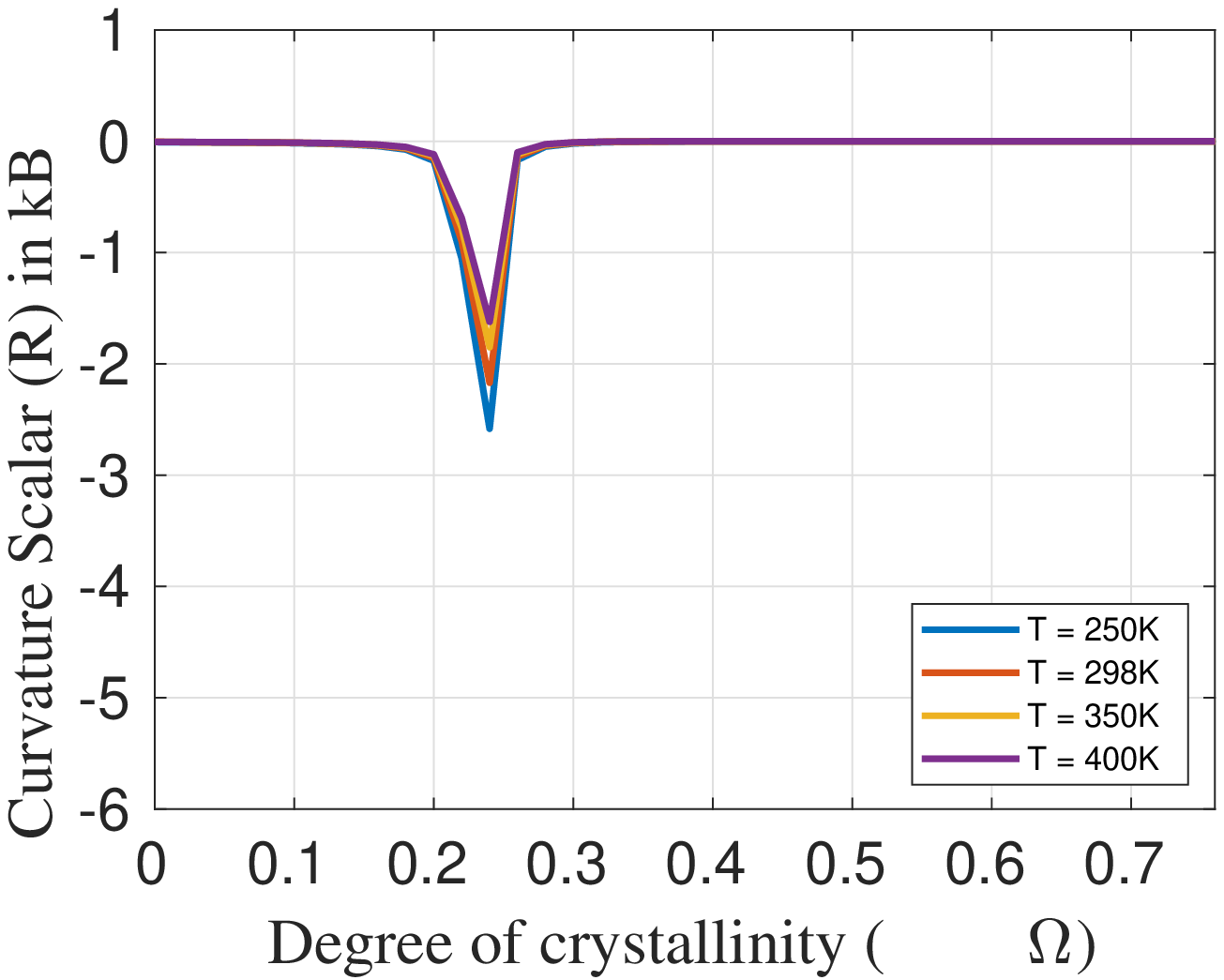}
         \caption{$\Gamma = 1$}
         \label{CurvaturNonisothermalTensioncrystallizationa}
     \end{subfigure}
     \hfill     
     \begin{subfigure}[b]{0.47\textwidth}
         \centering
         \includegraphics[width=0.7\textwidth]{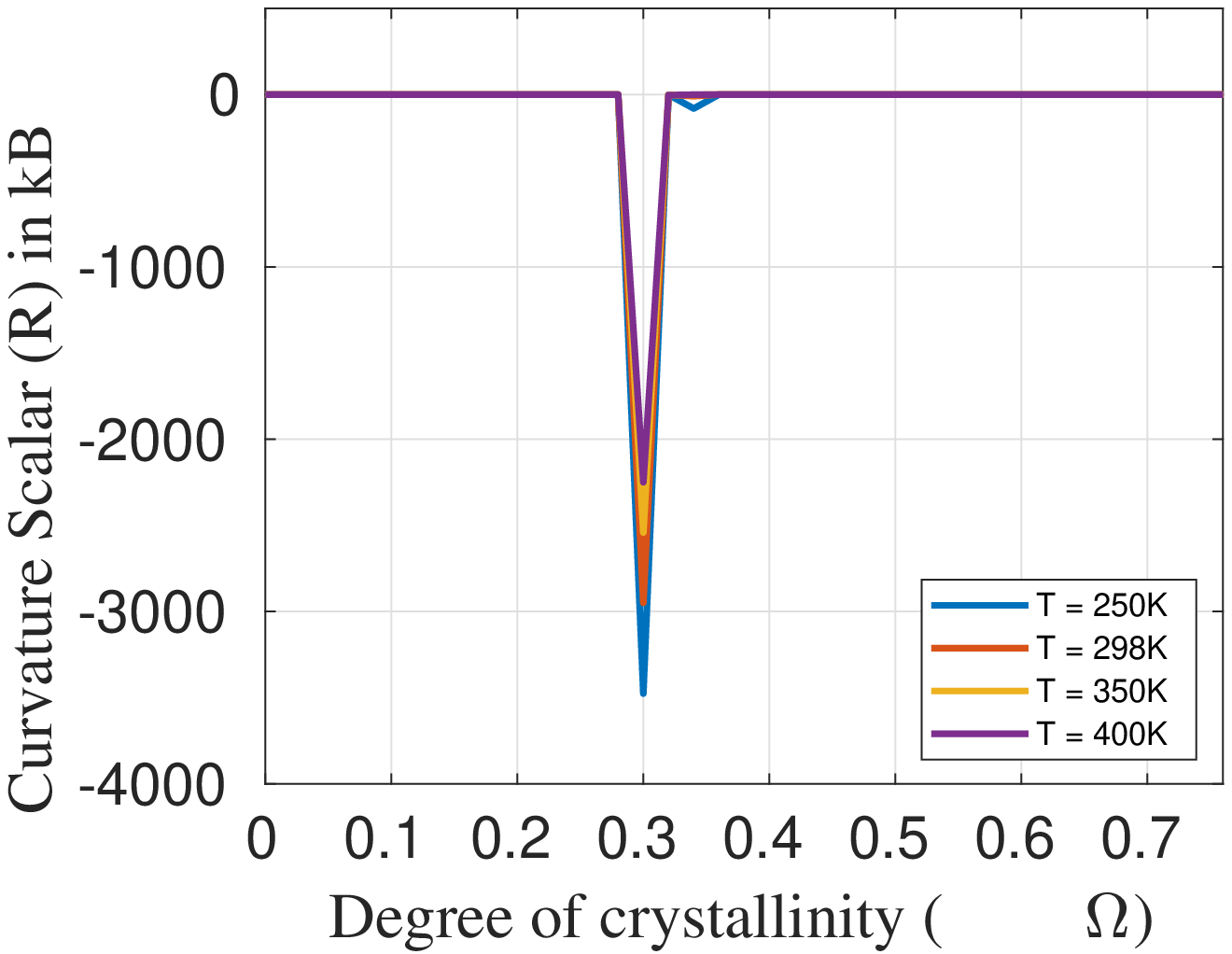}
         \caption{$\Gamma = 2$}
         \label{CurvaturNonisothermalTensioncrystallizationb}
     \end{subfigure}     
     \hfill     
     \begin{subfigure}[b]{0.47\textwidth} 
         \centering
         \includegraphics[width=0.7\textwidth]{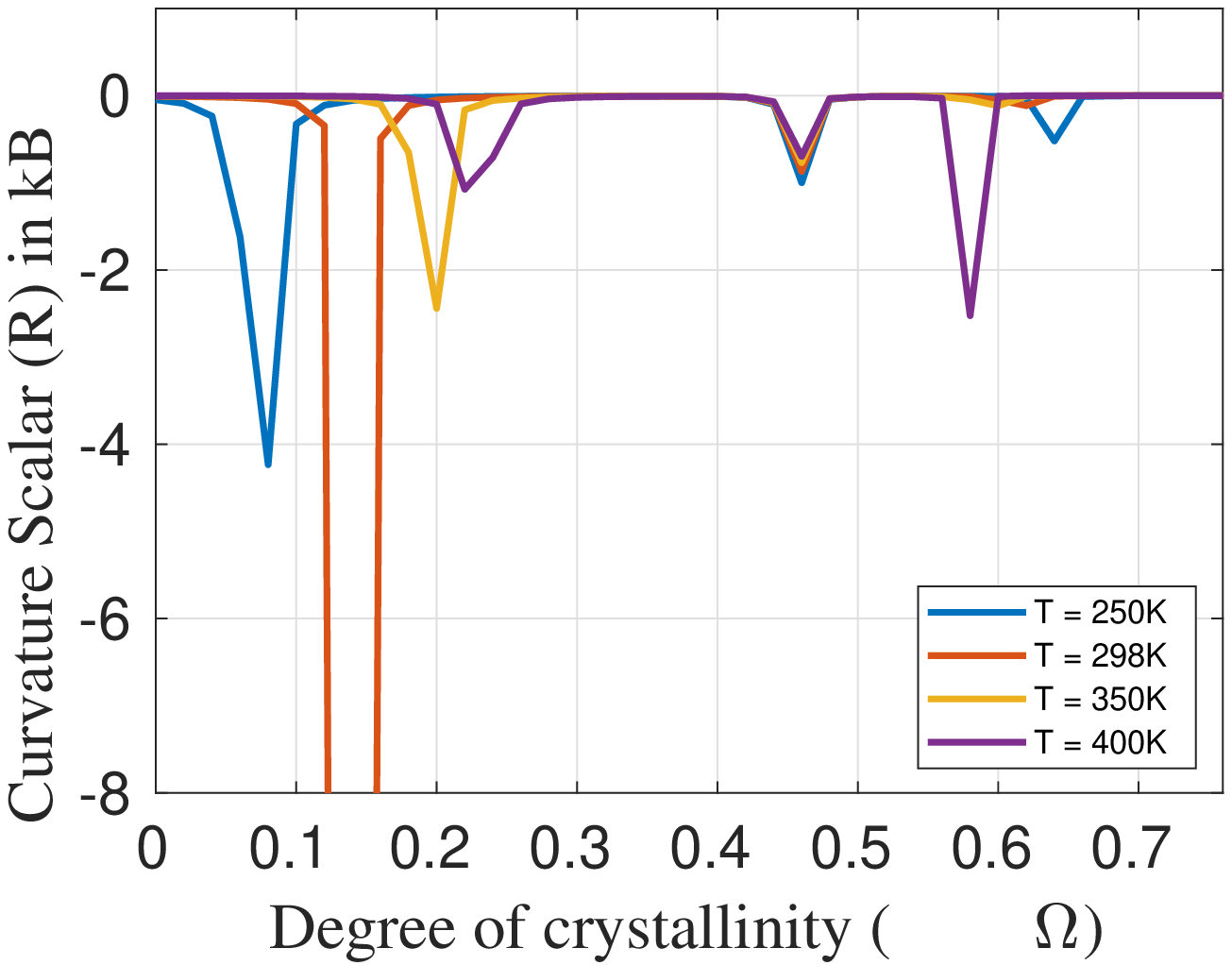}
         \caption{$\Gamma = 3.3$}
         \label{CurvaturNonisothermalTensioncrystallizationc}
     \end{subfigure}
        \caption{Plots of scalar curvature $R$ with $\Omega$ for varying values of temperature and tensile stretch }
        \label{CurvaturNonisothermalTensioncrystallization}
\end{figure*}

\begin{figure*}
    \centering
    \includegraphics[width=0.75\textwidth]{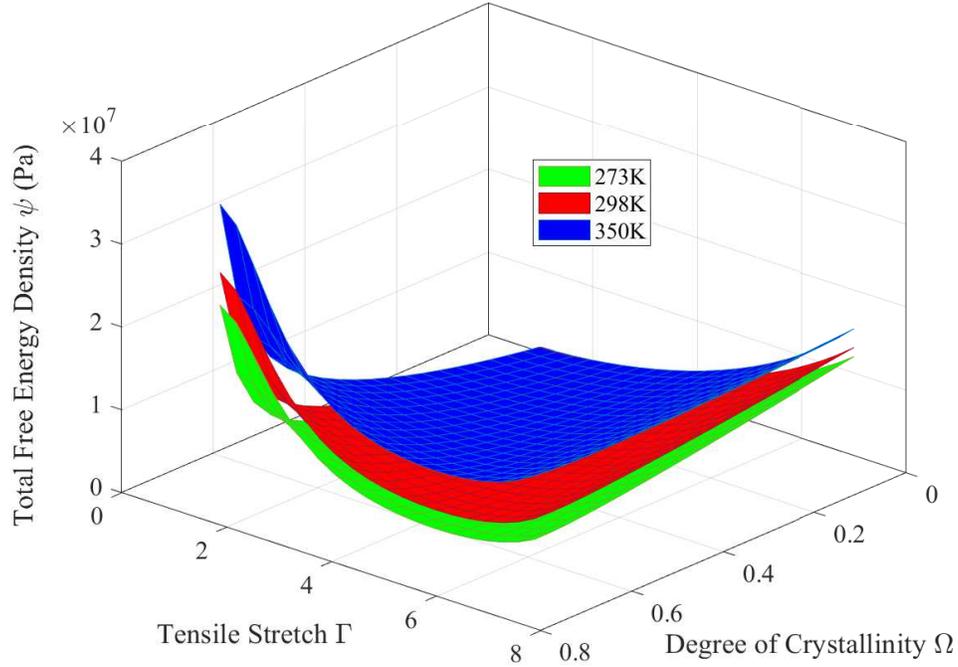}
    \caption{Free energy surfaces for different temperatures $T$ under uniaxial tension in an incompressible material}
    \label{FreeEnergy3DTension}
\end{figure*}

\begin{figure*}
    \centering
    \includegraphics[width=0.75\textwidth]{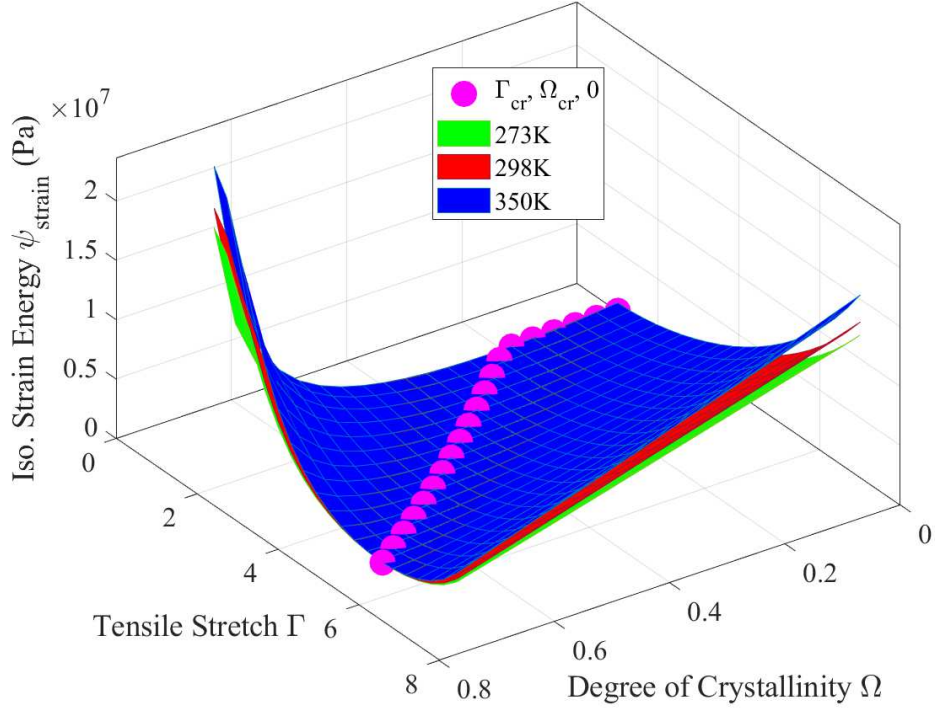}
    \caption{Strain energy surfaces for different temperatures $T$ for uniaxial tension in an incompressible hyperelastic material}
    \label{DeviatoricFreeEnergy3DTension}
\end{figure*}

\begin{figure*}
     \centering
     \begin{subfigure}[b]{0.47\textwidth}
         \centering
         \includegraphics[width=0.7\textwidth]{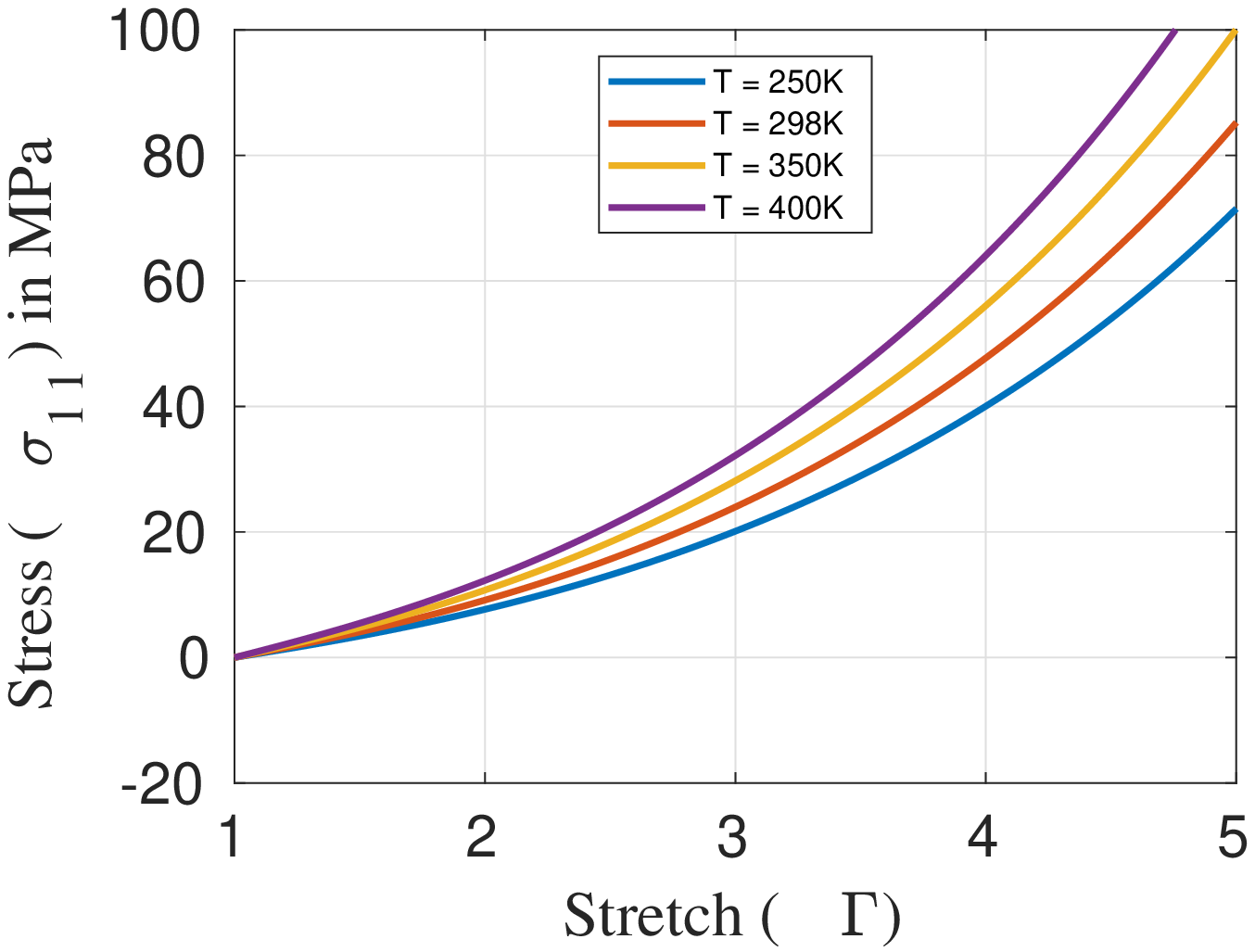}
         \caption{$\Omega = 0$}
         \label{sigma11vsstretchfortemperaturea}
     \end{subfigure}
     \hfill     
     \begin{subfigure}[b]{0.47\textwidth}
         \centering
         \includegraphics[width=0.7\textwidth]{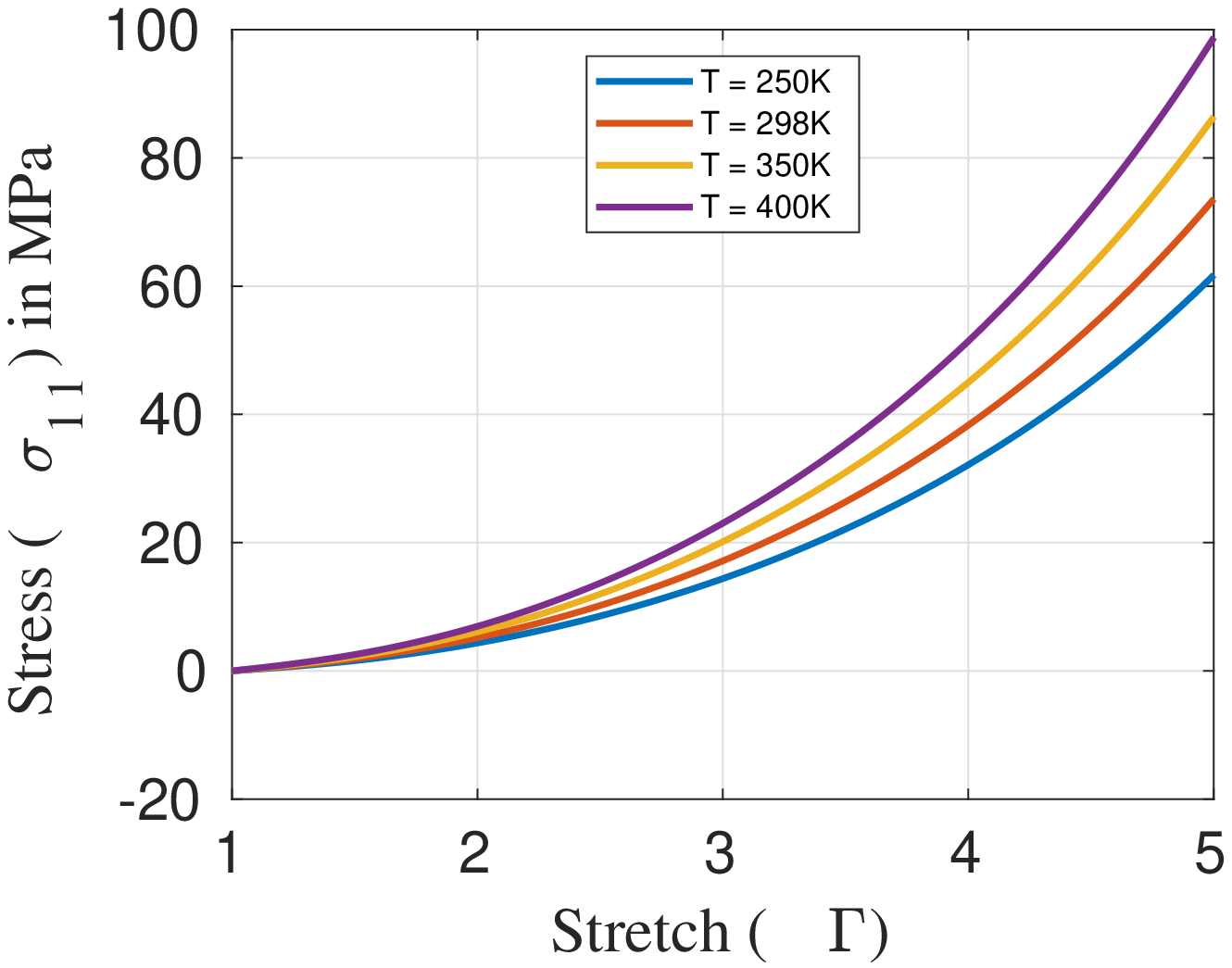}
         \caption{$\Omega = 0.15$}
         \label{sigma11vsstretchfortemperatureb}
     \end{subfigure}
     \hfill     
     \begin{subfigure}[b]{0.47\textwidth}
         \centering
         \includegraphics[width=0.7\textwidth]{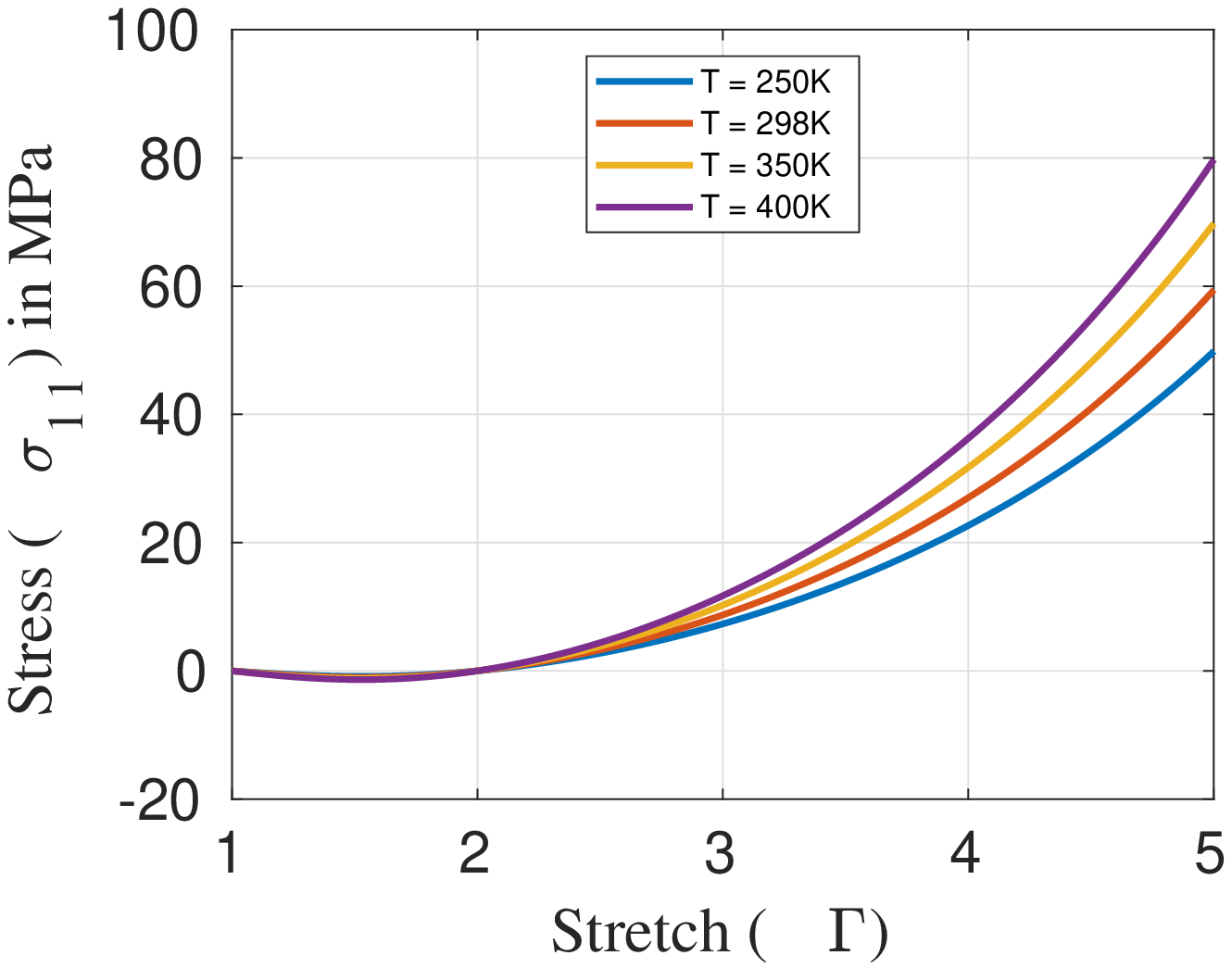}
         \caption{$\Omega = 0.30$}
         \label{sigma11vsstretchfortemperaturec}
     \end{subfigure}
     \hfill     
     \begin{subfigure}[b]{0.47\textwidth}
         \centering
         \includegraphics[width=0.7\textwidth]{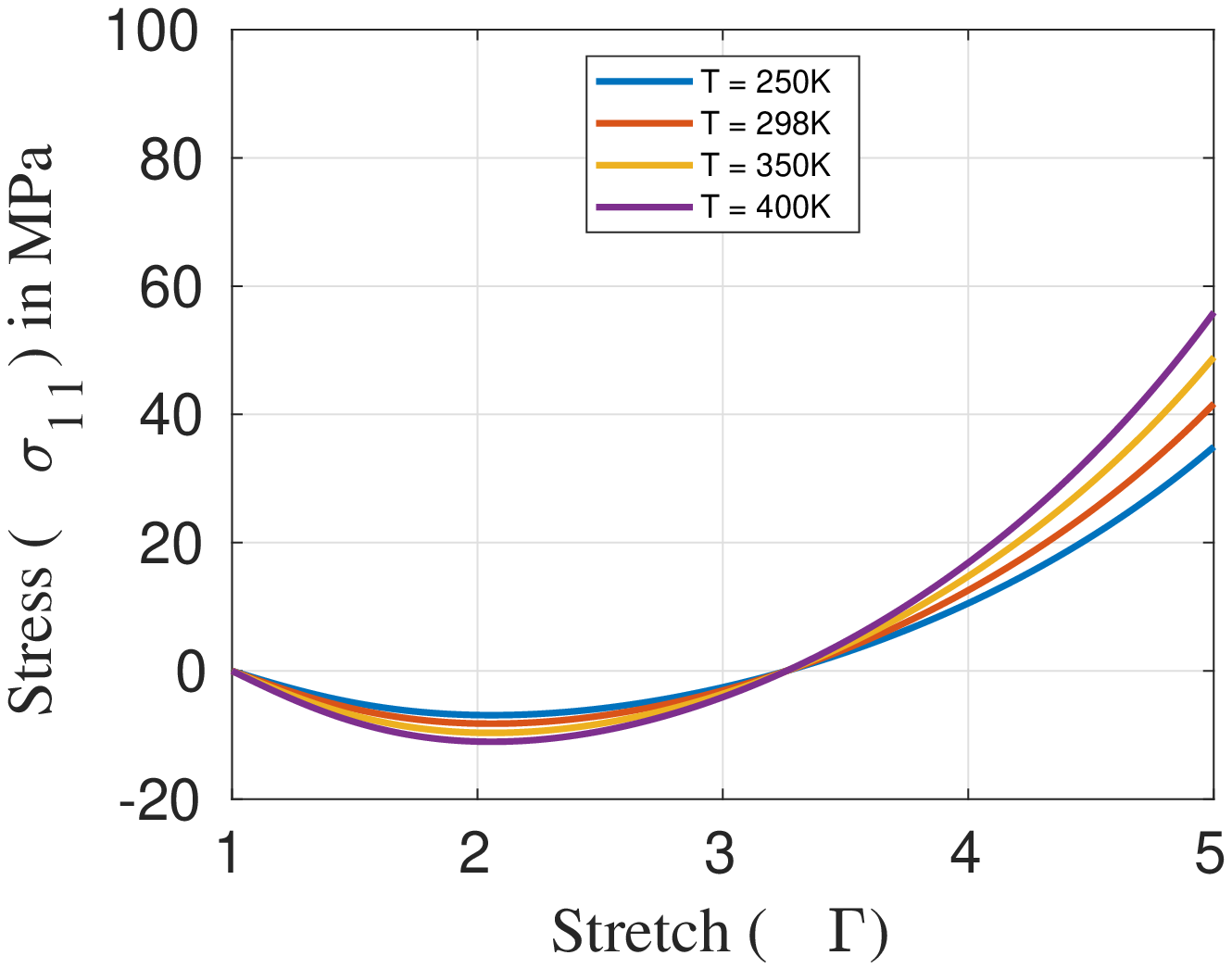}
         \caption{$\Omega = 0.45$}
         \label{sigma11vsstretchfortemperatured}
     \end{subfigure}     
        \caption{Plots of $\mathbf{\sigma_{11}}$ with tensile stretch $\Gamma$ for varying values of temperature for various crystallization ratios}
        \label{sigma11vsstretchfortemperature}
\end{figure*}

\begin{figure*}
     \centering
     \begin{subfigure}[b]{0.47\textwidth}
         \centering
         \includegraphics[width=0.7\textwidth]{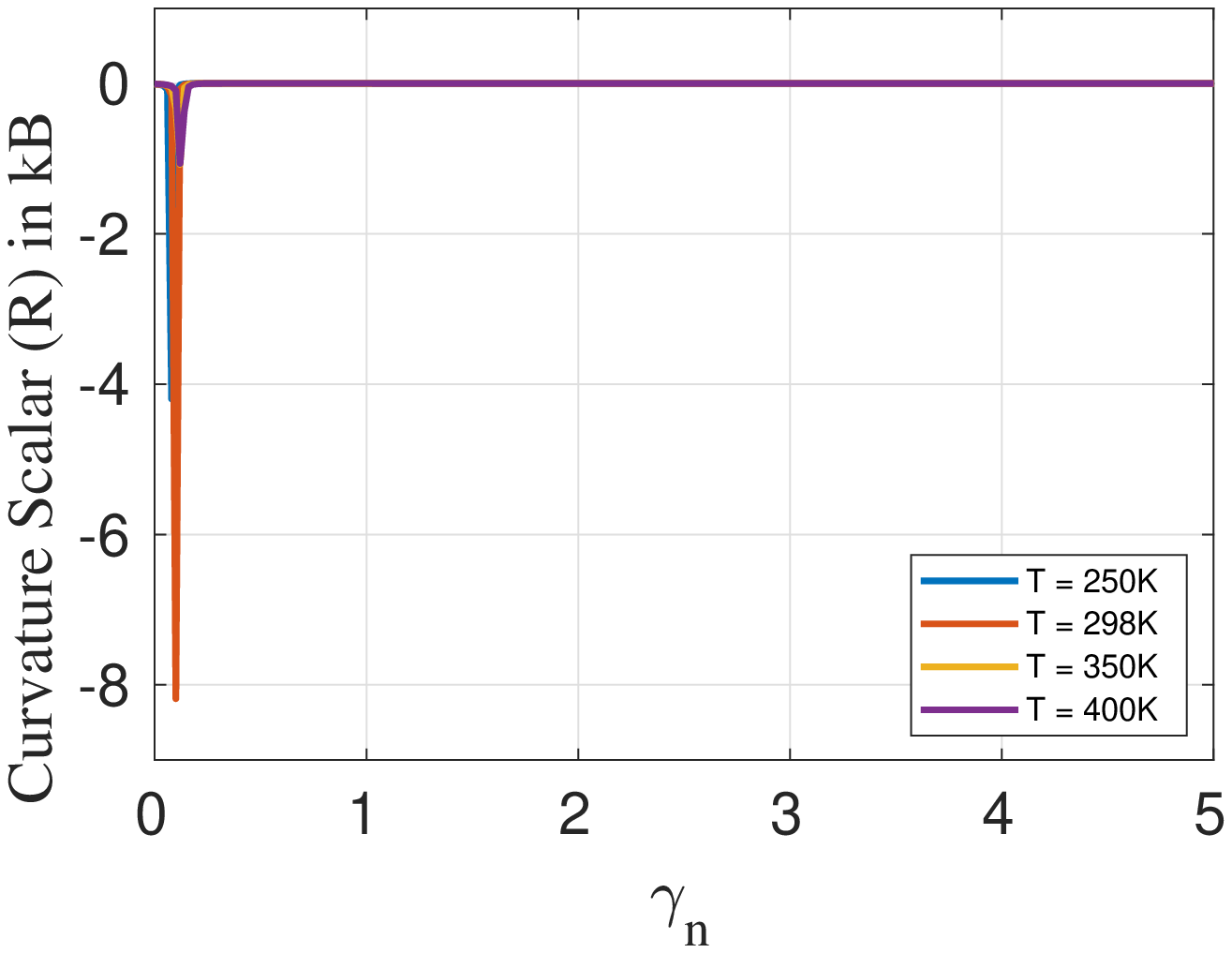}
         \caption{$\Omega = 0$}
         \label{Curvaturevsisostrainfortemperaturea}
     \end{subfigure}
     \hfill     
     \begin{subfigure}[b]{0.47\textwidth}
         \centering
         \includegraphics[width=0.7\textwidth]{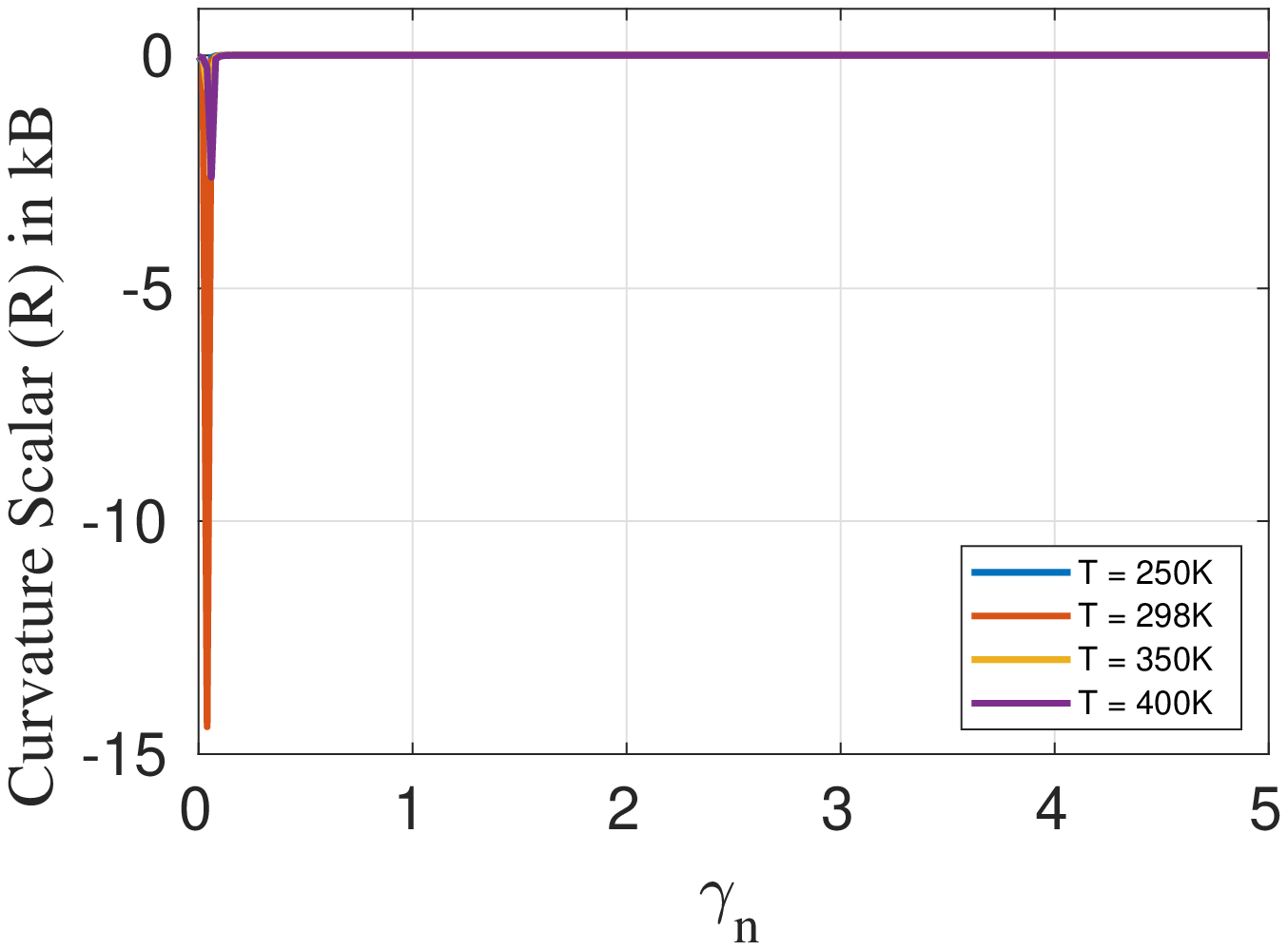}
         \caption{$\Omega = 0.15$}
         \label{Curvaturevsisostrainfortemperatureb}
     \end{subfigure}
     \hfill  
     \begin{subfigure}[b]{0.47\textwidth}
         \centering
         \includegraphics[width=0.7\textwidth]{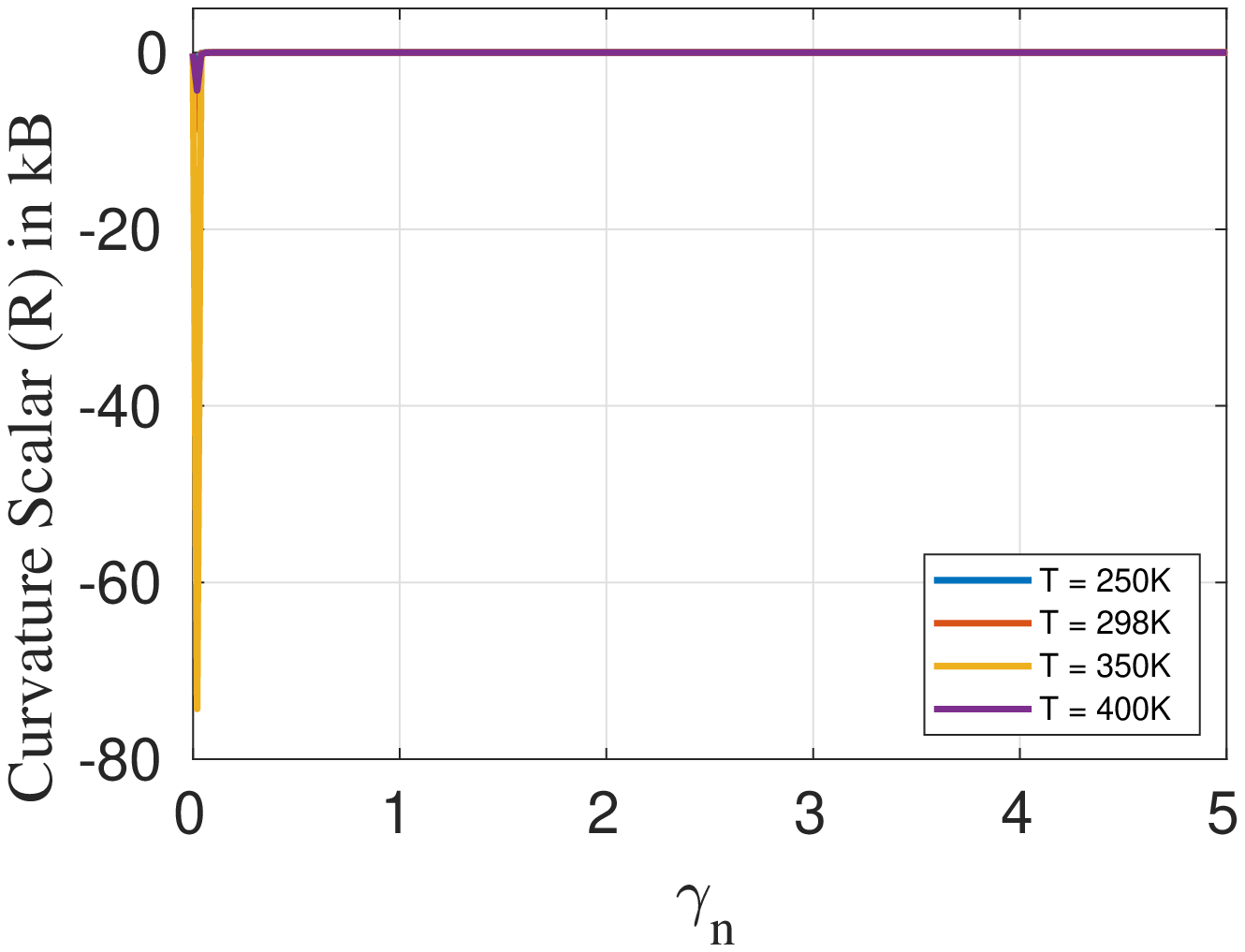}
         \caption{$\Omega = 0.20$}
         \label{Curvaturevsisostrainfortemperaturec}
     \end{subfigure}     
     \hfill  
     \begin{subfigure}[b]{0.47\textwidth}
         \centering
         \includegraphics[width=0.7\textwidth]{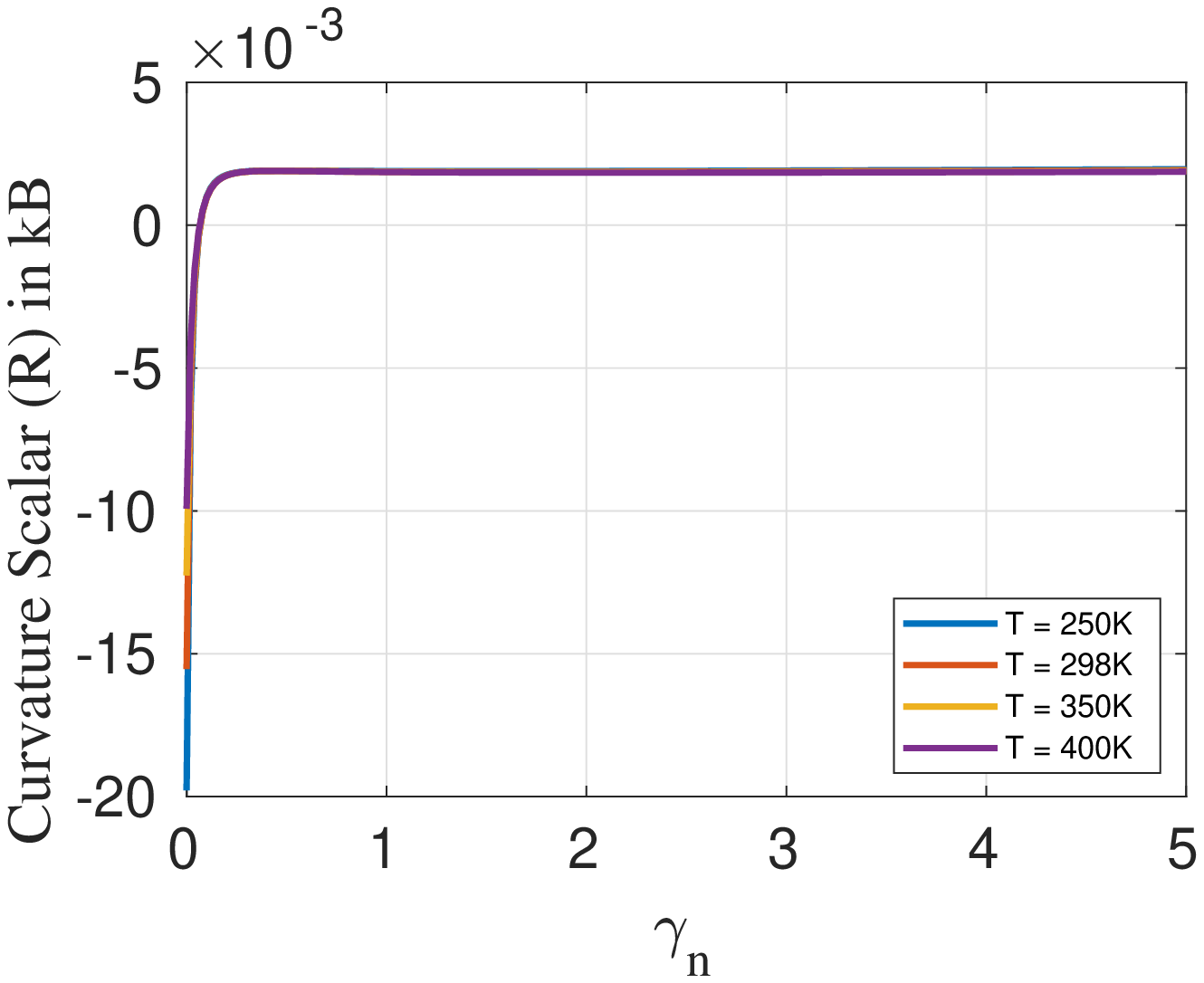}
         \caption{$\Omega = 0.30$}
         \label{Curvaturevsisostrainfortemperatured}
     \end{subfigure}
     \hfill     
     \begin{subfigure}[b]{0.47\textwidth}
         \centering
         \includegraphics[width=0.7\textwidth]{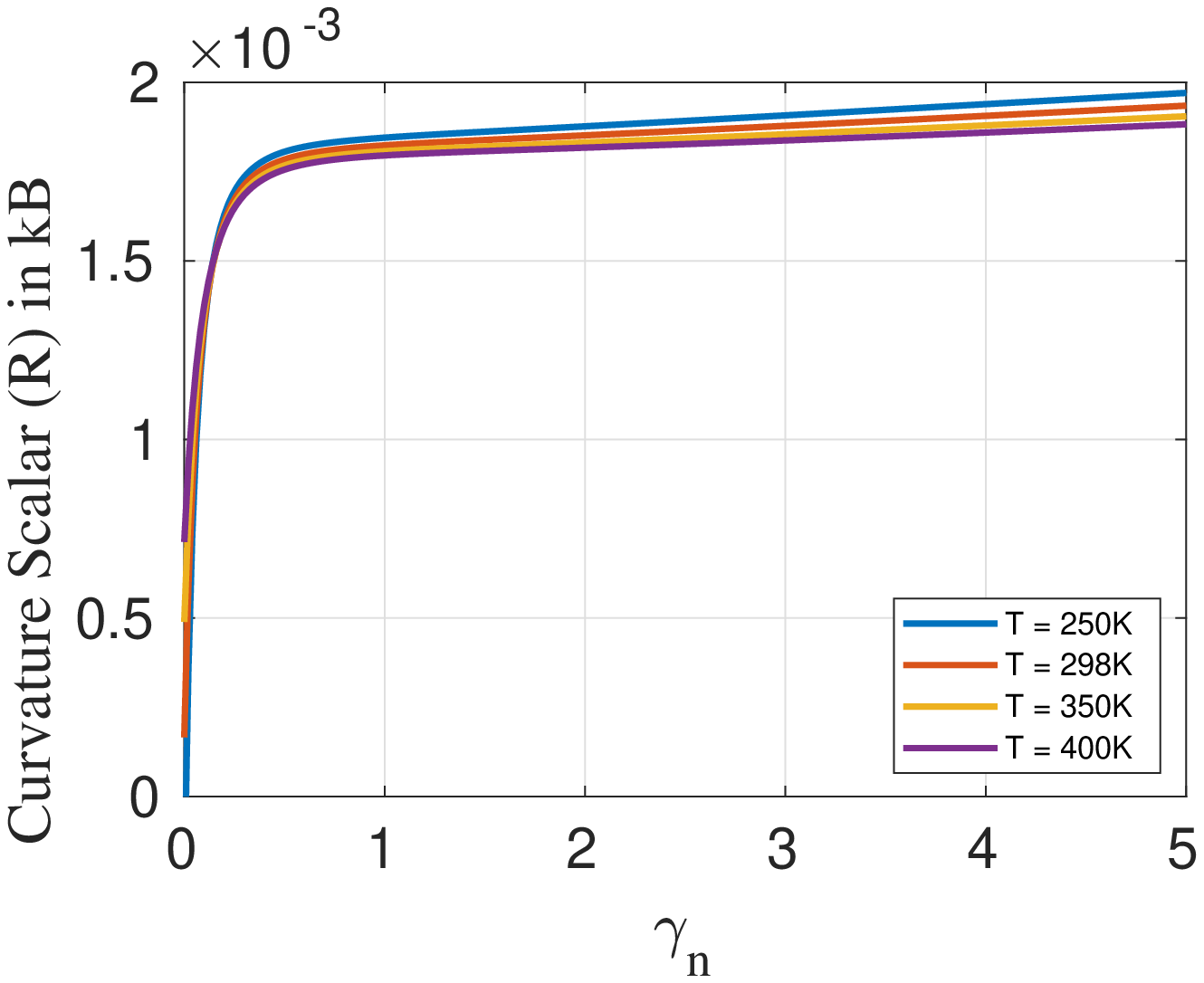}
         \caption{$\Omega = 0.40$}
         \label{Curvaturevsisostrainfortemperaturee}
     \end{subfigure}     
     \hfill     
     \begin{subfigure}[b]{0.47\textwidth}
         \centering
         \includegraphics[width=0.7\textwidth]{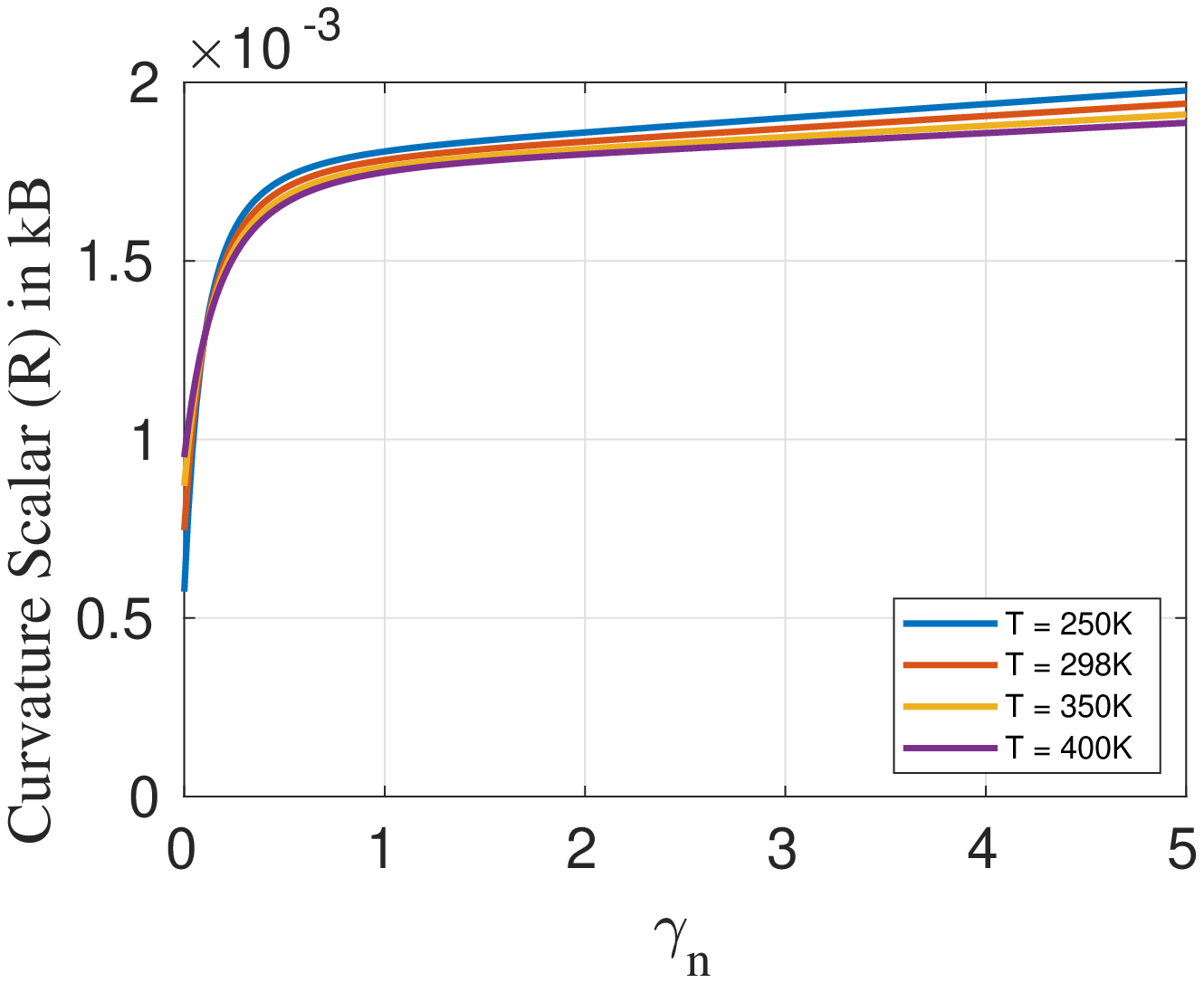}
         \caption{$\Omega = 0.45$}
         \label{Curvaturevsisostrainfortemperaturef}
     \end{subfigure}   
     \hfill     
     \begin{subfigure}[b]{0.47\textwidth}
         \centering
         \includegraphics[width=0.7\textwidth]{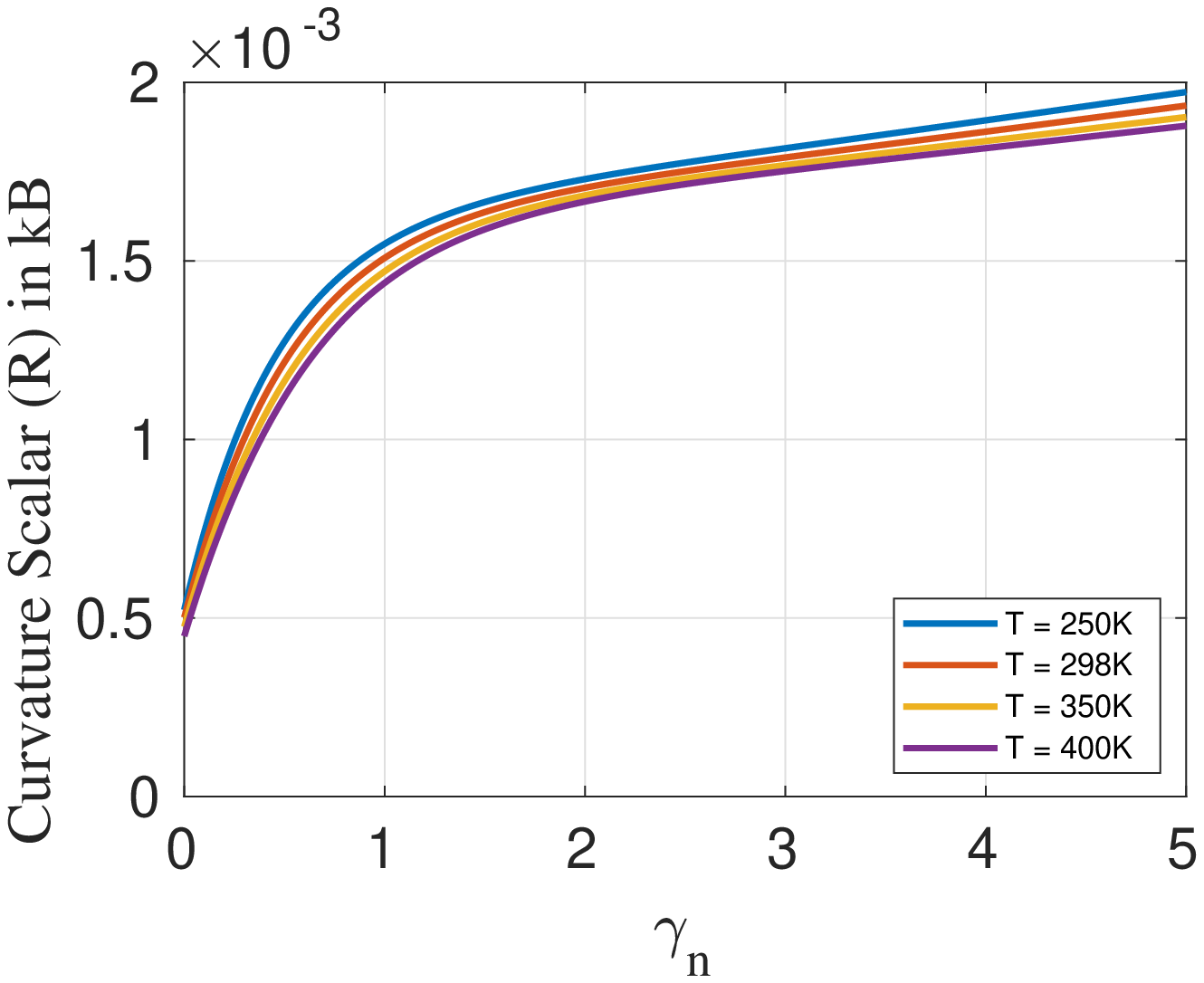}
         \caption{$\Omega = 0.60$}
         \label{Curvaturevsisostrainfortemperatureg}
     \end{subfigure}
     \hfill     
     \begin{subfigure}[b]{0.47\textwidth}
         \centering
         \includegraphics[width=0.7\textwidth]{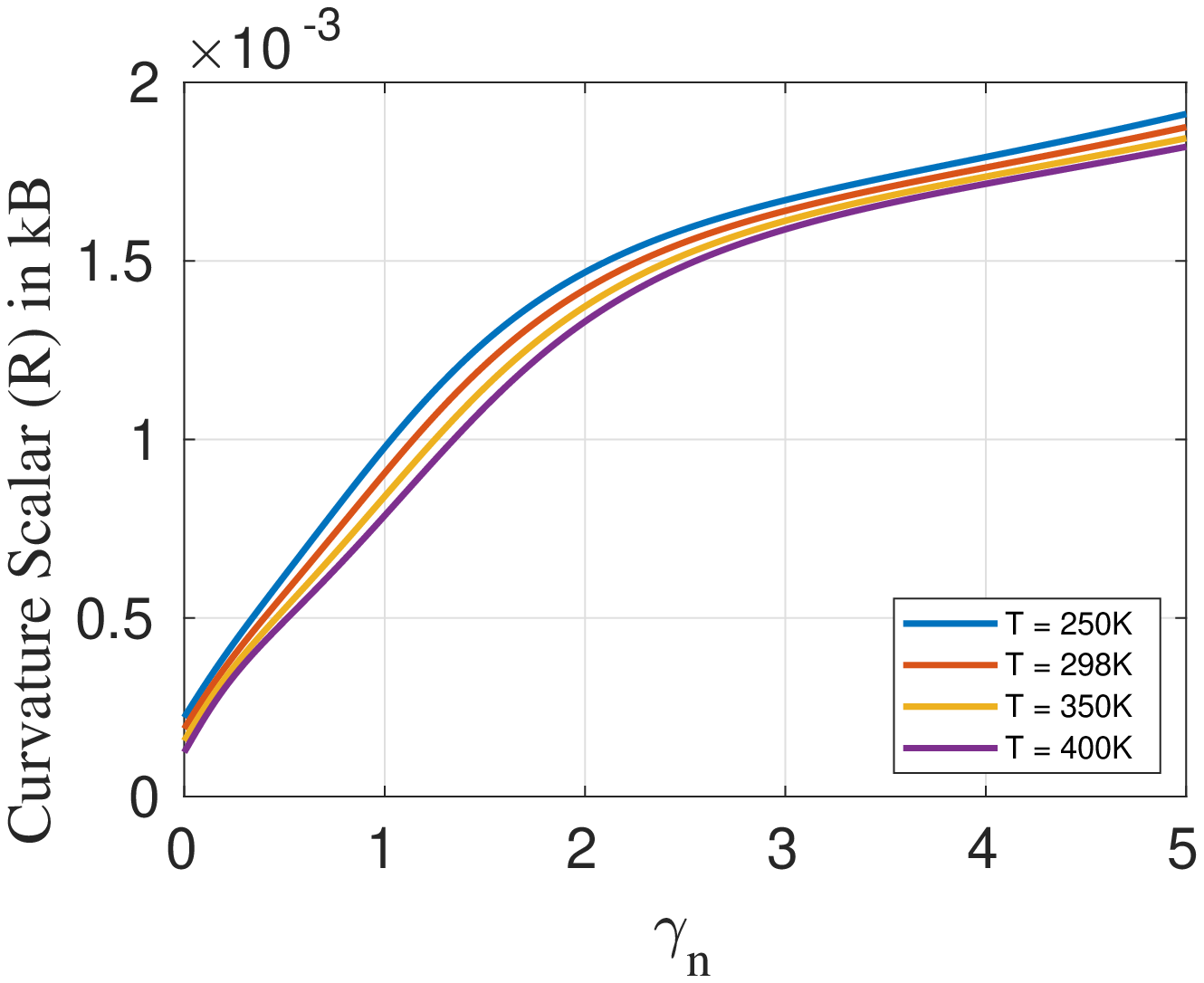}
         \caption{$\Omega = 0.70$}
         \label{Curvaturevsisostrainfortemperatureh}
     \end{subfigure}      
        \caption{Plots of scalar curvature $R$ with stretch $\gamma_n$ for varying temperature and crystallization ratio}
        \label{CurvatureNonisochorictemperature}
\end{figure*}

\begin{figure*}
     \centering
     \begin{subfigure}[b]{0.47\textwidth}
         \centering
         \includegraphics[width=0.8\textwidth]{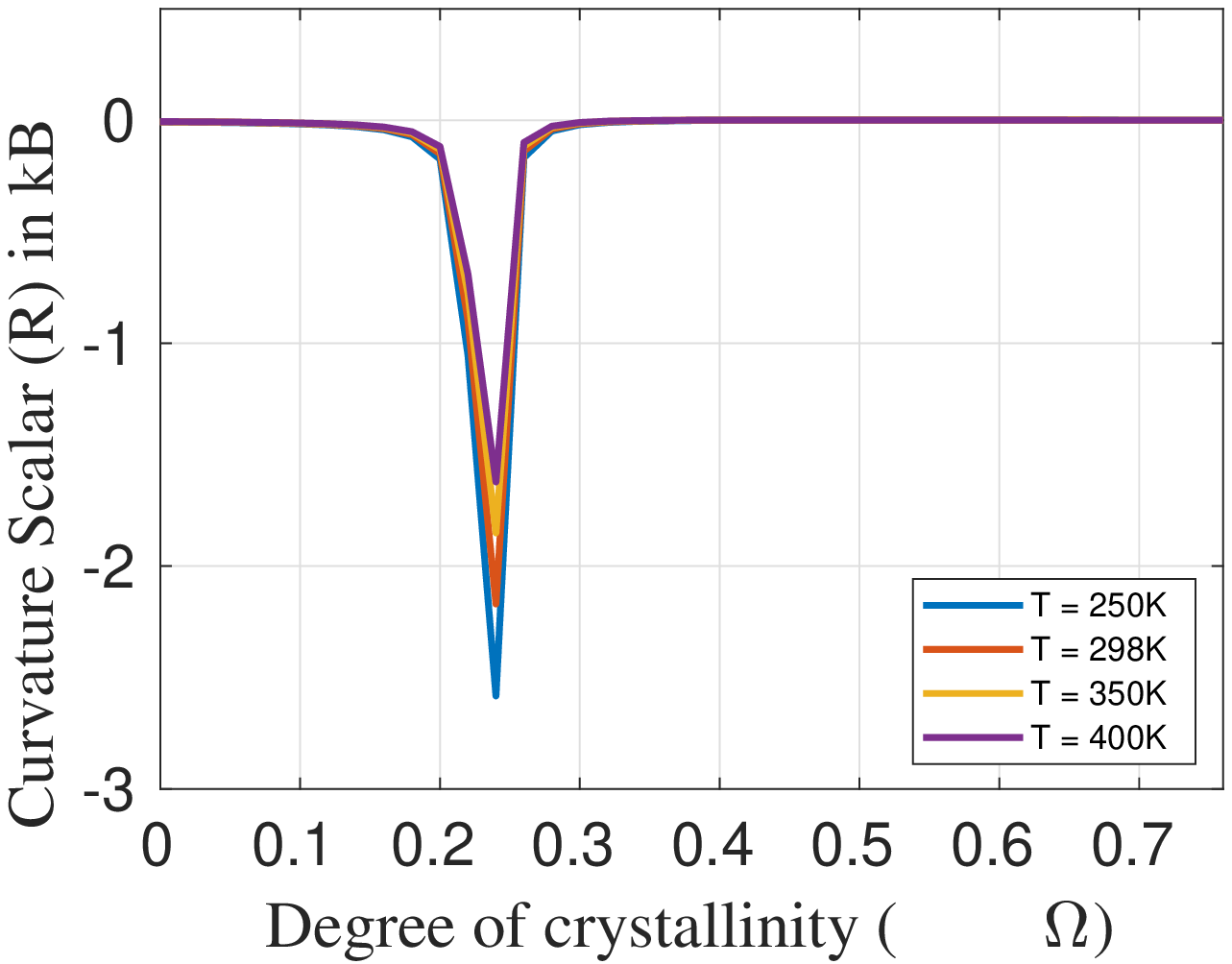}
         \caption{$\gamma_n = 0$}
         \label{CurvaturevsisostrainforGammaa}
     \end{subfigure}
     \hfill     
     \begin{subfigure}[b]{0.47\textwidth}
         \centering
         \includegraphics[width=0.8\textwidth]{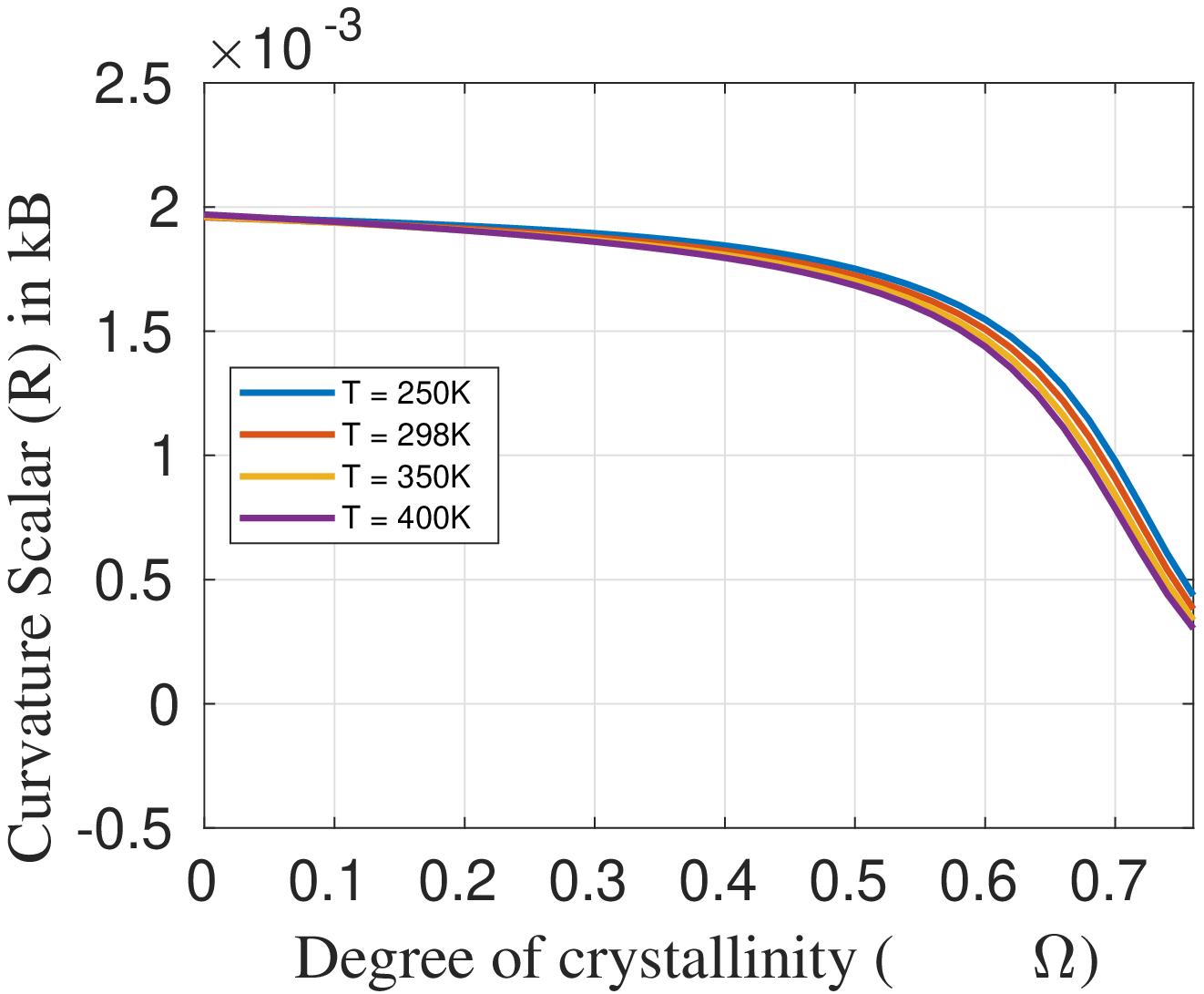}
         \caption{$\gamma_n = 1$}
         \label{CurvaturevsisostrainforGammab}
     \end{subfigure}
        \caption{Plots of scalar curvature $R$ vs. degree of crystallization $\Omega$ for varying temperature and $\gamma_n$}
        \label{CurvatureNonisochoricGamma}
\end{figure*}

Unlike the case of simple shear, we observe multiple negative peaks in each plot of Fig.~\ref{CurvaturNonisothermalTensiontemperature} which shows the variation of scalar curvature $R$ with tensile stretch $\Gamma$ for some chosen $\Omega$ values. Considering each plot, out of all the $\Gamma$ values at which the multiple peaks occur, we identify a single $\Gamma$ at which a peak is present for all temperatures. Thus this $\Gamma$ is a temperature independent state at which the curvature consistently displays criticality for a given $\Omega$. This is denoted as $\Gamma_{cr}$. The same can be said about the plots in Fig.~\ref{CurvaturNonisothermalTensioncrystallization} which shows the variation of scalar curvature $R$ with $\Omega$ for selected $\Gamma$ values and there are temperature independent $\Omega$s for all the $\Gamma$ values considered, which we denote as $\Omega_{cr}$.

Thus using the above plots, we can again determine pairs of ($\Gamma_{cr}$,$\Omega_{cr}$). The pairs thus graphically obtained are: ($1\text{,}0$), ($1\text{,}0.15$), ($1\text{,}0.22$), ($2\text{,}0.3$), ($3.26\text{,}0.45$), ($4.4\text{,}0.6$), ($5.2\text{,}0.7$). $\Gamma_{cr}$ for $\Omega<\frac{1}{N}$ is again $1$. The critical points satisfy condition Eq.~\eqref{conditionshear} as well. Also, the strain energy component in the free energy reduces to zero as evident in Fig.~\ref{FreeEnergy3DTension} and Fig.~\ref{DeviatoricFreeEnergy3DTension}. Hence we conclude that even for isochoric uniaxial tension, the critical state retains the same meaning as that in simple shear and that the infeasibility of any state ($\Gamma<\Gamma_{cr}$) for a given $\Omega_{cr}$ calls for the consideration of a spurious conformational energy and hence a subsequent reduction in the total free energy.

\subsection{Non-isochoric Deformation in \\a Hyperelastic Solid}\label{CurvatureTensionNonIsochoric}

For the previous cases, the volumetric component of free energy and the associated stress were absent either due to the applied deformation being isochoric (in simple shear) or a combination of conditions on the material response and the deformation (uniaxial tension in an incompressible solid). In both of them the critical pairs could be as well identified from the stress plots (Fig.~\ref{sigma12vsshearfortemperaturenoniso} and Fig.~\ref{sigma11vsstretchfortemperature}) as the stress at those states reduces to zero. 

We now consider a case where the volumetric component contributes to the total free energy and stress and investigate the information imparted by the scalar curvature. So, this case corresponds to non-isochoric deformation in a compressible solid. Let the deformation gradient be given by, 

\begin{equation}
\mathbf{F} = 
\begin{bmatrix}
1+\gamma_n & 0 & 0 \\
0 & 1 & 0 \\
0 & 0 & 1
\end{bmatrix}
\end{equation}
The 8-dimensional space is reduced to a 3-dimensional manifold similar to the earlier cases and the components of the right Cauchy-Green tensor are related to strain $\gamma_n$ through $C_{11} = (1+\gamma_n)^{2}$, $C_{22} = C_{33} = 1$. The other components of the tensor are $0$. Similar to the isochoric cases, we determine the variation of scalar curvature across the phase space through the plots in Fig.~\ref{CurvatureNonisochorictemperature}.

\begin{figure*}
    \centering
    \includegraphics[width=0.85\textwidth]{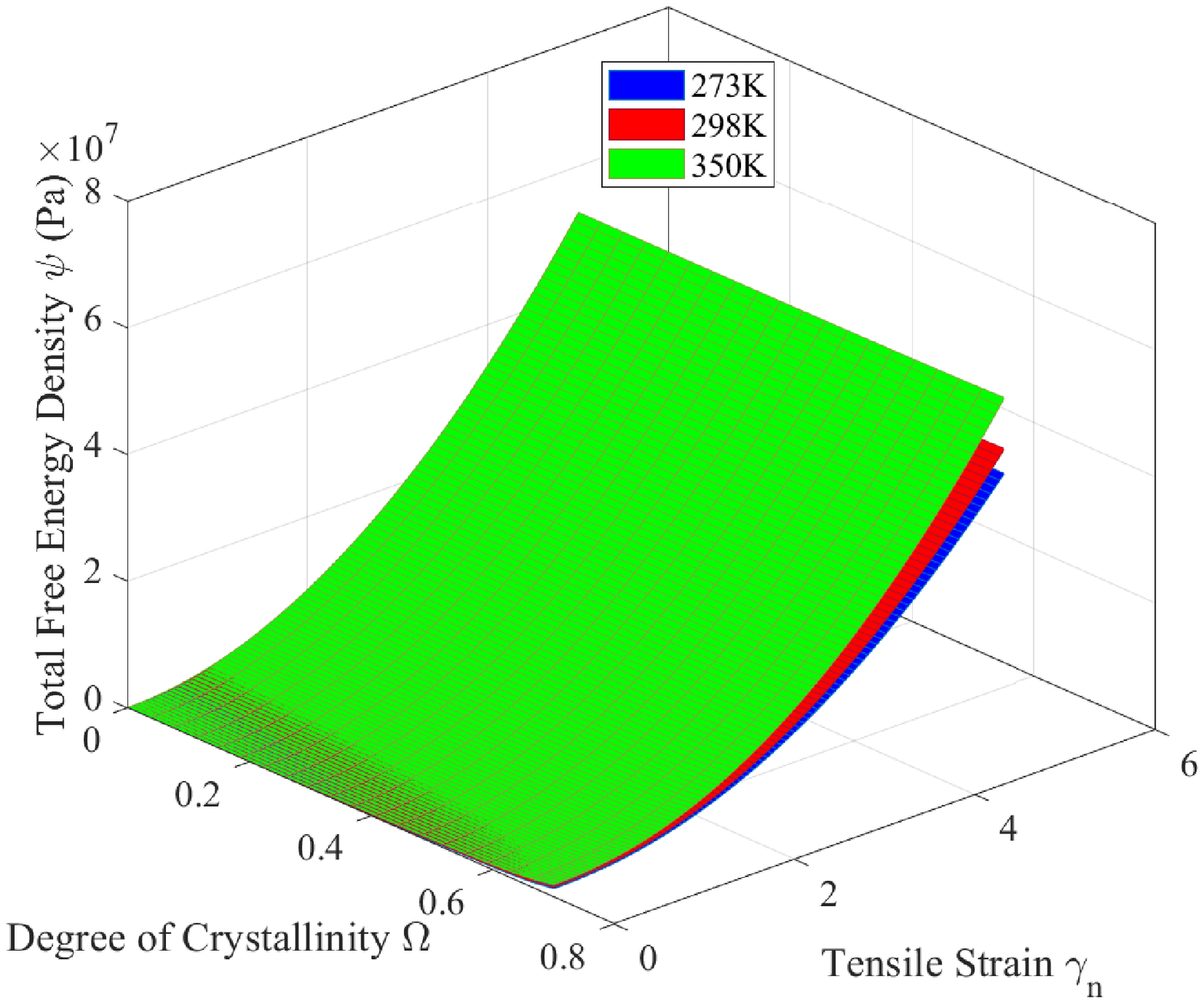}
    \caption{Free energy surfaces for different temperatures $T$ for nonisochoric deformation}
    \label{NonisochoricFreeEnergy3D}
\end{figure*}

\begin{figure*}
    \centering
    \includegraphics[width=0.8\textwidth]{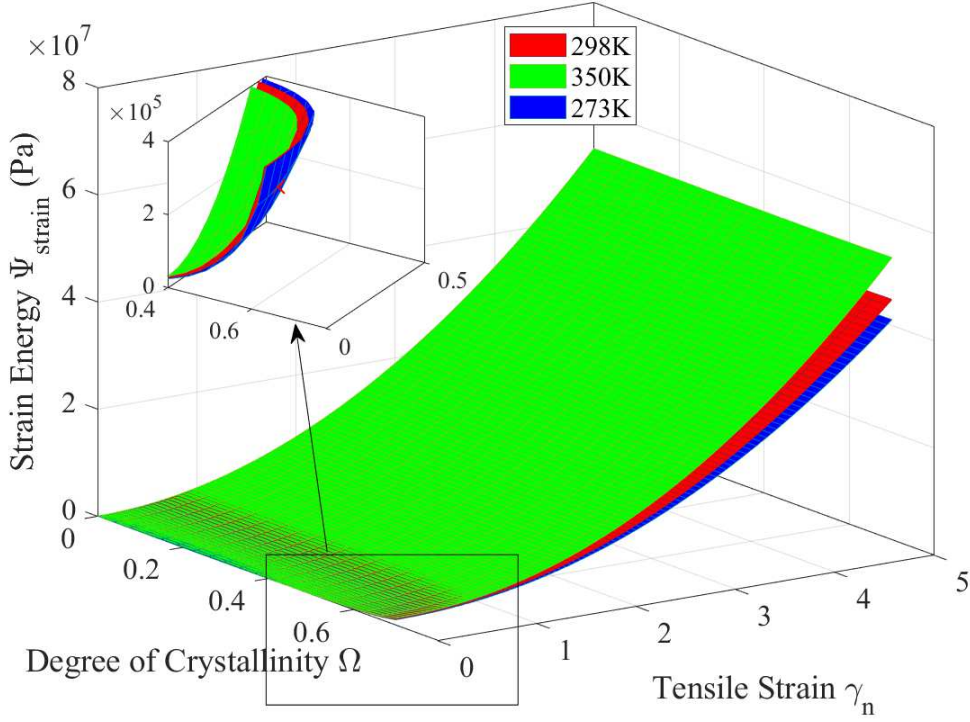}
    \caption{Strain energy surfaces for different temperatures $T$ for nonisochoric deformation}
    \label{NonisochoricStrainEnergy3D}
\end{figure*}

\begin{figure*}
     \centering
     \begin{subfigure}[b]{0.47\textwidth}
         \centering
         \includegraphics[width=0.7\textwidth]{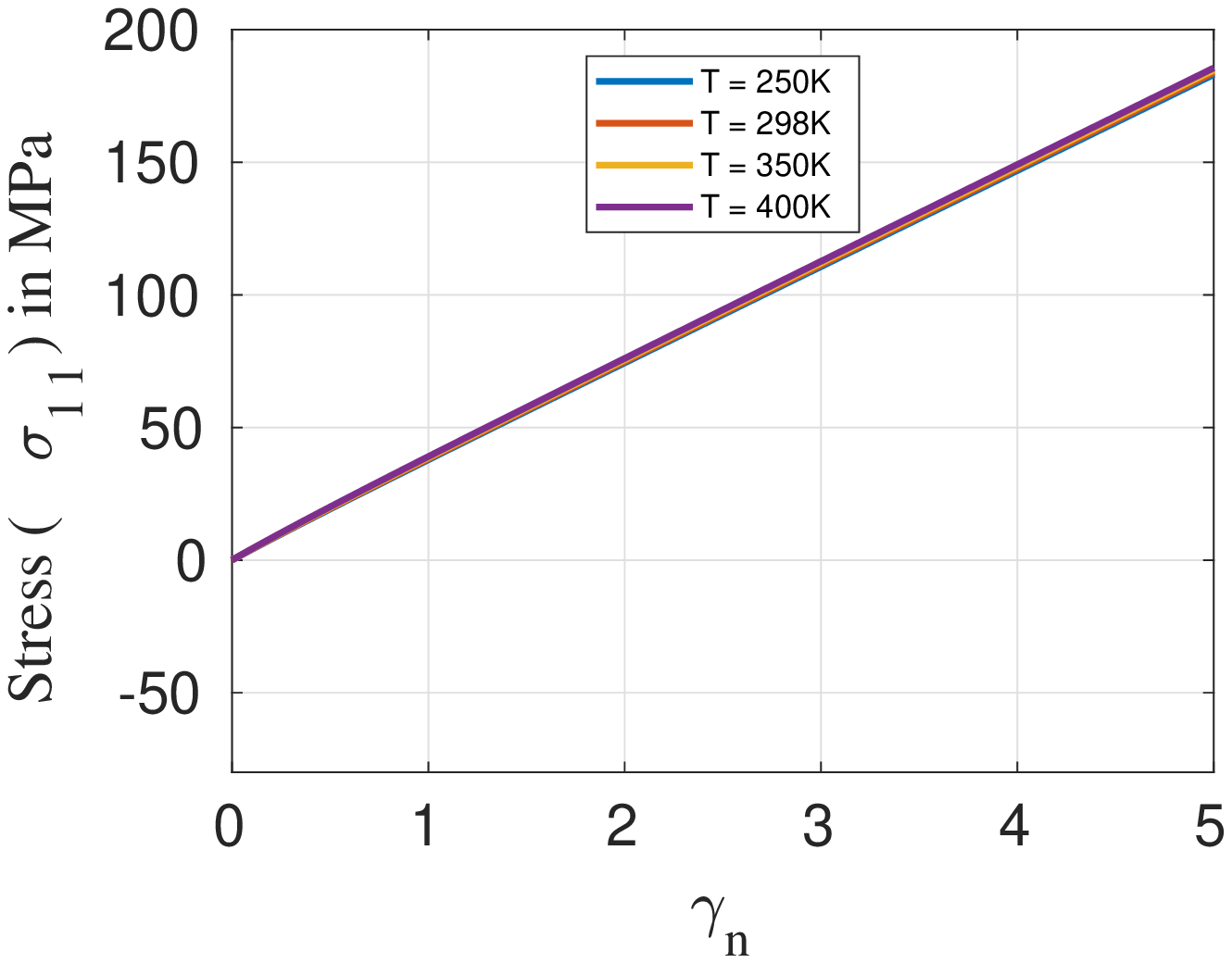}
         \caption{$\Omega = 0$}
         \label{Stressvsnonisostrainfortemperaturea}
     \end{subfigure}
     \hfill     
     \begin{subfigure}[b]{0.47\textwidth}
         \centering
         \includegraphics[width=0.7\textwidth]{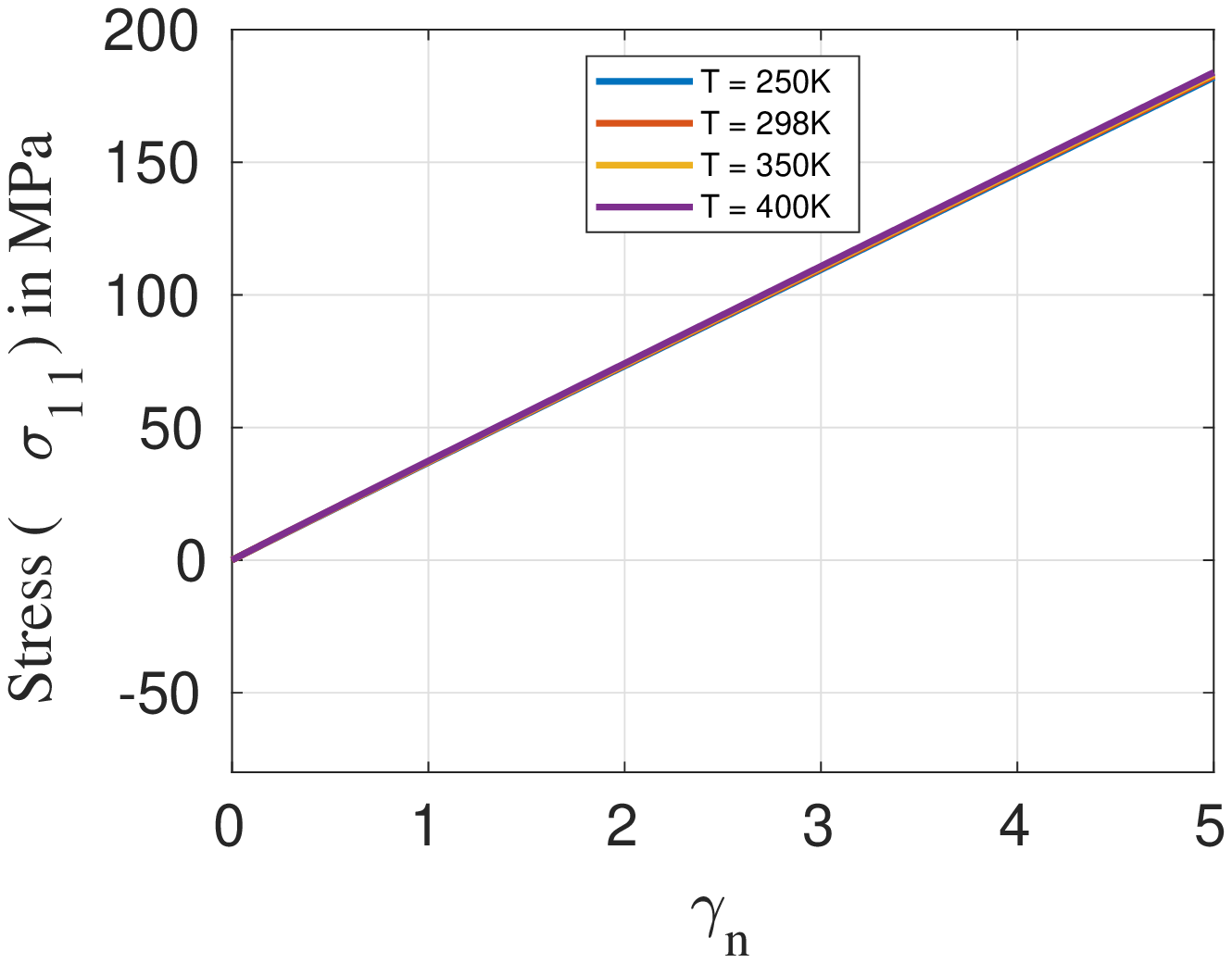}
         \caption{$\Omega = 0.15$}
         \label{Stressvsnonisostrainfortemperatureb}
     \end{subfigure}
     \hfill      
     \begin{subfigure}[b]{0.47\textwidth}
         \centering
         \includegraphics[width=0.7\textwidth]{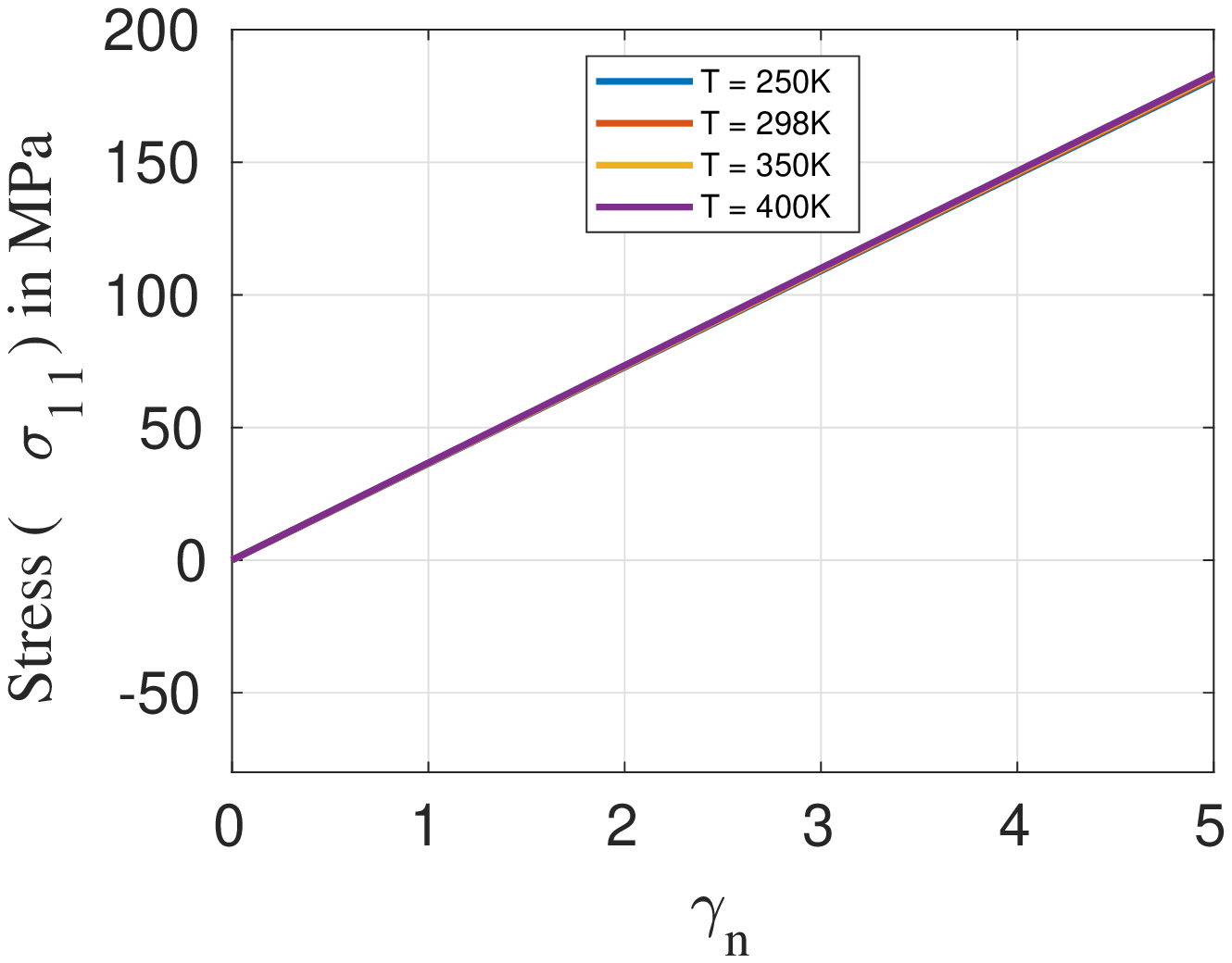}
         \caption{$\Omega = 0.20$}
         \label{Stressvsnonisostrainfortemperaturec}
     \end{subfigure}     
     \hfill      
     \begin{subfigure}[b]{0.47\textwidth}
         \centering
         \includegraphics[width=0.7\textwidth]{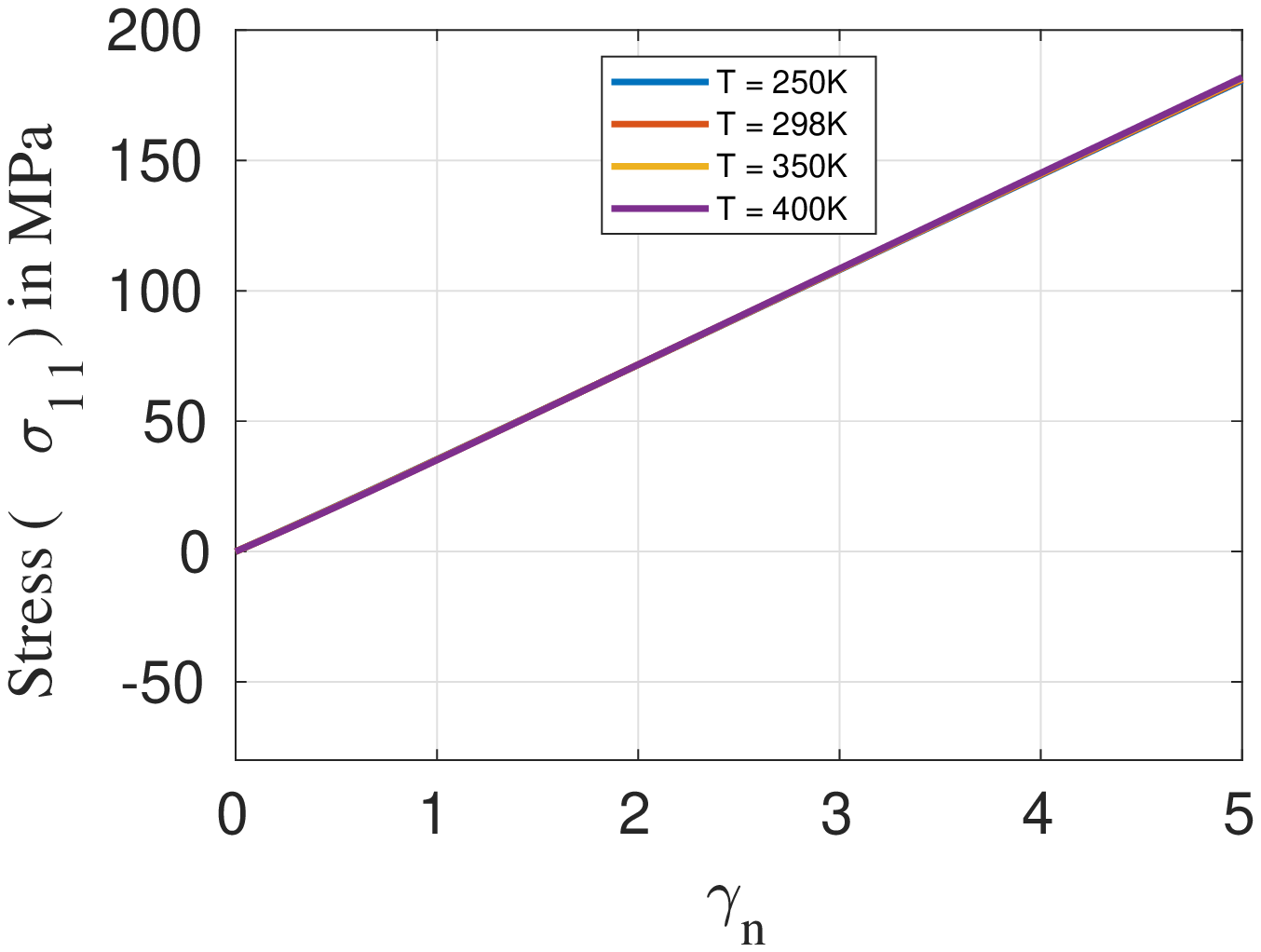}
         \caption{$\Omega = 0.30$}
         \label{Stressvsnonisostrainfortemperatured}
     \end{subfigure}
     \hfill     
     \begin{subfigure}[b]{0.47\textwidth}
         \centering
         \includegraphics[width=0.7\textwidth]{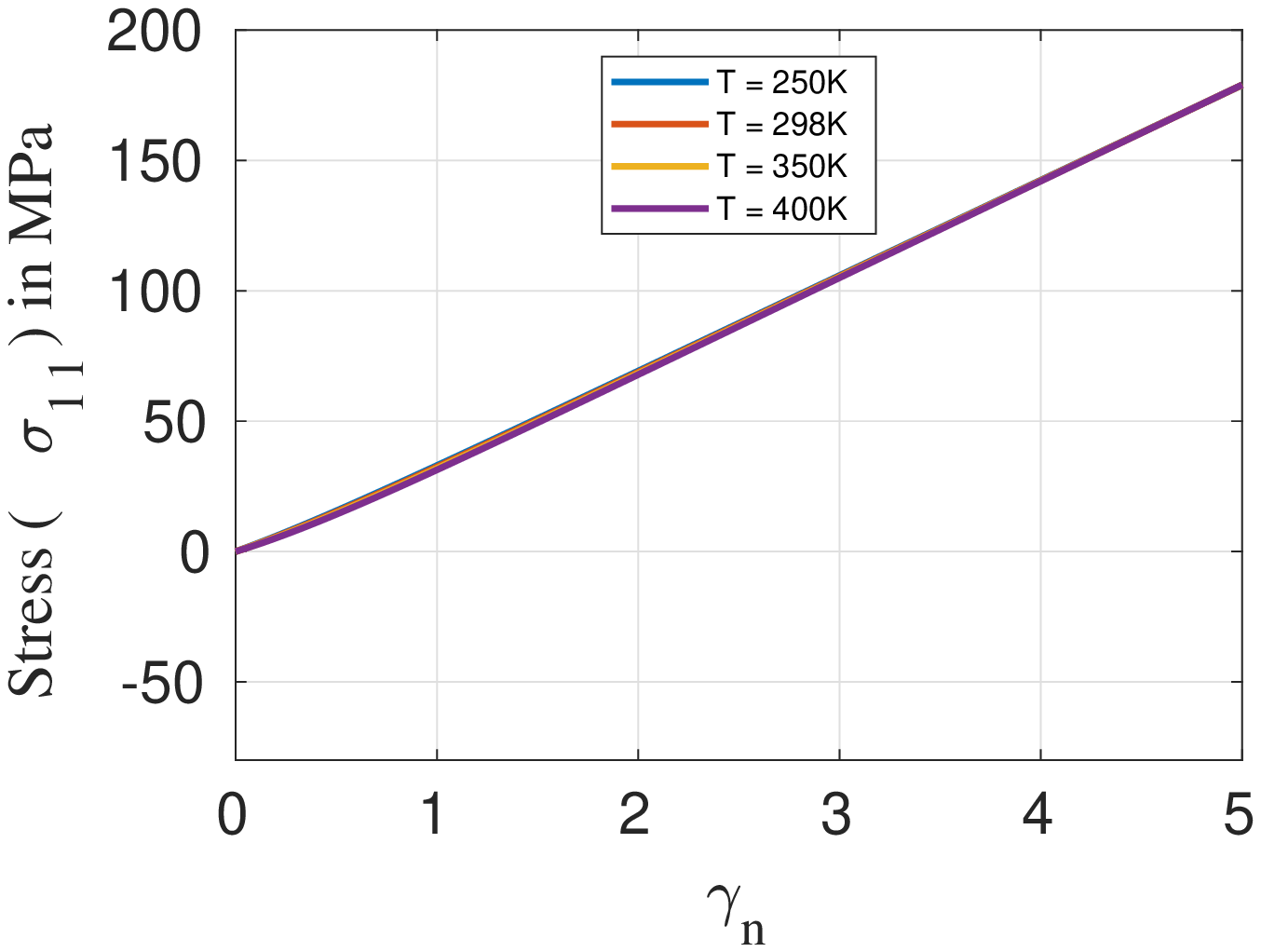}
         \caption{$\Omega = 0.45$}
         \label{Stressvsnonisostrainfortemperaturee}
     \end{subfigure}   
     \hfill     
     \begin{subfigure}[b]{0.47\textwidth}
         \centering
         \includegraphics[width=0.7\textwidth]{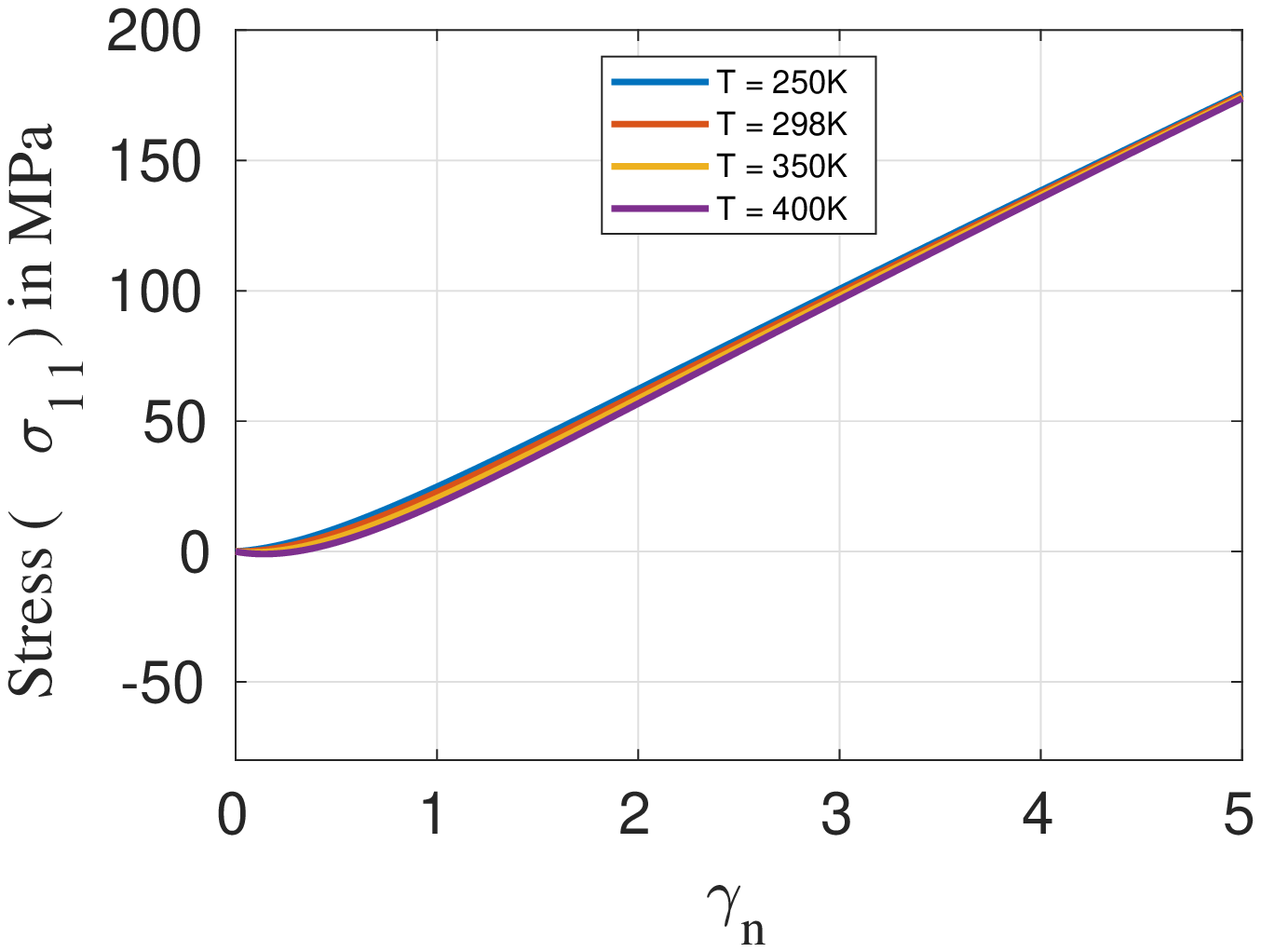}
         \caption{$\Omega = 0.60$}
         \label{Stressvsnonisostrainfortemperaturef}
     \end{subfigure} 
     \hfill     
     \begin{subfigure}[b]{0.47\textwidth}
         \centering
         \includegraphics[width=0.7\textwidth]{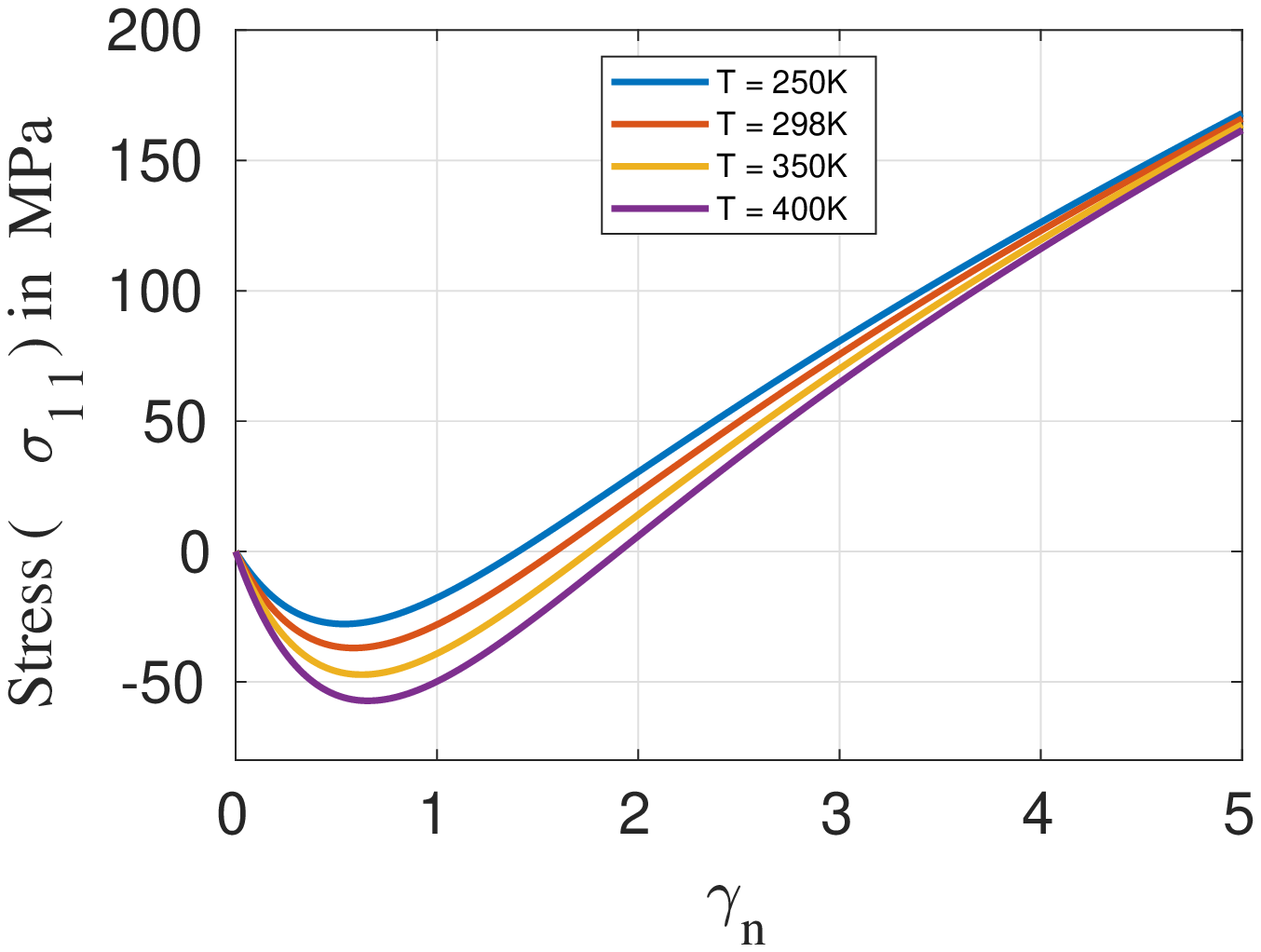}
         \caption{$\Omega = 0.70$}
         \label{Stressvsnonisostrainfortemperatureg}
     \end{subfigure}      
        \caption{Plots of $\mathbf{\sigma_{11}}$ vs. stretch $\gamma_n$ for varying  temperature and crystallization ratio}
        \label{StressNonisochorictemperature}
\end{figure*}

\begin{figure*}
    \centering
    \includegraphics[width=0.8\textwidth]{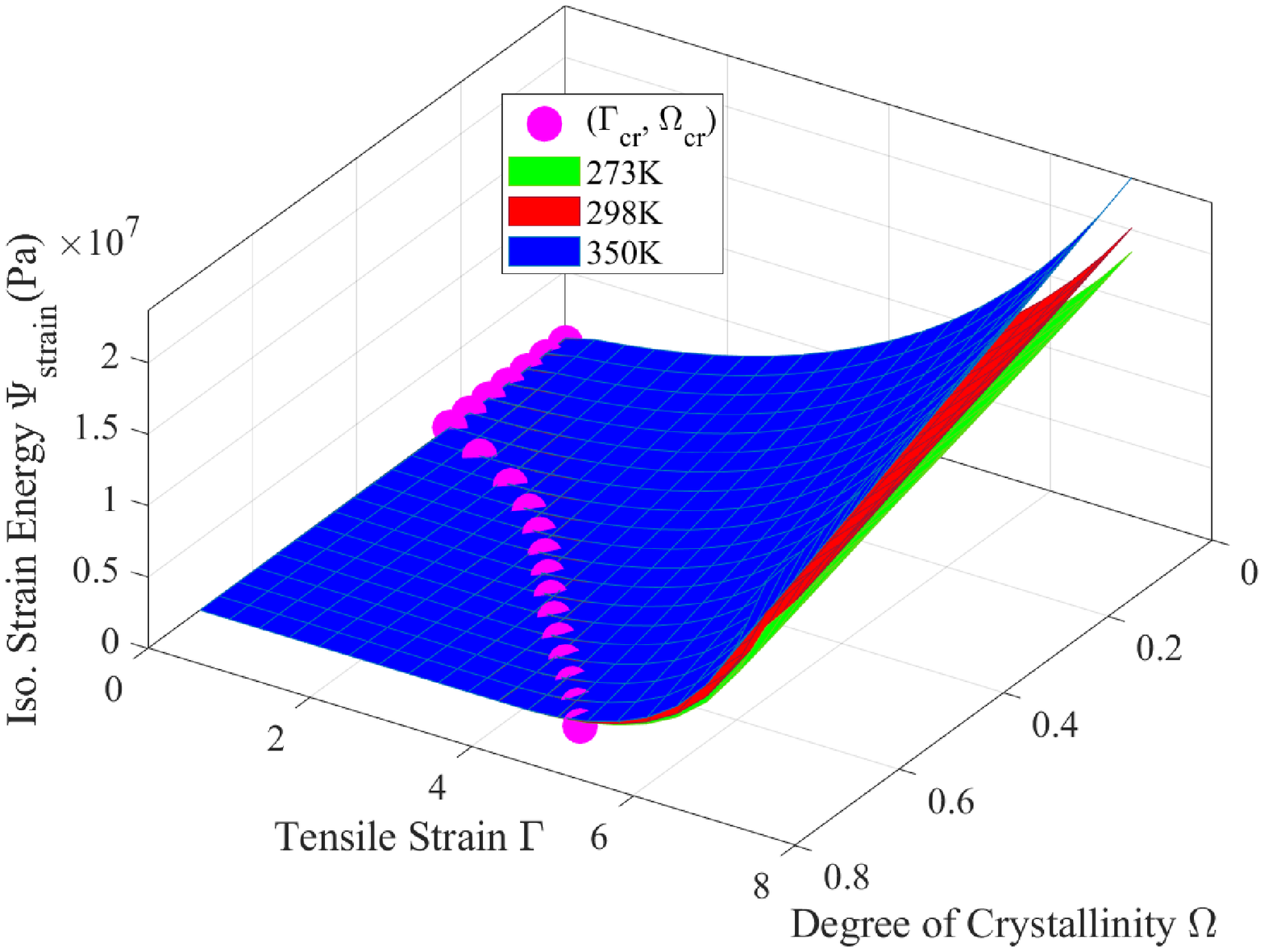}
    \caption{Strain energy for isochoric uniaxial tension after removal of spurious component}
    \label{DeviatoricFreeEnergy3DtruncatedIsochoric}
\end{figure*}

In the plots of Fig.~\ref{CurvatureNonisochorictemperature}, non-zero curvature peaks are observed only in the vicinity of or at $\gamma_n=0$ for all $\Omega$ values.The strain upto which the curvature remains non-zero is dependent on $\Omega$ and increases with it. From the plot corresponding to $\gamma_n=0$ in Fig.~\ref{CurvatureNonisochoricGamma} we obtain a negative peak at $\Omega=0.23$. Peaks are observed at this location irrespective of the temperature. Recall that in isochoric cases, we had obtained peaks at the same $\Omega$ value in plots of scalar curvature vs $\Omega$ for $\gamma=0$ and $\Gamma=1$ in Fig.~\ref{CurvaturNonisothermalShearcrystallization} and Fig.~\ref{CurvaturNonisothermalTensioncrystallization} respectively. We had inferred that when $\Omega<\frac{1}{\sqrt{N}}$, the critical values of the strain or the stretch should be $\gamma_{cr}=0$ and $\Gamma_{cr}=1$ respectively. Invoking the same rationale, we can conclude that the critical value of strain in this case must be 0, i.e. $\gamma_{ncr}=0$.

But, we cannot identify the critical strain for $\Omega>\frac{1}{\sqrt{N}}$ from the curvature plots in Fig.~\ref{CurvatureNonisochorictemperature} and Fig.~\ref{CurvatureNonisochoricGamma} alone. In order to remove the spurious energy that evidently is present as indicated by the non-zero curvature and strain energy at zero strain (See Fig.~\ref{CurvatureNonisochorictemperature} and  Fig.~\ref{NonisochoricStrainEnergy3D}), we must resort to other conditions such as those derived from the mathematical expression of the $\Psi_{strain}$. Such conditions cannot be derived graphically.

 Upon examination of the expressions of $\Psi_{strain}$, $\Psi_{vol}$ and $\Psi_{dev}$, it is evident that the spurious energy is of the same form as $\Psi_{dev}$ as $\Psi_{vol}$ is identically zero. The condition under which $\Psi_{dev}=0$, is same as that stated in Eq.~\eqref{conditionshear}. Thus, for the non-isochoric case as well, we may assume that Eq.~\eqref{conditionshear} defines the bounds on the spurious energy. Any pair of ($\gamma_{n}$,$\Omega$) which satisfies condition Eq.~\eqref{conditionshear} may thus be regarded as ($\gamma_{ncr}$,$\Omega_{cr}$).

It should be noted that the uniaxial stress $\sigma_{11}$ shown in Fig.~\ref{StressNonisochorictemperature} does not impart any significant information on the bounds. This is probably due to the dominating contribution of the volumetric component of stress. 

In both isochoric and non-isochoric deformation for a chosen $\Omega$, there exists a $\gamma_{cr}$ or a $\Gamma_{cr}$ or a $\gamma_{ncr}$, such that strain in excess of the critical value only contributes to the conformational isochoric strain energy. Having clarified the physical significance of these critical states, we must focus on how the spurious energy must functionally depend on the strain measure involving $\gamma_{cr}$, $\Gamma_{cr}$ or $\gamma_{ncr}$. As already emphasised, this energy is conformational and must be purely of the form $\Psi_{dev}$ to effect a nonzero strain energy at zero strains. 

\section{Modified Free Energy}\label{SIC_Modified_Free_Energy}

For both isochoric and non-isochoric cases, based on the arguments in the preceding sections, any strain leading to $\lambda_a<\lambda_{acr}$ contributes to the spurious energy where $\lambda_{acr}$ satisfies the following condition, 
\begin{equation}\begin{split}\label{IsochoricCondition}
(\Omega \sqrt{N}-\lambda_{acr})H(\Omega-\frac{1}{\sqrt{N}})+(1-\lambda_{acr})H(\frac{1}{\sqrt{N}}-\Omega)=0
\end{split}\end{equation}
The spurious energy which is the isochoric contribution due to $\lambda_a<\lambda_{acr}$ accordingly takes the form  
\begin{equation}\begin{split}\label{IsochoricSpurious}
\Psi_{spur} = k_{B}TNn(1-\Omega)H(\lambda_{acr}-\lambda_a)\\\times\left(\dfrac{3}{2}\Lambda_a^2+\dfrac{9}{20}\Lambda_a^4+\dfrac{99}{350}\Lambda_a^6\right)
\end{split}\end{equation}
The final free energy density is thus given by:

\begin{equation}\label{ModifiedFreeEnergy}
\Psi = \Psi_{vol}+\Psi_{dev}-\Psi_{spur}+\Psi_{cr}+\Psi_{surr}
\end{equation}

\begin{figure*}
    \centering
    \includegraphics[width=0.8\textwidth]{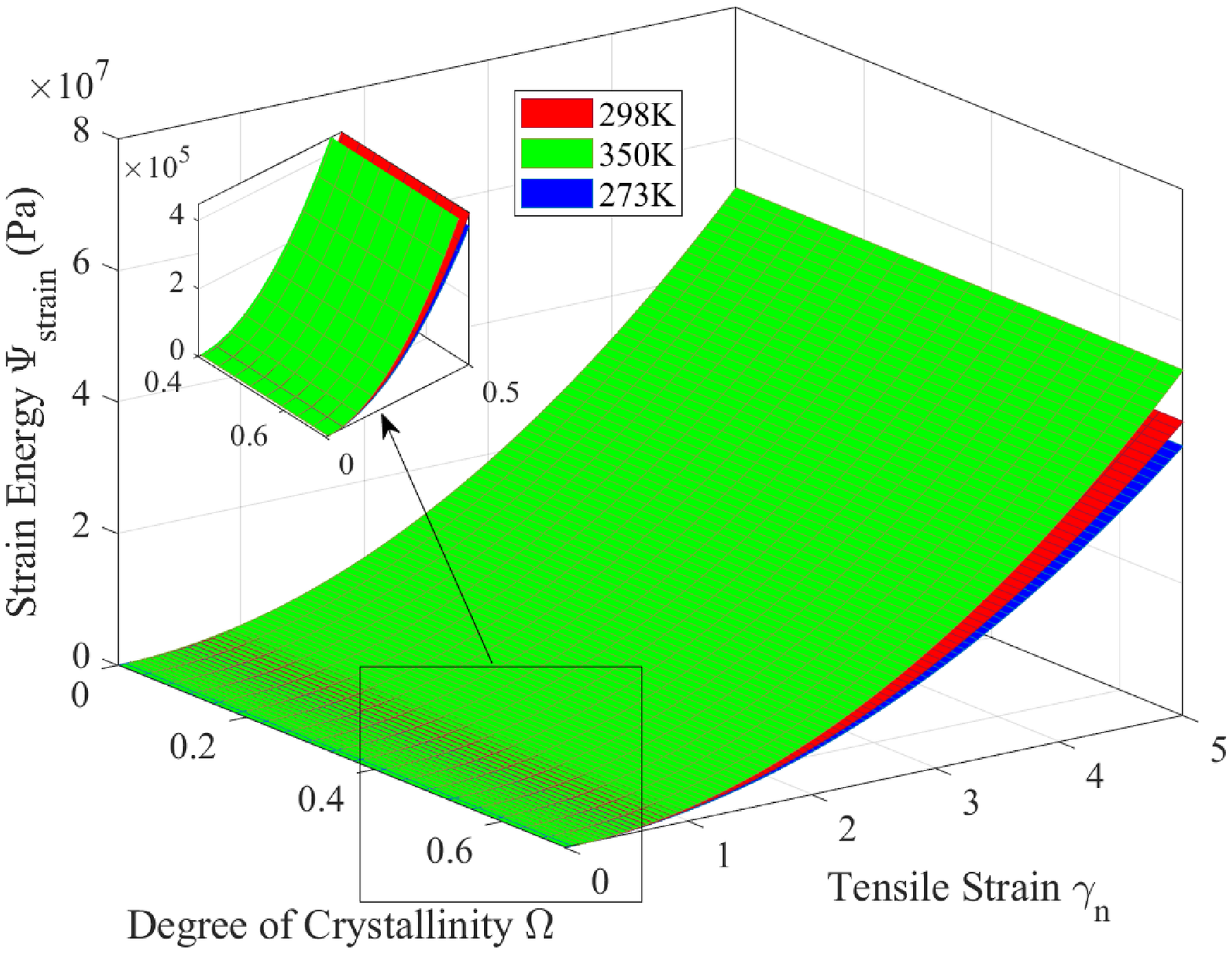}
    \caption{Strain energy for uniaxial tension of a compressible solid after removal of spurious component}
    \label{DeviatoricFreeEnergy3DtruncatedNonIsochoric}
\end{figure*}

The strain energy plots in Fig.~\ref{DeviatoricFreeEnergy3DtruncatedIsochoric} and Fig.~\ref{DeviatoricFreeEnergy3DtruncatedNonIsochoric} depict the corrected strain energy variation, where the spurious energy has been removed. This free energy may now be used to construct a constitutive theory for strain-induced crystallization in elastomers where stresses due to unphysical conformational energy are absent. Also, the energy driving the crystallization will not lead to an exaggerated crystallization ratio. In previous theories for strain induced crystallization such as \cite{rastak2018non}, no such physical and systematic approach has been used to eliminate these unphysical components. The fundamental equations in these theories do not include any provisions for correcting the quantities, thus rendering the constitutive theories incomplete.

\section{Conclusion}\label{Conclusion}

The classical fluctuation theory reveals significant information on underlying chain conformations in equilibrated polymers undergoing elastic deformation. However, in the case of inelastic deformation where relaxation occurs slowly enough for the hypothesis of local equilibrium to hold, the fluctuation theory may be still used to extract microstructural information regarding the thermodynamic system. One such information would be regarding the proliferation and transformation of defects or incompatibilities responsible for the non-equilibrium response of the system. Additionally, if the fluctuation theory in consideration should be geometrically consistent in the sense of \cite{ruppeiner1979thermodynamics}, an additional quantity is at our disposal - the Ricci curvature. 

In this article we have tried to apply such a fluctuation theory to a thermodynamic system describing strain-induced-crystallization in elastomers. Instead of the fluctuation theorem, we are interested in the Riemannian curvature and the information it has to offer. For this thermodynamic system, the curvature imparts vital information on a spurious, crystallization-induced strain that results in a spurious isochoric energy. In the absence of a suitable correction, this energy leads to an unphysical stress and an erroneous evolution of crystallization. We determine this spurious strain from the curvature plots for both isochoric and non-isochoric strain states, though in the case of non-isochoric strains the graphically obtained information is not complete. We establish their unphysicality through an examination of the free energy and strain energy landscapes. Using this strain, we construct a modified free energy. 

It is evident from our study that the curvature does prove to be a vital quantity containing essential information on the physical viability of the thermodynamic states across the thermodynamic space even in non-equilibrium processes. In the case of crystallization, the curvature points to states with unphysical stretch components attributable to the geometric inconsistency whilst defining the stretch measures, which contributes to spurious strain energies. Thus the curvature yields constraints, extraneous to the constitutive relations of the thermodynamic forces and the evolution equations of internal variables and temperature. These constraints then need to be satisfied along with the balance laws plus the constitutive laws. The utility of curvature however is applicable to non-equilibrium processes where local equilibrium hypothesis can be applied. Hence one needs to be cautious regarding the timescale associated with the process.

Our future efforts would be to investigate the importance of curvature in other non-equilibrium phenomena. Specifically, we would like to study the implications of such a geometrically motivated fluctuation theory for glassy polymers undergoing viscoplastic deformation.



\nocite{*}

\bibliography{apssamp}
\end{document}